\documentclass[letterpaper,twocolumn,10pt]{article}
\usepackage{zhanggroup}

\usepackage{tikz}
\usepackage{amsmath}
\usepackage{amssymb}
\usepackage{multirow}
\usepackage{color}
\usepackage{xcolor}
\usepackage{url}
\usepackage{caption}
\usepackage{subcaption}
\usepackage{float}
\usepackage{balance}
\usepackage{graphicx}
\usepackage{textcomp}
\usepackage{booktabs}
\usepackage{colortbl}
\usepackage[ruled,linesnumbered]{algorithm2e}

\usepackage[absolute]{textpos}

\newcommand{\mypara}[1]{\smallskip\noindent\textbf{#1}}
\newcommand{\MIA}{\mathcal{I}}
\newcommand{\MLModel}{\mathcal{M}}

\newcommand{\DataPoint}{x}

\newcommand{\AttackModel}{\mathcal{A}}

\newcommand{\system}{\textsc{TrajectoryMIA}\xspace}

\begin{document}

\begin{textblock}{15}(1.9,1)
To Appear in 2022 ACM SIGSAC Conference on Computer and Communications Security, November 2022
\end{textblock}

\title{\Large \bf Membership Inference Attacks by Exploiting Loss Trajectory}

\author{
Yiyong Liu \ \ \ Zhengyu Zhao \ \ \ Michael Backes \ \ \ Yang Zhang \ \ \
\\
\\
CISPA Helmholtz Center for Information Security \ \ \ 
}

\date{}

\maketitle

\begin{abstract}
Machine learning models are vulnerable to membership inference attacks in which an adversary aims to predict whether or not a particular sample was contained in the target model's training dataset.
Existing attack methods have commonly exploited the output information (mostly, losses) solely from the given target model.
As a result, in practical scenarios where both the member and non-member samples yield similarly small losses, these methods are naturally unable to differentiate between them. 
To address this limitation, in this paper, we propose a new attack method, called \system, which can exploit the membership information from the whole training process of the target model for improving the attack performance.
To mount the attack in the common black-box setting, we leverage knowledge distillation, and represent the membership information by the losses evaluated on a sequence of intermediate models at different distillation epochs, namely \emph{distilled loss trajectory}, together with the loss from the given target model.
Experimental results over different datasets and model architectures demonstrate the great advantage of our attack in terms of different metrics.
For example, on CINIC-10, our attack achieves at least 6$\times$ higher true-positive rate at a low false-positive rate of 0.1\% than existing methods.
Further analysis demonstrates the general effectiveness of our attack in more strict scenarios.\footnote{The source code of our experiments can be found at \url{https://github.com/DennisLiu2022/Membership-Inference-Attacks-by-Exploiting-Loss-Trajectory}.}
\end{abstract}

\section{Introduction}

Recent machine learning (ML) tasks have involved sensitive data such as healthy records in model training.
However, prior studies~\cite{FJR15,SSSS17,CLEKS19} have shown that most of the training data can be memorized by the ML models, which incurs the risk of privacy leakage. 
Membership inference attack (MIA)~\cite{SSSS17} is one of the privacy attacks against ML models whereby an adversary aims to infer whether or not a target sample was used to train a specific ML model. 
As of today, MIA is the de facto standard for evaluating ML models' privacy risks.

\begin{figure}[t]
\centering
\subfloat[Loss Distribution]{\includegraphics[width=0.5\linewidth]{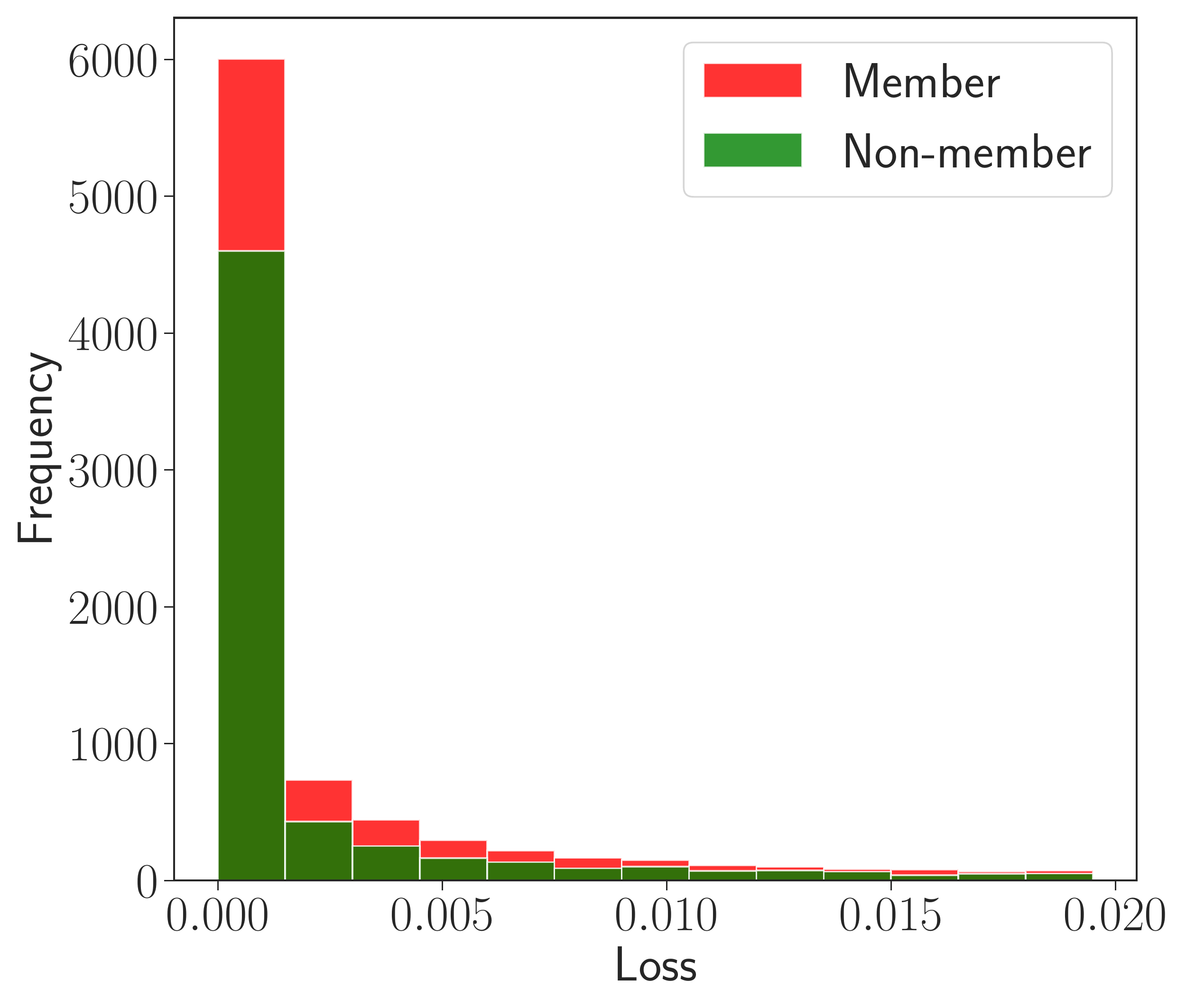}\label{loss_distribution}}
\subfloat[Loss Trajectory]{\includegraphics[width=0.5\linewidth]{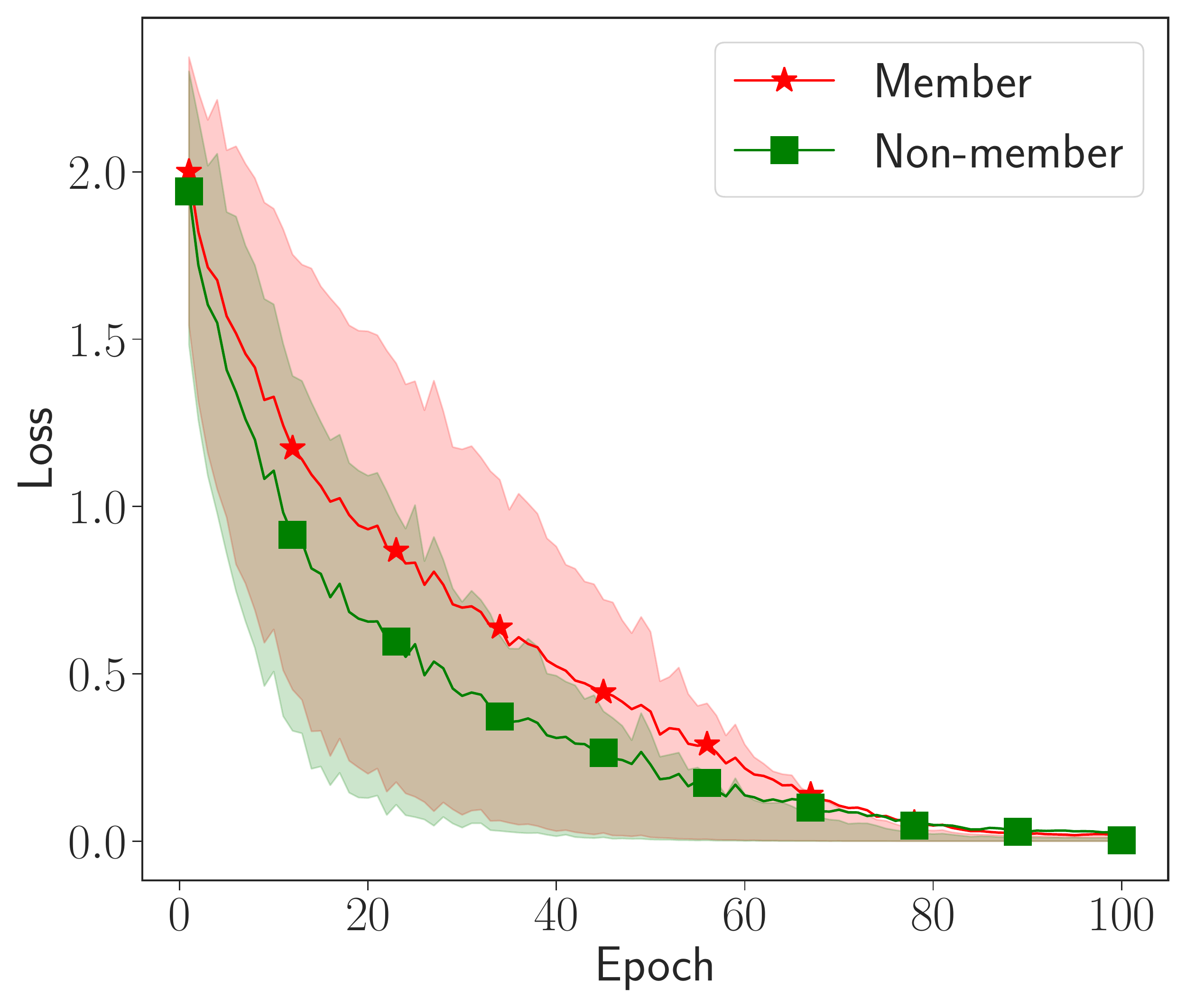}\label{loss_trajectory}}
\caption{(a) the loss distribution of all target member and non-member samples on the target model; (b) the \textit{loss trajectory} for specific member and non-member samples that have similarly small (< 0.02) losses on the target model. 
Using loss trajectory can help differentiate between these specific member and non-member samples.}
\label{fig:loss_distribution_and_loss_trajectory}
\end{figure}

In order to infer the membership of a given target sample, most of the current MIA methods have used its losses (or posteriors) obtained from the target model as their inputs~\cite{SSSS17,SZHBFB19,YGFJ18}. 
The general assumption of these MIAs is that member samples have overall smaller losses than non-member samples~\cite{YGFJ18}. 
These MIAs are effective in terms of average-case metrics (e.g. balanced accuracy and ROC-AUC); however, they cannot differentiate between member and non-member samples that have similar losses, while the fact is most non-member samples have similarly small losses as member samples, as illustrated in \autoref{loss_distribution}.
This is also the reason that the current MIAs suffer a relatively high false-positive rate~\cite{CCNSTT21}. 
In this paper, we investigate whether an adversary can leverage other signals to improve membership inference performance, in particular, reducing the attack's false-positive rate. 

Although a non-member sample might have a small loss like a member sample on a target model, since it does not participate in the training process of the model, it might exhibit a different \emph{loss trajectory}, i.e., its losses evaluated on the target model at its different training epochs, than a member sample.
Our general hypothesis is that an adversary can exploit a sample's loss trajectory to differentiate between member and non-member samples.
As can be seen from \autoref{loss_trajectory}, there are indeed substantial differences between the loss trajectory of member vs. non-member samples even when both have similarly small losses on the target model (the losses at its last epoch).
Specifically, the loss trajectory of non-member samples is lower than that of member samples. 
This can be explained based on the hypothesis of sample hardness~\cite{SDSOJ19,WGCS21,CCNSTT21,TSCTBG19,HCW20}, which suggests these non-member samples with small losses are essentially easy samples while member samples are gradually learned through the whole training process to eventually reach similarly small losses. 
See more detailed discussion in \autoref{sec:des}.

In this paper, we propose a novel membership inference attack against machine learning models, namely \system, which leverages target samples' loss trajectory to differentiate members from non-members.
We focus on machine learning classifiers in the image domain as most of the MIAs are evaluated in the same setting.
However, we emphasize that \system is general and can be directly applied to ML models in other domains.

To mount \system, the first step is to obtain the target sample's loss trajectory.
However, in practical scenarios, the adversary can only observe the final trained target model instead of all the intermediate models during the target model's training process.
To solve this, we leverage knowledge distillation~\cite{HVD15}.
Concretely, the adversary first performs a black-box model distillation to the target model to obtain a distilled model.
During the process, they keep all the intermediate versions of the distilled model locally.
Here, different versions correspond to different training epochs.
Then, the adversary evaluates the target sample's loss on each of the intermediate distilled models to obtain the target sample's loss trajectory, namely \emph{distilled loss trajectory}.
In the end, the attack model, which is a membership classifier, takes as input the target sample's distilled loss trajectory as well as its loss on the original target model to infer the sample's membership status.
Note that \system only needs to perform model distillation once and keeps on reusing the distilled intermediate models for all target samples at inference time.

We evaluate \system on a comprehensive suite of benchmark datasets, with extensive comparisons to other advanced attack methods.
Following the recent recommendation on evaluating MIAs~\cite{CCNSTT21}, we mainly focus on the metric that measures \textit{True-Positive Rate at low False-Positive Rate} (TPR at low FPR), but also report results in terms of other average-case metrics, including balanced accuracy and ROC-AUC.
Experimental results show that \system is able to achieve 5.3\% TPR at 0.1\% FPR on the CINIC-10 dataset, at least 6$\times$ better than other considered advanced attacks.
Furthermore, we evaluate the attack performance at a more fine-grained level by calculating the TPR at low FPR for separate groups of target samples that have varied loss values on the target model.
The evaluation results demonstrate that \system achieves strong performance in all the settings.
We conduct extensive ablation studies to analyze the impact of different important factors on the attack success of \system, e.g., the size of the dataset used for training the target model and for the knowledge distillation, as well as the number of epochs used in the distillation process.
We further explore \system in more strict scenarios with relaxed assumptions about the knowledge of the adversary, including different architectures between the target model and local models, and data distribution shift.
Finally, we provide additional insights into understanding the characteristics of \system, by discussing the importance of its main components.
In general, this paper makes the following contributions:

\begin{itemize}
\item We take the first step to exploit the information from the training process of the target model to conduct membership inference attacks, and propose a novel attack method \system based on knowledge distillation.
\item We demonstrate that \system consistently outperforms other advanced attack methods in common scenarios, but also in more strict scenarios with relaxed assumptions.
\item We provide in-depth analyses about the impact of each component of \system and other important factors on the attack performance.
\end{itemize}

\mypara{Roadmap.}
In \autoref{sec:pre}, we introduce the preliminary knowledge about machine learning, membership inference attacks, and knowledge distillation. 
\autoref{sec:traject} presents the threat model, design intuition, and the details of our attack method. 
We conduct extensive experiments to show the effectiveness of our attack in \autoref{sec:Eval}, and the impact of important factors on the attack performance in \autoref{sec:Ablation}. 
In \autoref{sec:discuss}, we provide an in-depth analysis on the impact of each component in our attack. 
We discuss the related work in \autoref{sec:related} and conclude the paper in \autoref{sec:con}.

\section{Preliminary}
\label{sec:pre}

\subsection{Machine Learning}

For machine learning classification tasks, a learned neural network $\MLModel_\theta$ is a function that maps each data sample from a dataset $\mathbf{X}$ to its class/label in a label set $\mathbf{Y}$. 
Given a sample $\DataPoint$, its output from $\MLModel_\theta$, denoted as $p=\MLModel_\theta(\DataPoint)$, is a vector that represents the prediction posteriors of the sample over different pre-defined classes. 
In order to train a ML model, a loss function $\mathcal{L}(y,p)$ is defined to determine the error between a sample's prediction posteriors and its corresponding label. 
Cross-entropy loss is one of the most common loss functions used for classification tasks, and it is defined as:
\begin{equation}
\label{equa:cross_entropy}
\mathcal{L}_{CE}(y,p)=-\sum_{i=1}^k y_i \log{p_{i}}
\end{equation}
where $k$ is the total number of classes. 
$y_i$ equals $1$ only if the sample belongs to class $i$ and otherwise $0$, and $p_i$ is the $i$-th element of the prediction posteriors.
The model training is implemented to minimize the empirical loss by stochastic gradient descent:
\begin{equation}
\label{equa:general_training}
\theta_{i+1} \leftarrow \theta_{i} - \epsilon\sum_{(x,y)\in\mathcal{B}}\nabla_{\theta}\mathcal{L}(y,\MLModel_{\theta_i}(\mathbf{x}))
\end{equation}
where $\mathcal{B}$ is a small batch of training samples and $\epsilon$ is the learning rate for iteratively updating the parameters $\theta$ of the neural network. 
The model will be trained for multiple epochs (times that the entire training set is passed to the model) in order to achieve a well-generalized model.
Normally, the intermediate models at different epochs can be preserved and the training process can be stopped at a specific epoch.

\subsection{Membership Inference Attacks}

The objective of membership inference attacks is to identify whether or not a target sample exists in the training set of the given model. 
First proposed by~\cite{SSSS17}, MIA has drawn great attention in various scenarios~\cite{CTWJHLRBSEOR20,SS19,CYZF20,HMDC19,HHB19,ZRWRCHZ21,MSCS19,NSH19}. 
What makes the MIA so important is its connection to privacy leakage due to the increasingly sensitive data for training ML models and also its simplicity to be deployed.

\mypara{Definition of MIA.}
Concretely, given a target sample $\DataPoint$, a trained ML model $\MLModel_\theta$ and some external knowledge of the adversary, denoted by $\MIA$, membership inference attacks $\AttackModel$ can be defined by the following function:
\begin{equation}
\AttackModel: \DataPoint, \MLModel_\theta, \MIA \rightarrow \{0, 1\}
\end{equation}
Here, 0 means $\DataPoint$ is not a member from $\MLModel_\theta$'s training set and 1 means $\DataPoint$ is a member.
Most of the current MIA attacks are based on training a binary classifier, which we follow as well in this paper.

\mypara{Adversary's Knowledge.}
Basically, the adversary is assumed to have black-box access to the target model, that is, they can only obtain the posteriors output by the target model for each query sample. 
In addition, they are able to leverage an auxiliary dataset that comes from the same distribution as the training set of the target model. 
In this way, they can train a shadow model to mimic the behavior of the target model and take as input the posteriors output by the shadow model for training a binary classifier, namely attack model.
The trained attack model is then used to infer the membership of any given target sample. 
To make the attack more efficient, some studies~\cite{SZHBFB19,YGFJ18} use the output from the target model directly as a signal to predict the membership status without training any shadow model. 

Recently, MIA has also been explored in other scenarios, including white-box~\cite{NSH19} and label-only~\cite{LZ21,CTCP21,ROF202}. 
The former scenario assumes the adversary has full access to the target model, which contains the model architecture and parameters. 
The latter scenario considers a more strict setting in which the adversary can only get the hard label predicted by the target model. 
As a result, the adversary mounts the attack by perturbing the sample to change its predicted label, and then measures the magnitude of the perturbation. 
If it is larger than a predefined threshold, the adversary will consider the sample as a member and otherwise a non-member.

\mypara{Defense Against MIA.}
The overfitting level of the ML model is one of the major factors that influence the success of the MIA as demonstrated in~\cite{SZHBFB19,SSSS17}.
For this reason, general regularization techniques, such as Dropout~\cite{SHKSS14} and confidence penalty~\cite{PTCKH17}, can be used to defend against MIAs. 
Besides, knowledge distillation is another effective tool for mitigating MIAs, such as PATE~\cite{PAEGT17} and DMP~\cite{SH21}.
MemGuard~\cite{JSBZG19} obfuscates the output from the target model to reduce the information that can be leveraged by the adversary. However, it fails when a label-only MIA is conducted.
DP-SGD~\cite{ACGMMTZ16}, one popular application of differential privacy (DP) in machine learning, provides a provable privacy guarantee for defending against membership inference attacks. 
It can achieve strong protection against MIAs, but with a sacrifice in severe accuracy degradation for the original classification tasks.

\subsection{Knowledge Distillation}

Knowledge distillation (KD) represents a class of methods that train a smaller student model to have better performance by learning based on the output of a larger teacher model.
The key idea of KD is that the soft information (i.e. output posteriors) from a larger teacher model contains a lot more information than the hard, ground-truth label.
Here we adopt the most classical KD framework proposed by Hinton et al.~\cite{HVD15}.
Given any input $\DataPoint$, the corresponding output from the teacher model is essentially a vector of posteriors, denoted as $p^t = [p^t_1,\cdots,p^t_C]$, $C$ is the number of classes, and the output from the student model is $p^s$. 
Normally, these posteriors are calculated through a softmax function, but in order to make them softer so as to extract more information from the teacher model, it is modified to:
\begin{equation}
\label{equa:modified_softmax}
\Tilde{p}^t_i = \frac{\exp^{s^t_i/\tau}}{\sum_j\exp^{s^t_j/\tau}}
\end{equation}
where $s^t_i$ is the logit value before the softmax function for the $i$-th class and $\tau$ is the temperature used for controlling the softness level.
To learn a student model through distillation, one only needs to submit a set of samples to the target model and obtains their posteriors.
Then, the student model is trained on the samples supervised by their posteriors using the loss that is a linear combination of the typical Cross-entropy $\mathcal{L}_{cls}$ and the knowledge distillation loss $\mathcal{L}_{KD}$:
\begin{equation}
\label{equa:loss_student_model}
\mathcal{L} = \alpha\mathcal{L}_{cls} + (1-\alpha)\mathcal{L}_{KD}
\end{equation}
Here, $\mathcal{L}_{KD}$ is calculated between the soft posteriors output from the teacher model $\Tilde{p}^t$ and student model $\Tilde{p}^s$ by Kullback-Leibler divergence loss:
\begin{equation}
\label{equa:KD_loss}
\mathcal{L}_{KD}(\Tilde{p}^t,\Tilde{p}^s) = \sum_{i=1}^k \Tilde{p}^t_i \log{\frac{\Tilde{p}^t_i}{\Tilde{p}^s_i}}
\end{equation}

In this paper, we aim to learn a student model that is as similar to the teacher model as possible, and thus we set $\alpha=0$ and $\tau=1$.
Note that knowledge distillation in this setting is similar to model stealing techniques~\cite{TZJRR16}.

\section{Attack Methodology}
\label{sec:traject}

In this section, we introduce the methodology of \system.
We start by defining the threat model. 
Then, we introduce the design intuition of our attack and explain why it works. 
Finally, the detailed pipeline of our attack is provided to illustrate how to conduct \system with knowledge distillation in practical scenarios with only black-box access to the target model.

\subsection{Threat Model}

In this paper, we focus on the commonly-adopted, black-box scenario of MIAs, in which the adversary only has access to the posterior output from the target model.
For a given target model, we assume that the adversary has an auxiliary dataset $\mathcal{D}^a$ that comes from the same distribution as the target model's training set $\mathcal{D}^t$. 
This follows the standard setting of most of the advanced MIAs~\cite{SSSS17,SZHBFB19,WGCS21,SM21,CCNSTT21}.
Both the data used to train the shadow model and used to distill the target/shadow model for obtaining the corresponding \textit{distilled loss trajectory} are sampled from this auxiliary dataset.
Furthermore, we assume the adversary knows the architecture of the target model. 
Later in \autoref{sec:Ablation}, we show that these two assumptions on the adversary's knowledge about the training data distribution and the architecture of the target model can be relaxed.

\subsection{Design Intuition}
\label{sec:des}

As ML models are trained to minimize the losses from their training samples, it is assumed that member samples are on average more likely to have smaller losses than non-member samples, which is also due to the overfitting of the model. 
Thus, losses (or posteriors) are used as common signals by most current methods to conduct membership inference. Although these loss-based attacks are effective in terms of the average-case metrics such as balanced accuracy and ROC-AUC; they actually fail to differentiate between member samples and non-member samples when both of them have small losses. 
We show it in \autoref{loss_distribution}, the loss distribution between member samples and non-member samples from the target model, that most non-member samples indeed get similar small losses as member samples. 
This causes a high false-positive rate in most existing MIAs and renders them unreliable in real scenarios. 
To this end, we propose to utilize other stronger signals to enhance membership inference and especially aim to reduce the attack's false-positive rate.

\begin{figure}[t]
\centering
\includegraphics[width=1\linewidth]{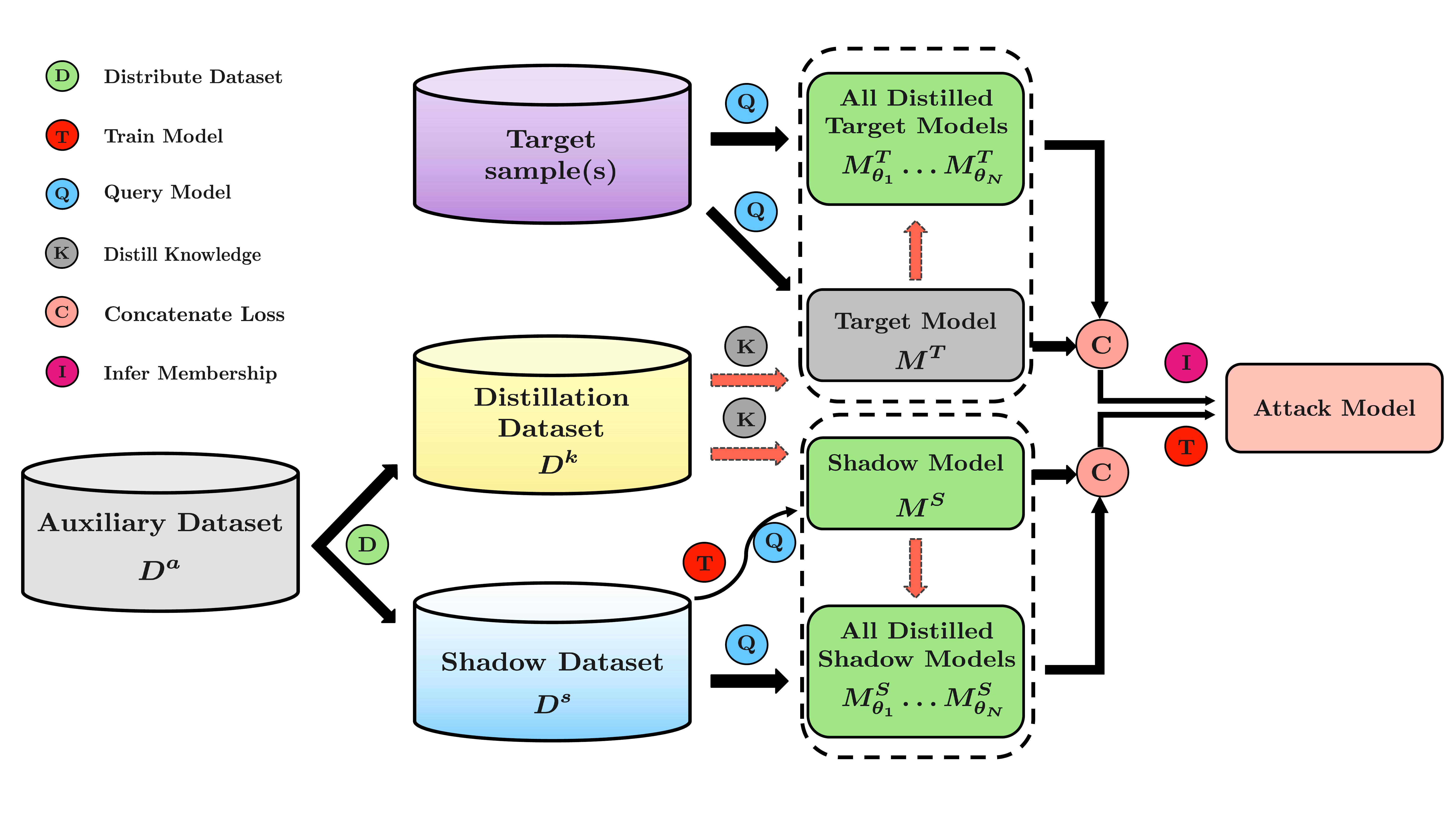}
\caption{The working pipeline of our method. Different from the conventional MIA pipeline, our method uses a proportion of the auxiliary dataset as the knowledge distillation dataset for obtaining distilled target/shadow models at their different epochs, and each sample is represented by the concatenation of the loss from the target/shadow model and the distilled loss trajectories from the distilled target/shadow models.}
\label{fig:pipeline}
\end{figure}

The main idea behind our attack is to leverage the major difference between member samples and non-member samples, that is the former indeed participates in the training of the target model while the latter does not. 
Our general hypothesis is that non-member samples should have a distinct changing pattern on the losses evaluated at each training epoch, namely \textit{loss trajectory}, compared to member samples. 
As illustrated in \autoref{loss_trajectory}, for those member and non-member samples that have similarly small losses from the model at its last training epoch (i.e., the given target model), they behave quite differently along the training process regarding the changing pattern of the loss trajectory. This difference has a close connection to sample hardness.
Concretely, we show that non-member samples are indeed easier samples in terms of two existing sample hardness metrics.
For the metric that defines the hardness as the epoch when the model does not change the prediction~\cite{TSCTBG19,HCW20}, the result for members vs. non-members is 35.4 vs. 26.1. 
For the metric that defines the hardness as the loss from arbitrary reference models~\cite{SDSOJ19,WGCS21,CCNSTT21}, the result is 0.006 vs. 0.005.
Thus, loss trajectory can provide a more detailed profile for data samples and can be used as a stronger signal to conduct MIA.

\subsection{Attack Method}

Inspired by the differences in the loss trajectory between the member and non-member samples, we propose a new membership inference attack, called \system. 
In order to mount \system, the adversary needs to get the loss trajectory from the training process of the target model. 
However, in practical scenarios, the adversary only has black-box access to the target model, i.e., only the target model at its last training epoch is directly accessible. 
To address this issue, we leverage knowledge distillation. Specifically, the adversary first conducts a standard model distillation to the target model and gets a distilled model.
By doing this, the adversary has full control of the distillation process and can preserve the distilled target models at different epochs. 
After distillation, the adversary can evaluate any given target sample on all intermediate distilled models to acquire its loss trajectory, which we call \textit{distilled loss trajectory}. 
Finally, the attack model takes as input this distilled loss trajectory together with the loss obtained from the original target model to infer the membership. 

\begin{table*}[t]
\centering
\caption{Training and testing accuracy for all model architectures on different datasets.}
\label{table:train_test_acc}
\setlength{\tabcolsep}{3pt}
\scalebox{0.75}{
\begin{tabular}{l|cccccccc}
\toprule
Target&  \multicolumn{2}{c}{CIFAR-10}&  \multicolumn{2}{c}{CIFAR-100}& \multicolumn{2}{c}{GTSRB}& \multicolumn{2}{c}{CINIC-10}\\
Model&Train acc&Test acc&Train acc&Test acc&Train acc&Test acc&Train acc&Test acc\\
\midrule
MibileNetV2&  {97.2\%}& 70.2\%&  {100.0\%}& 32.4\%&  {100.0\%}& 78.4\%&  {98.8\%}& 52.0\%\\
VGG-16&  {100.0\%}& 82.6\%&  {99.9\%}& 47.5\%&  {100.0\%}& 84.1\%&  {99.9\%}& 65.8\%\\
ResNet-56&  {98.3\%}& 76.2\%&  {94.1\%}& 42.2\%&  {100.0\%}& 88.1\%&  {97.4\%}& 60.7\%\\
WideResNet-32&  {93.6\%}& 77.1\%&  {94.5\%}& 45.1\%&  {100.0\%}& 85.4\%&  {90.5\%}& 62.8\%\\
\bottomrule
\end{tabular}
}
\end{table*}

The detailed pipeline of \system is illustrated in \autoref{fig:pipeline}.
It consists of four stages: shadow model training, model distillation, attack model training, and membership inference.
In particular, the model distillation stage is newly introduced by our \system, and the other three stages follow the common MIA pipeline except that the input to the attack model is different.

\mypara{Shadow Model Training.}
As aforementioned, the adversary has an auxiliary dataset $\mathcal{D}^a$ drawn from the same distribution as the training dataset $\mathcal{D}^t$ of the target model $\mathcal{M}^{T}$. 
This auxiliary dataset is split into two disjoint subsets. 
One subset is used as the shadow dataset $\mathcal{D}^s$ to train a shadow model $\mathcal{M}^S$ in conjunction with classical training techniques. 
Following the common practice, we use the same architecture of the target model to build our shadow model. 

\mypara{Model Distillation.}
The other subset of the auxiliary dataset is used as the knowledge distillation dataset $\mathcal{D}^k$ to distill the trained shadow model and the target model in order to obtain their distilled loss trajectory. 
Specifically, in order to make the distilled model more similar to the original model (target model or trained shadow model), we only use the posteriors output from the original model and distilled model to calculate a Kullback-Leibler divergence loss in the distillation process, and do not consider the ground truth label.
One thing needs to mention, for the shadow model, although we already have the loss trajectory from its actual training process, we still conduct the distillation to it in order to make sure its distilled loss trajectory is better aligned with the distilled loss trajectory from the target model.
This is reasonable as we cannot access the actual loss trajectory from the target model and the distilled loss trajectory of the target model may behave differently.
More discussion regarding the advantage of this specific design choice can be found in \autoref{sec:discuss}.
The distillation process for our shadow model can be derived from \autoref{equa:general_training} and \autoref{equa:KD_loss} as:
\begin{equation}
\label{equa:shadow_model_stealing}
\theta_{i+1} \leftarrow \theta_{i} - \epsilon\sum_{(x,y)\in\mathcal{D}^k}\nabla_{\theta}\mathcal{L}_{KD}(\mathcal{M}^{S}(\mathbf{x}),\MLModel_{\theta_i}^{S}(\mathbf{x}))
\end{equation}
where $i=1,2\cdots,N$ as $N$ is the number of epochs for model distillation, meaning that we can obtain $N$ intermediate shadow models.
Similarly, we use the same knowledge distillation dataset $\mathcal{D}^k$ to distill the target model and get $N$ distilled target models in total.

\mypara{Attack Model Training.}
The adversary trains an attack model on the shadow dataset $\mathcal{D}^s$ as common MIA methods do, but the only difference is that the input to the attack model becomes the concatenation of the loss trajectory from all distilled shadow models and the loss from the original shadow model:
\begin{equation}
\label{equa:attack_model_input}
\hat\DataPoint = \mathcal{L}(\mathcal{M}^S_{\theta_1}(\DataPoint),y)\oplus\cdots\oplus\mathcal{L}(\mathcal{M}^S_{\theta_N}(\DataPoint),y)\oplus\mathcal{L}(\mathcal{M}^S(\DataPoint),y)
\end{equation}
Here $\hat\DataPoint$ is the input to the attack model and the corresponding label is $1$ when $\DataPoint$ is used for training the shadow model and otherwise $0$.

\mypara{Membership Inference.}
Finally, the adversary can infer the membership of each given target sample by feeding the concatenation of its losses obtained from $N+1$ target models (including one original target model and all $N$ distilled target models) to the trained attack model. 

For label-only attacks, we do not have the posteriors but only the hard labels predicted by the target model. 
Thus, we can view the predicted hard label as the ground truth to calculate the Cross-entropy loss instead of the Kullback-Leibler divergence loss in the distillation process.
Accordingly, the loss from the target model will be replaced by the HopSkipJump boundary distance, and the same is done for the shadow model as we want to better align the training and testing phase of the attack model. 

\section{Evaluation}
\label{sec:Eval}

In this section, we conduct extensive experiments to evaluate our \system on diverse model architectures and benchmarking datasets, with comparisons to other representative attack baselines.
We focus on the commonly-adopted, black-box attack scenario, but also explore the more challenging, label-only scenario.

\begin{figure*}[t]
\centering
{\includegraphics[width=0.245\linewidth]{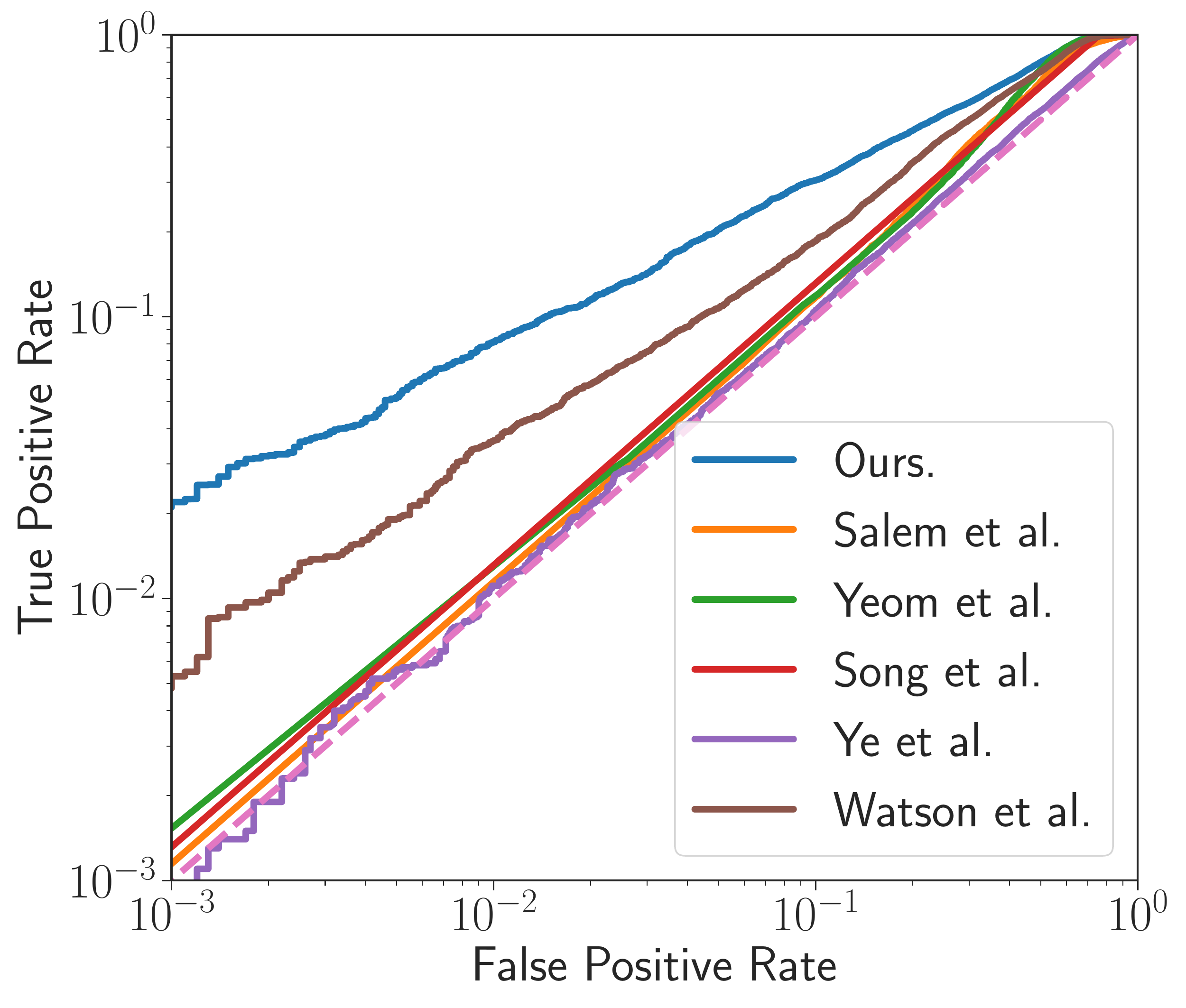}
\includegraphics[width=0.245\linewidth]{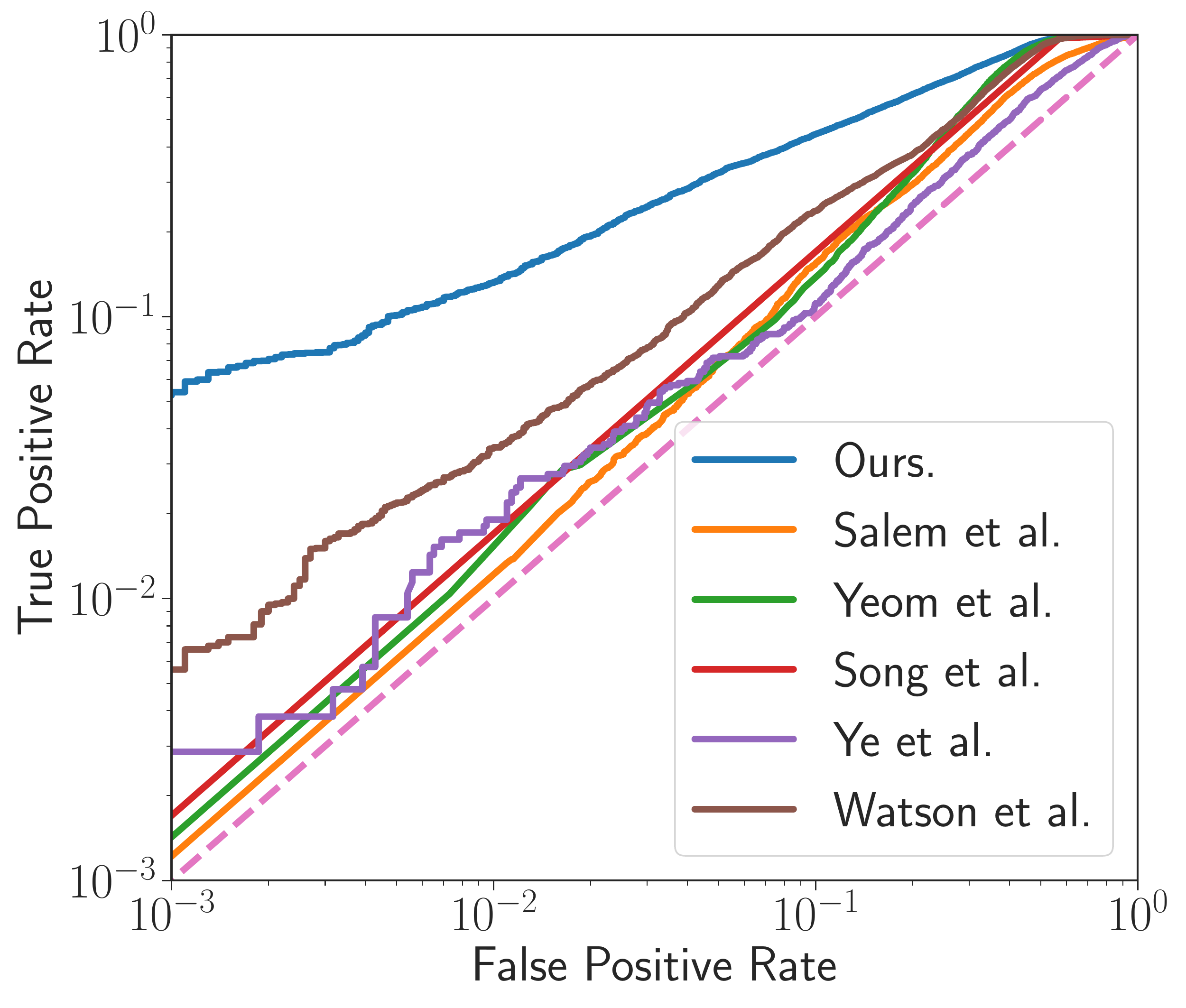}
\includegraphics[width=0.245\linewidth]{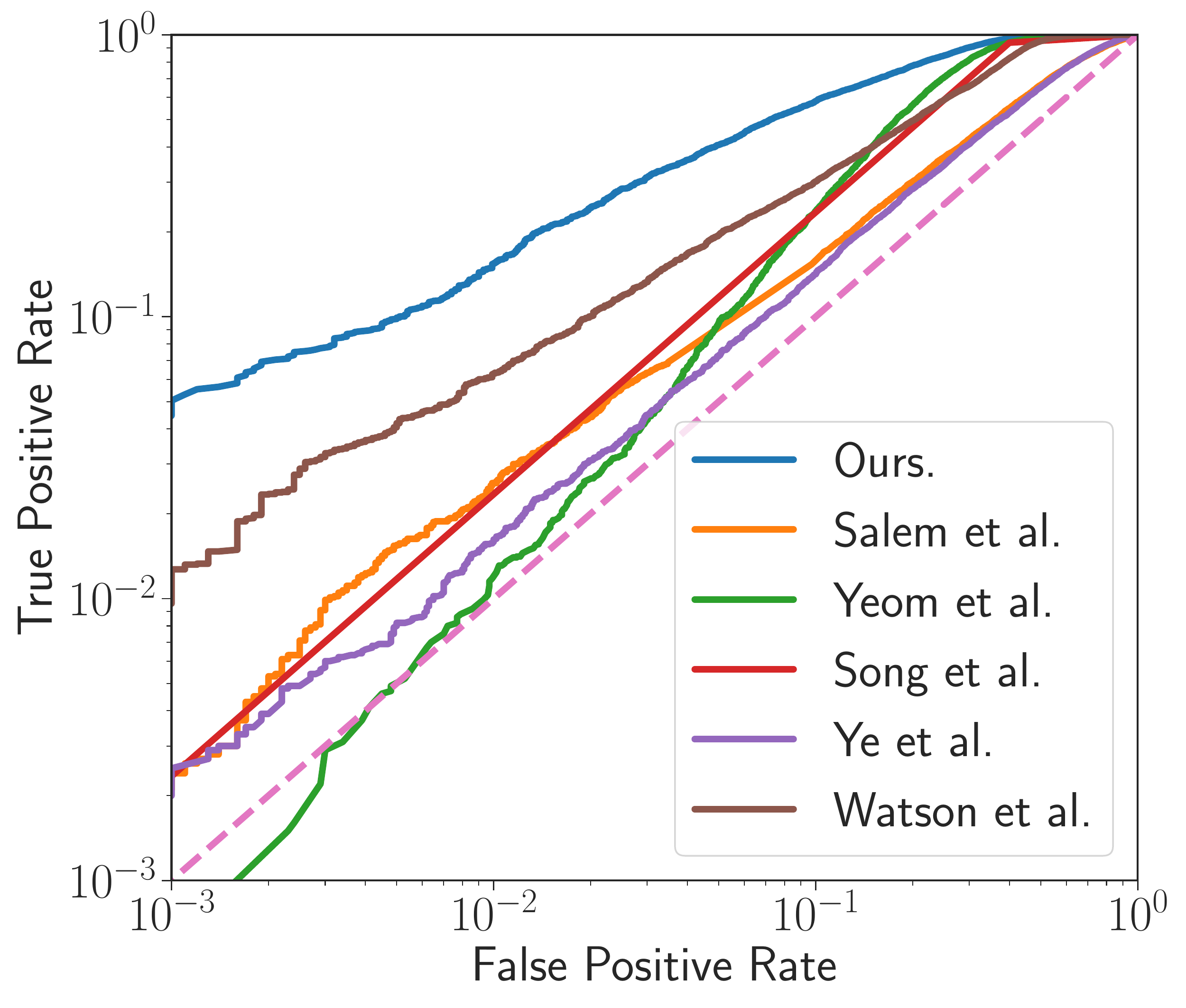}
\includegraphics[width=0.245\linewidth]{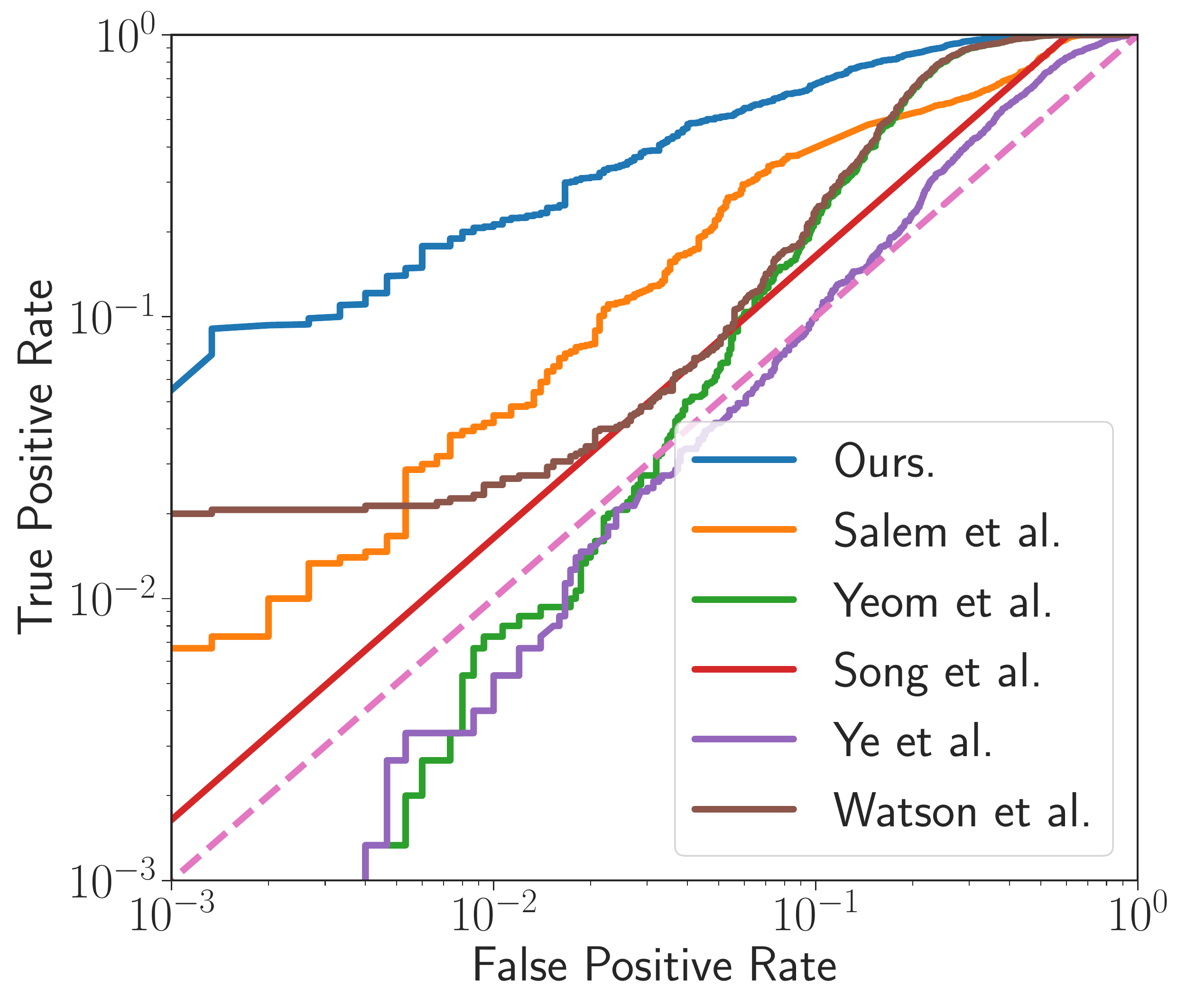}\label{resnet_log_auc}}

{\includegraphics[width=0.245\linewidth]{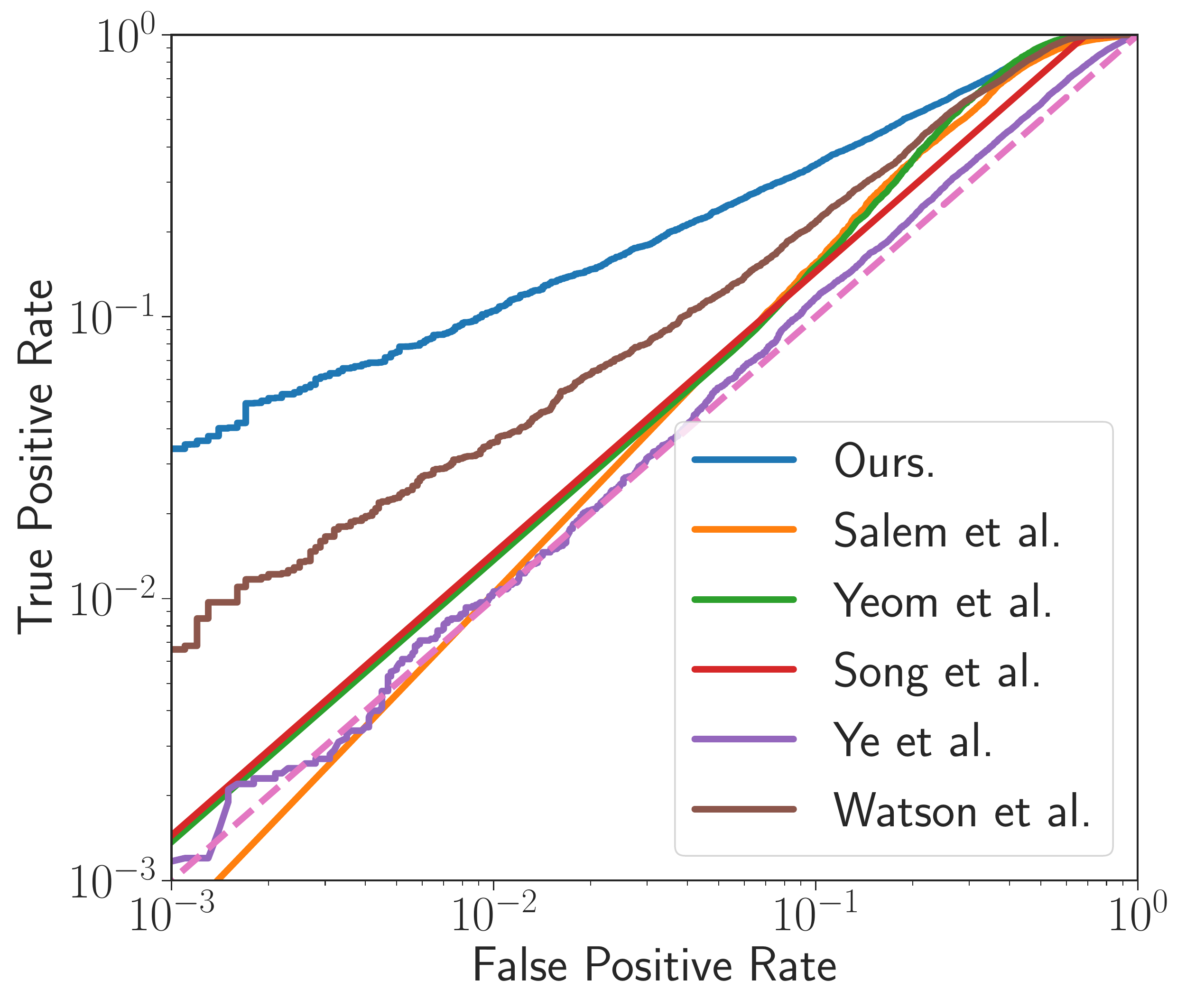}
\includegraphics[width=0.245\linewidth]{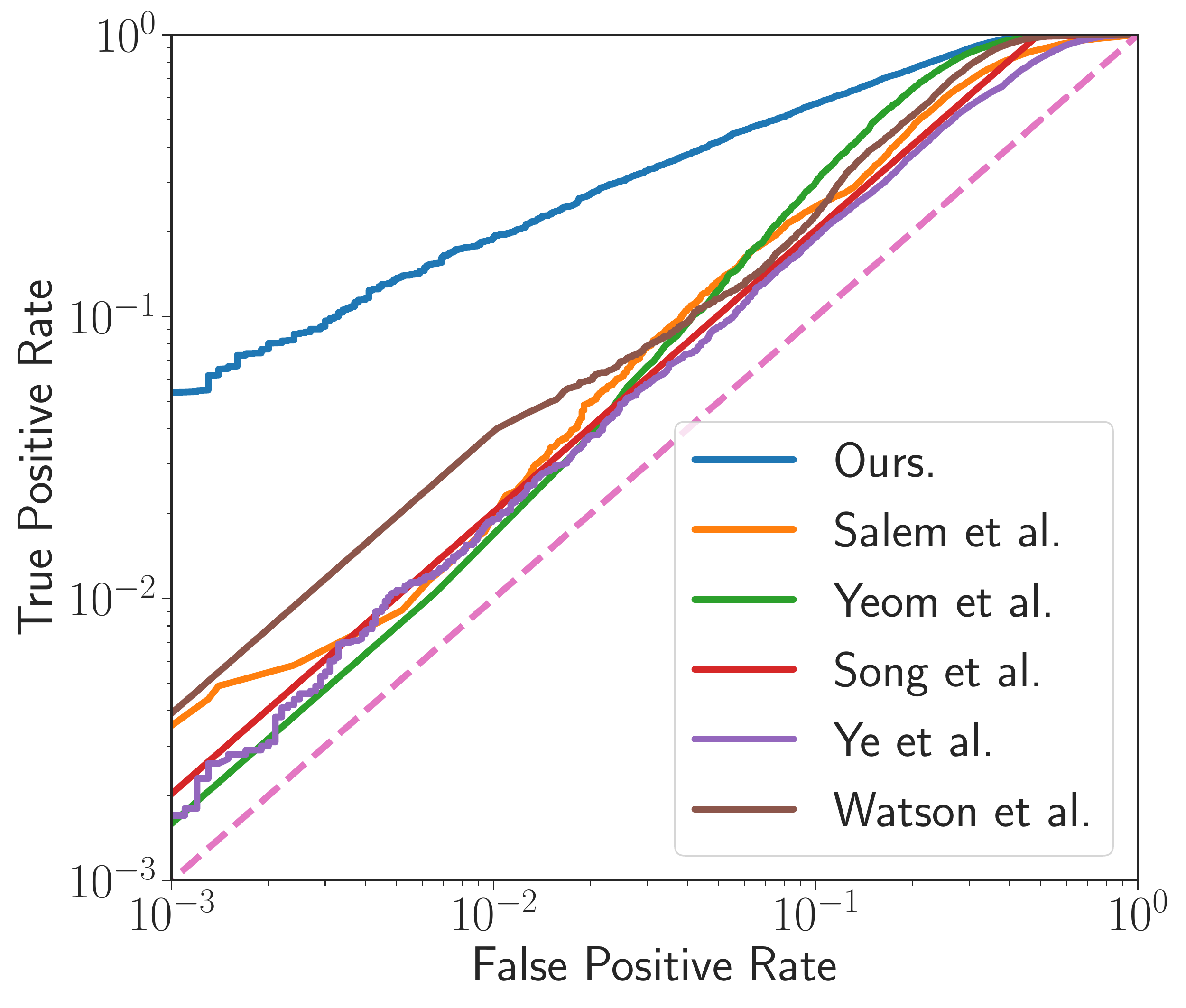}
\includegraphics[width=0.245\linewidth]{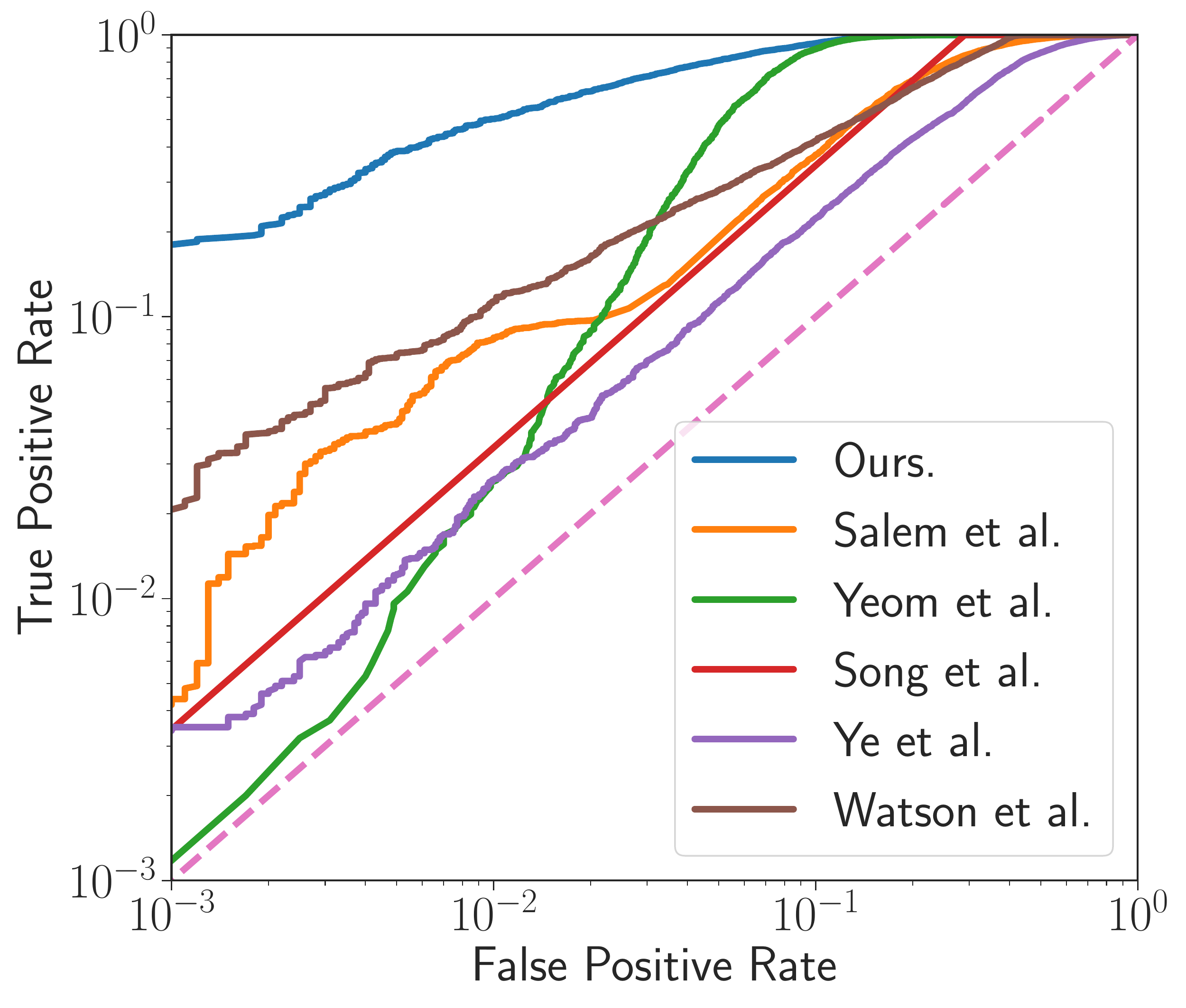}
\includegraphics[width=0.245\linewidth]{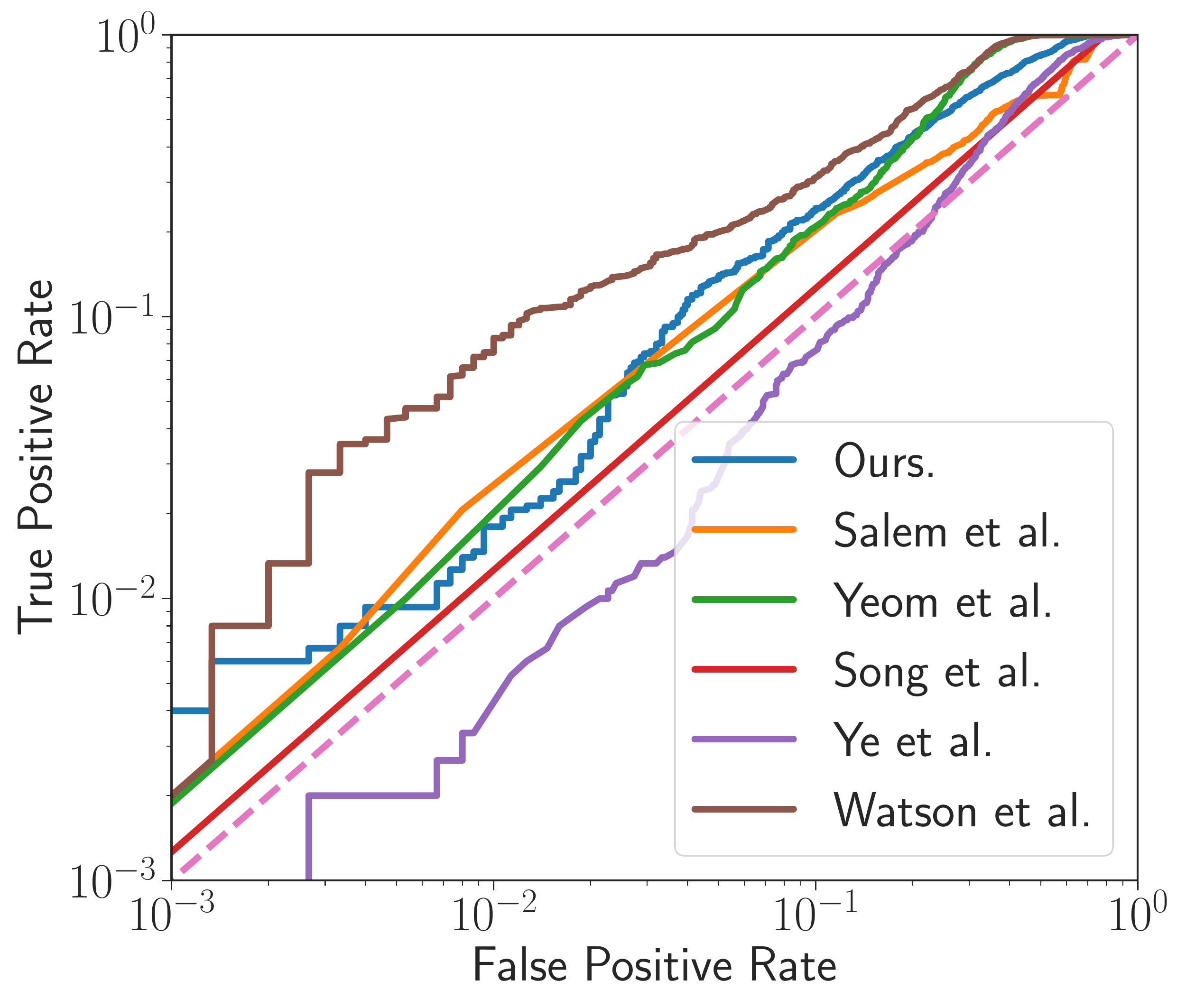}\label{mobilenet_log_auc}}

{\includegraphics[width=0.245\linewidth]{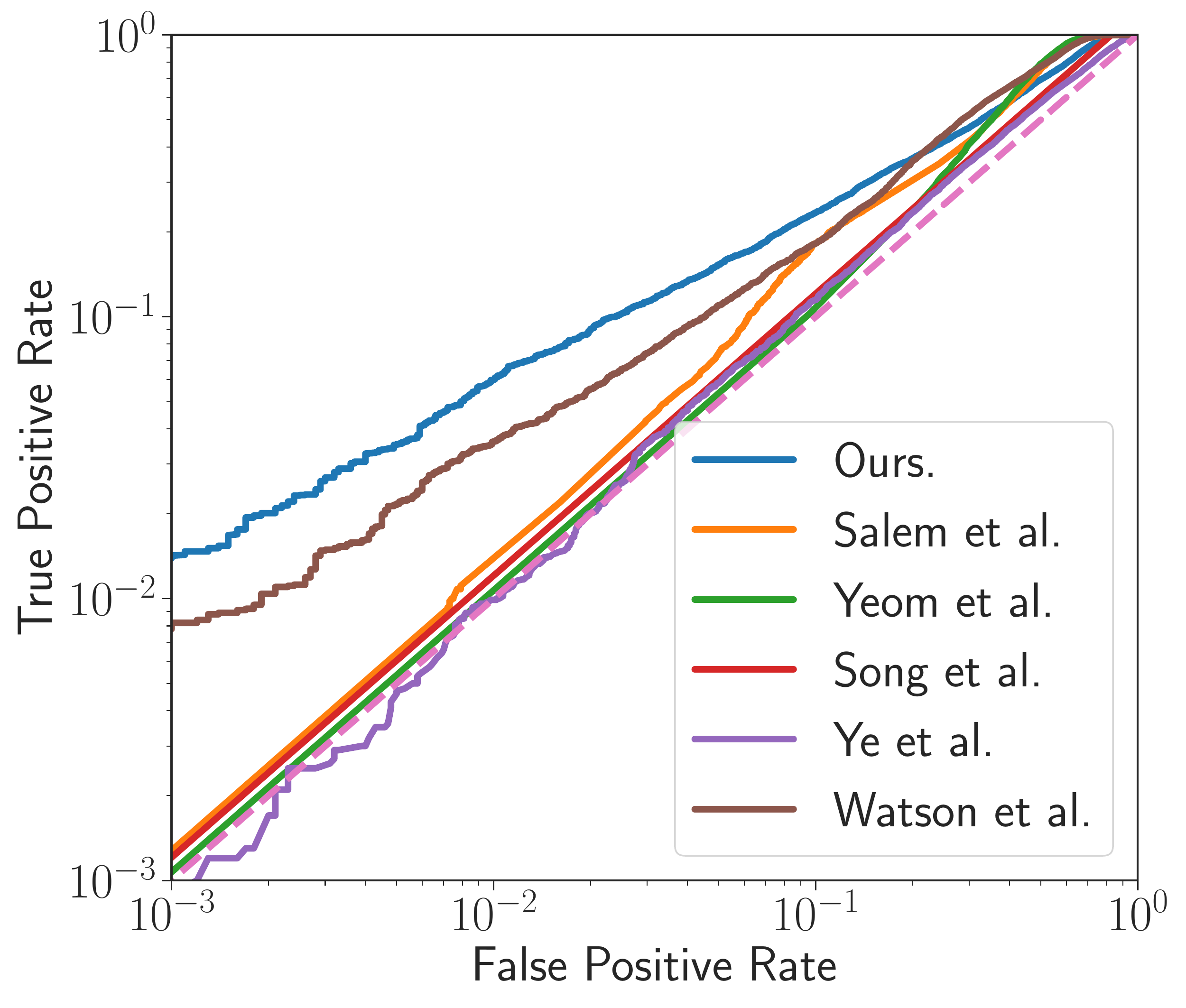}
\includegraphics[width=0.245\linewidth]{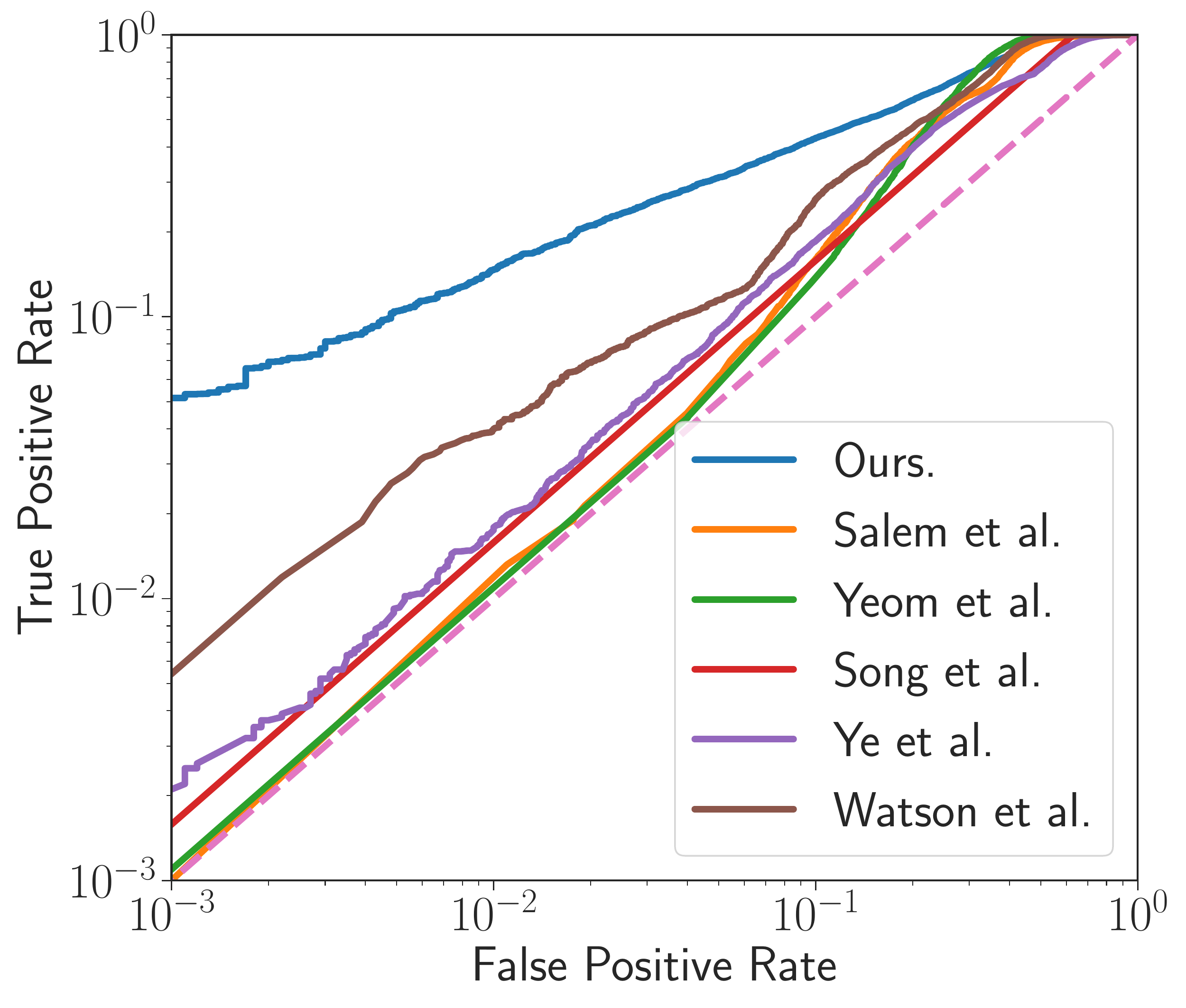}
\includegraphics[width=0.245\linewidth]{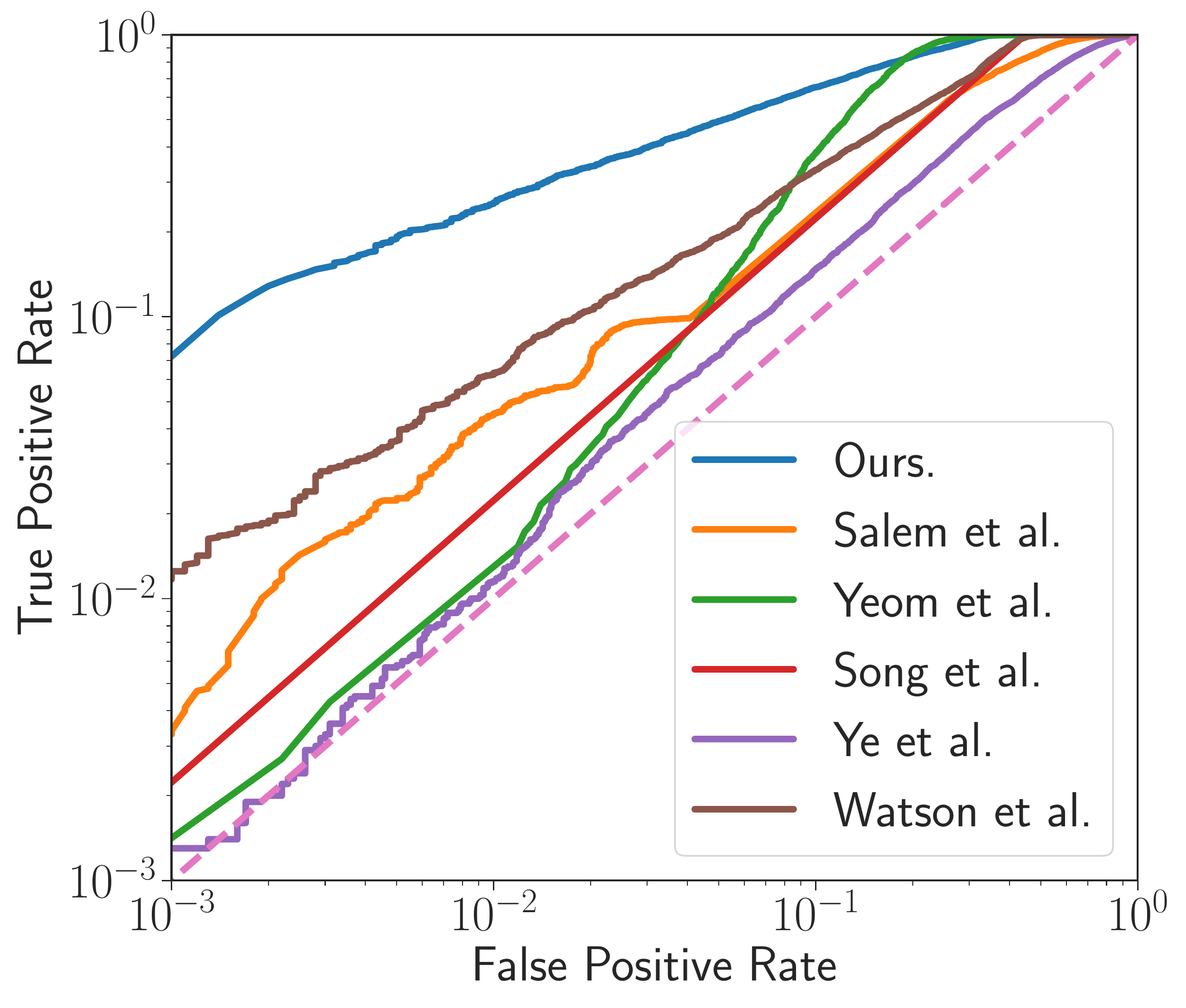}
\includegraphics[width=0.245\linewidth]{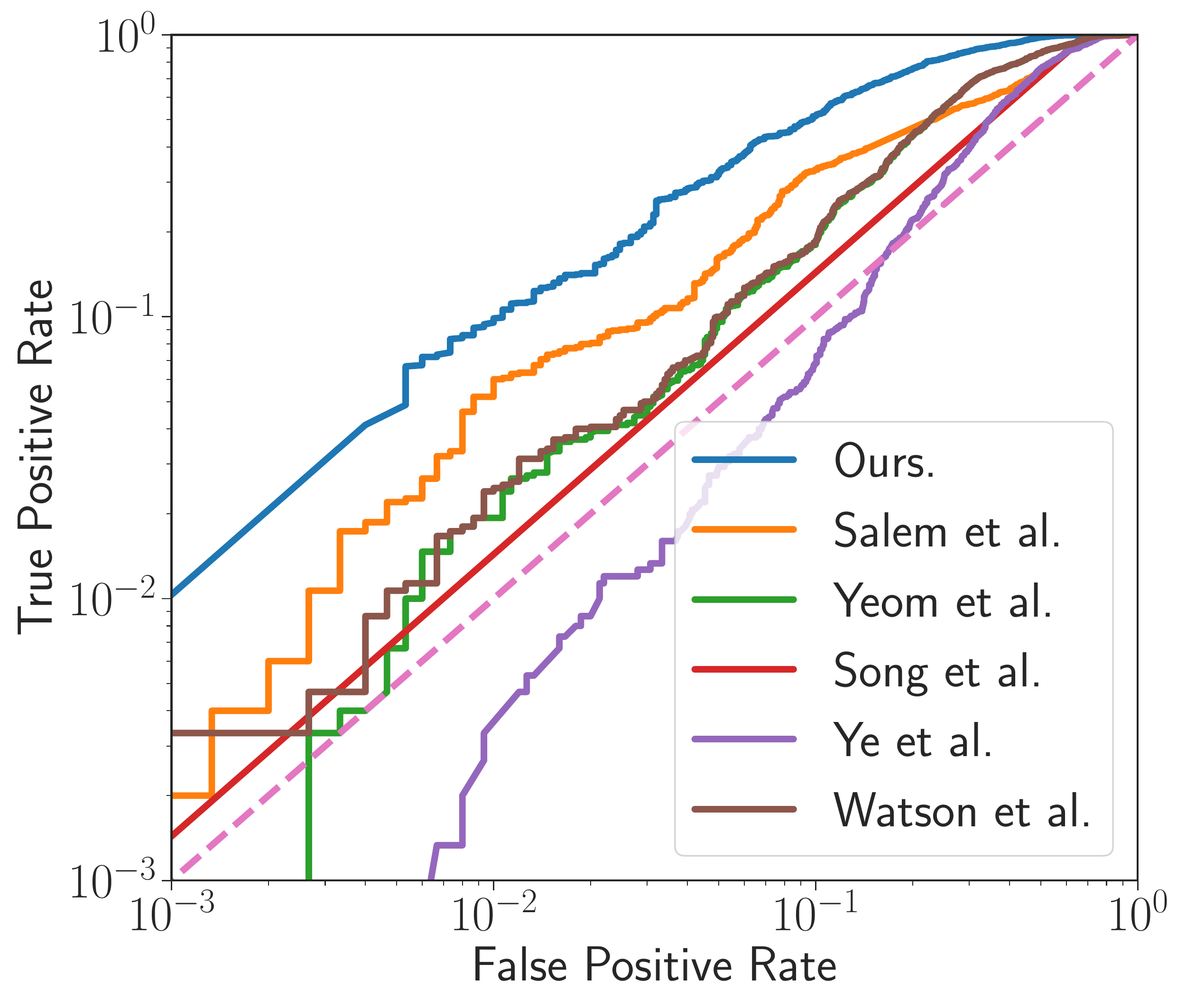}\label{vgg_log_auc}}

\subfloat[CIFAR-10]{\includegraphics[width=0.245\linewidth]{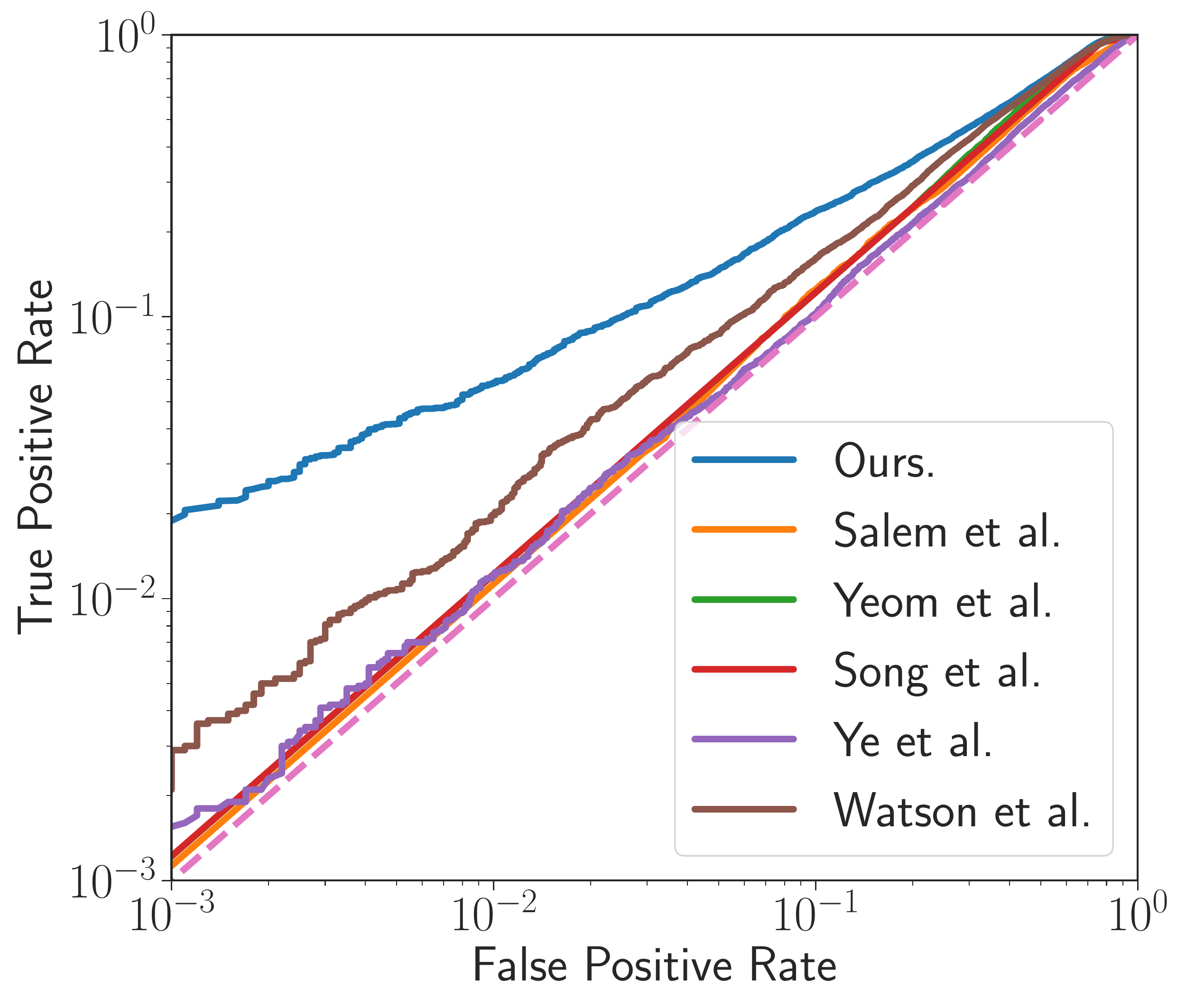}}\hspace{0.1mm}
\subfloat[CINIC-10]{\includegraphics[width=0.245\linewidth]{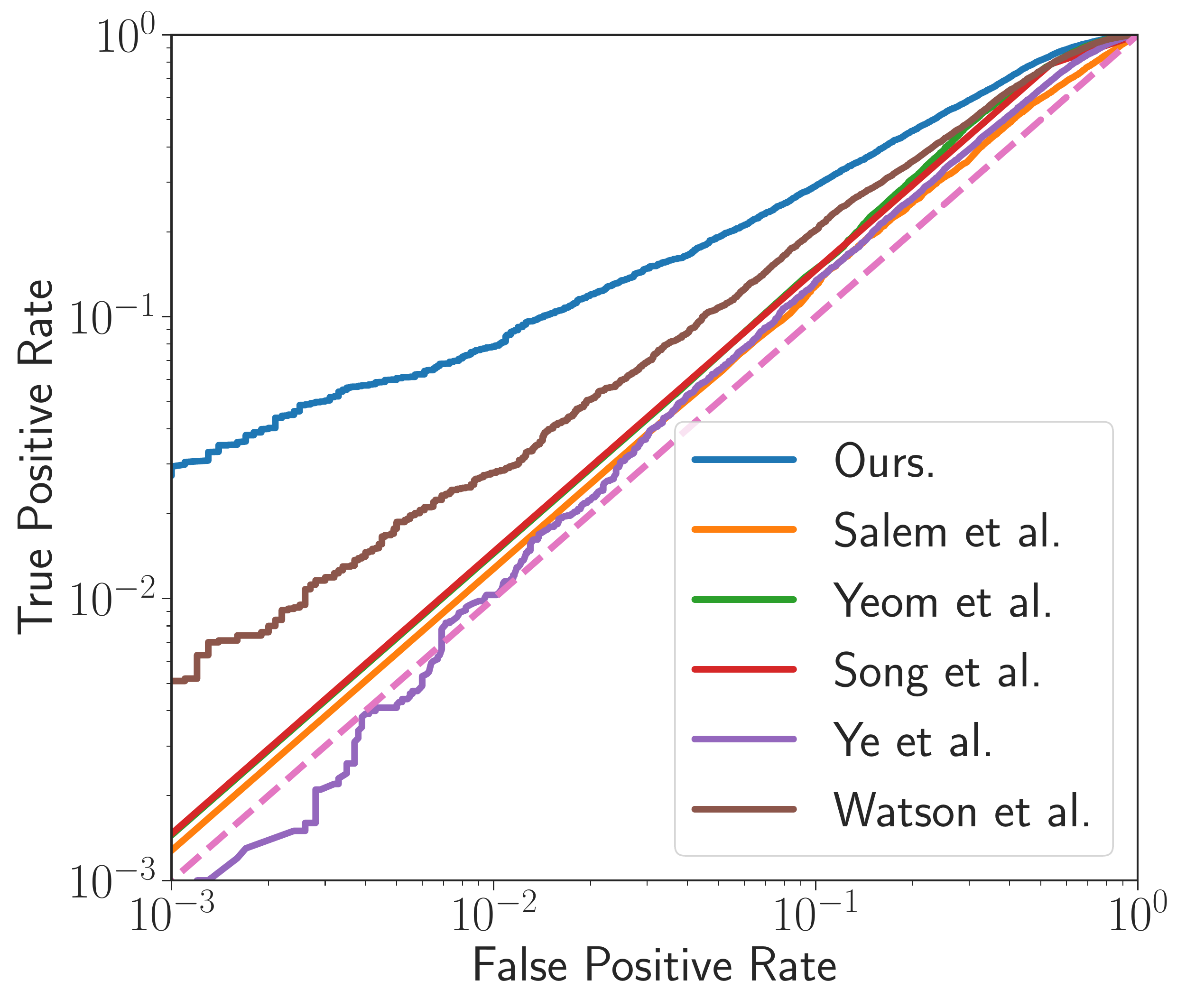}}\hspace{0.1mm}
\subfloat[CIFAR-100]{\includegraphics[width=0.245\linewidth]{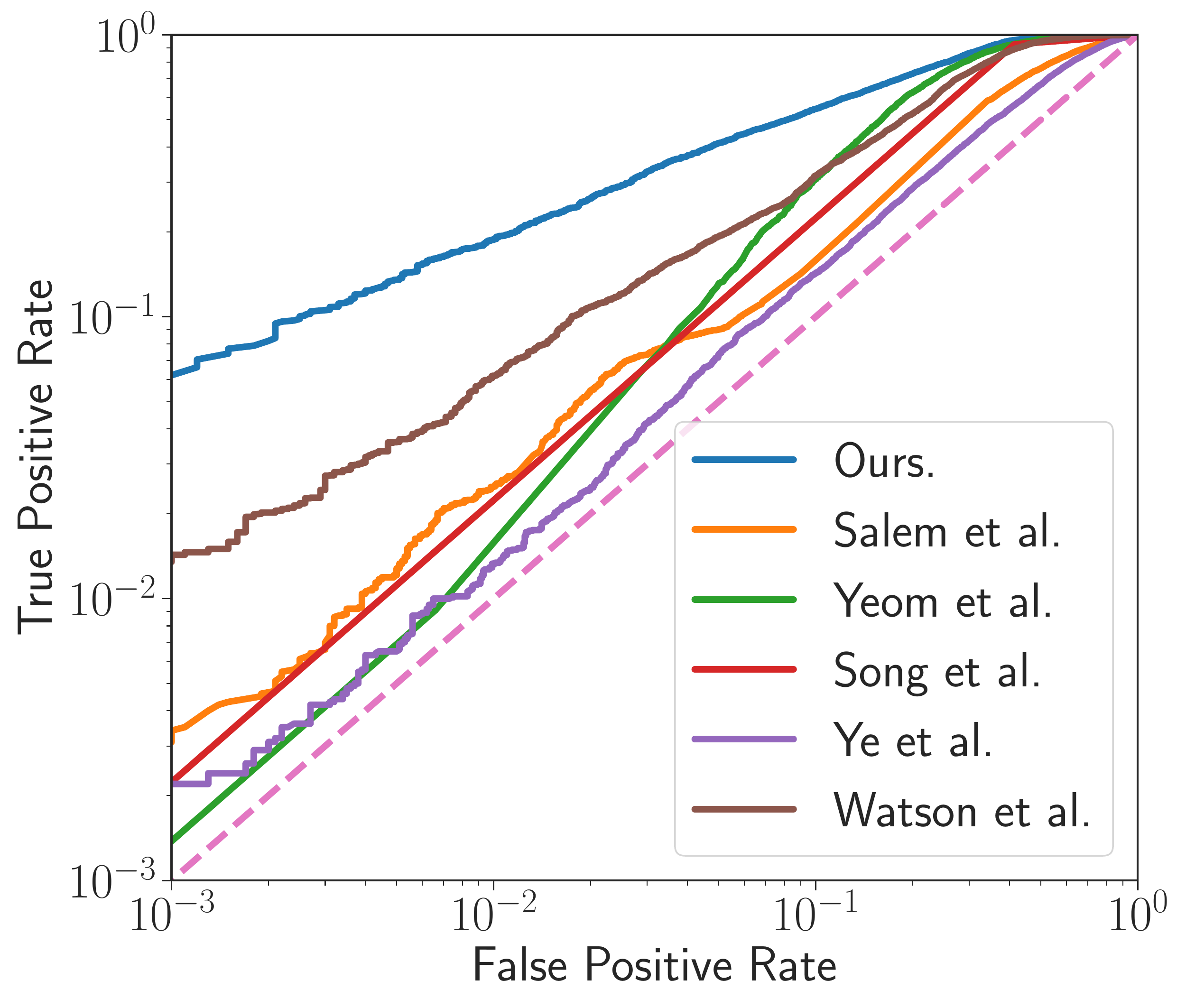}}\hspace{0.1mm}
\subfloat[GTSRB]{\includegraphics[width=0.245\linewidth]{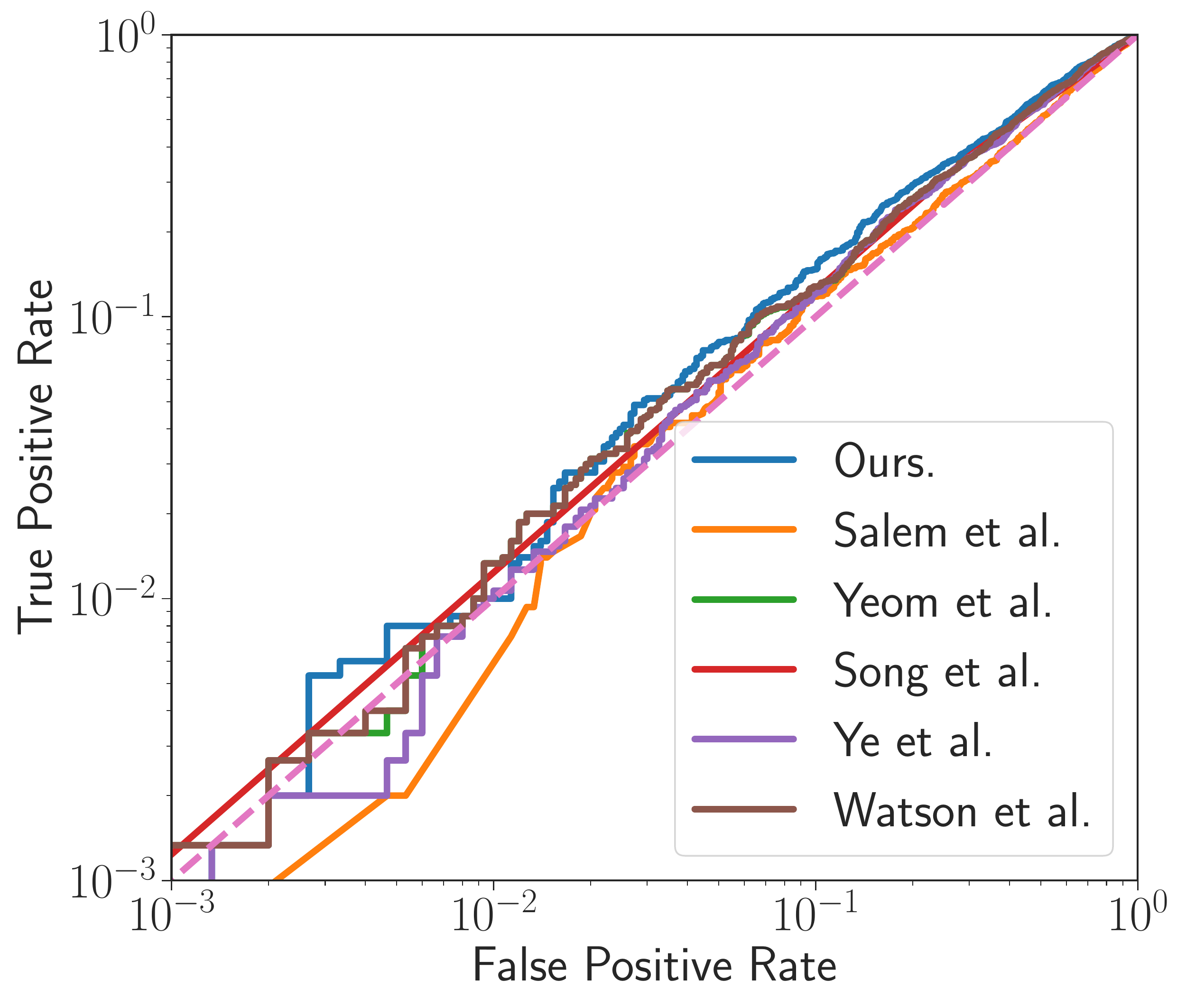}\label{wideresnet_log_auc}}
\caption{ROC curves for attacks on four different datasets and four model architectures (from top to bottom: ResNet-56, MobileNetV2, VGG-16, and WideResNet-32).}
\label{fig:log_auc}
\end{figure*}

\begin{table*}[!t]
\centering
\caption{Attack performance of different attacks against ResNet-56 trained on four datasets. 
Additional results for the other three model architectures with a similar pattern can be found in \autoref{app:model}.}
\label{table:attack_performance_on_resnet}
\setlength{\tabcolsep}{4.0pt}
\scalebox{0.75}{
\begin{tabular}{l|cccccccccccc}
\toprule
\rowcolor{white}
Attack & \multicolumn{4}{c}{TPR at 0.1\% FPR}&  \multicolumn{4}{c}{Balanced accuracy}& \multicolumn{4}{c}{AUC}\\
\cmidrule(l{5pt}r{5pt}){2-5}\cmidrule(l{5pt}r{5pt}){6-9}\cmidrule(l{5pt}r{0pt}){10-13}
method& CIFAR-10& CINIC-10& CIFAR-100& GTSRB&CIFAR-10& CINIC-10& CIFAR-100& GTSRB&CIFAR-10& CINIC-10& CIFAR-100& GTSRB\\
\midrule
Salem et al.~\cite{SZHBFB19}&0.1\%&0.0\%&0.2\%&0.3\%&0.610&0.623&0.577&0.677&0.628&0.646&0.612&0.755\\
Yeom et al.~\cite{YGFJ18}&0.1\%&0.2\%&0.1\%&0.0\%&0.647&0.705&0.772&0.797&0.646&0.755&0.804&0.818\\
Song et al.~\cite{SM21}&0.1\%&0.2\%&0.1\%&0.0\%&0.650&0.707&0.773&0.681&0.644&0.728&0.804&0.820\\
Ye et al.~\cite{YMMS21}&0.0\%&0.1\%&0.2\%&0.0\%&0.527&0.603&0.578&0.606&0.531&0.632&0.605&0.618\\
Watson et al.~\cite{WGCS21}&0.5\%&0.6\%&0.9\%&1.5\%&0.631&0.698&0.727&0.798&0.677&0.735&0.778&0.822\\
\midrule
Ours&\textbf{2.1\%}&\textbf{5.3\%}&\textbf{4.9\%}&\textbf{7.3\%}&\textbf{0.650}&\textbf{0.730}&\textbf{0.800}&\textbf{0.839}&\textbf{0.724}&\textbf{0.819}&\textbf{0.886}&\textbf{0.914}\\
\bottomrule
\end{tabular}
}
\end{table*}

\subsection{Experimental Setup}
\label{setup}

\mypara{Datasets.}
For the main experiments, we consider the following four image datasets:
\begin{itemize}
\item \textbf{CIFAR-10~\cite{CIFAR}.} The CIFAR-10 is a benchmark dataset used for classification tasks, which has totally 60000 images with 10 classes and each class has the same number of samples. 
Each sample has a size of 32$\times$32$\times$3.
\item \textbf{CINIC-10~\cite{DCAS18}.} CINIC-10 is an extension of CIFAR-10 via the addition of downsampled ImageNet~\cite{DDSLLF09} images for the same classes in CIFAR-10. The number of classes is 10 as well but with 270000 images in total. 
The size of each sample is 32$\times$32$\times$3.
\item \textbf{CIFAR-100~\cite{CIFAR}.} Similar to CIFAR-10 dataset, CIFAR-100 also contains 60000 images with a size of 32$\times$32$\times$3. 
And it has 100 classes with 600 images for each class.
\item \textbf{GTSRB~\cite{GTSRB}.} German Traffic Sign Recognition Benchmark (GTSRB) is a classification benchmark with 51839 images for all 43-category traffic signs. 
Since the size of each sample varies, we resize them to 32$\times$32$\times$3.
\end{itemize}

\begin{figure*}[t]
\centering
\subfloat[CIFAR-10]{\includegraphics[width=0.250\linewidth]{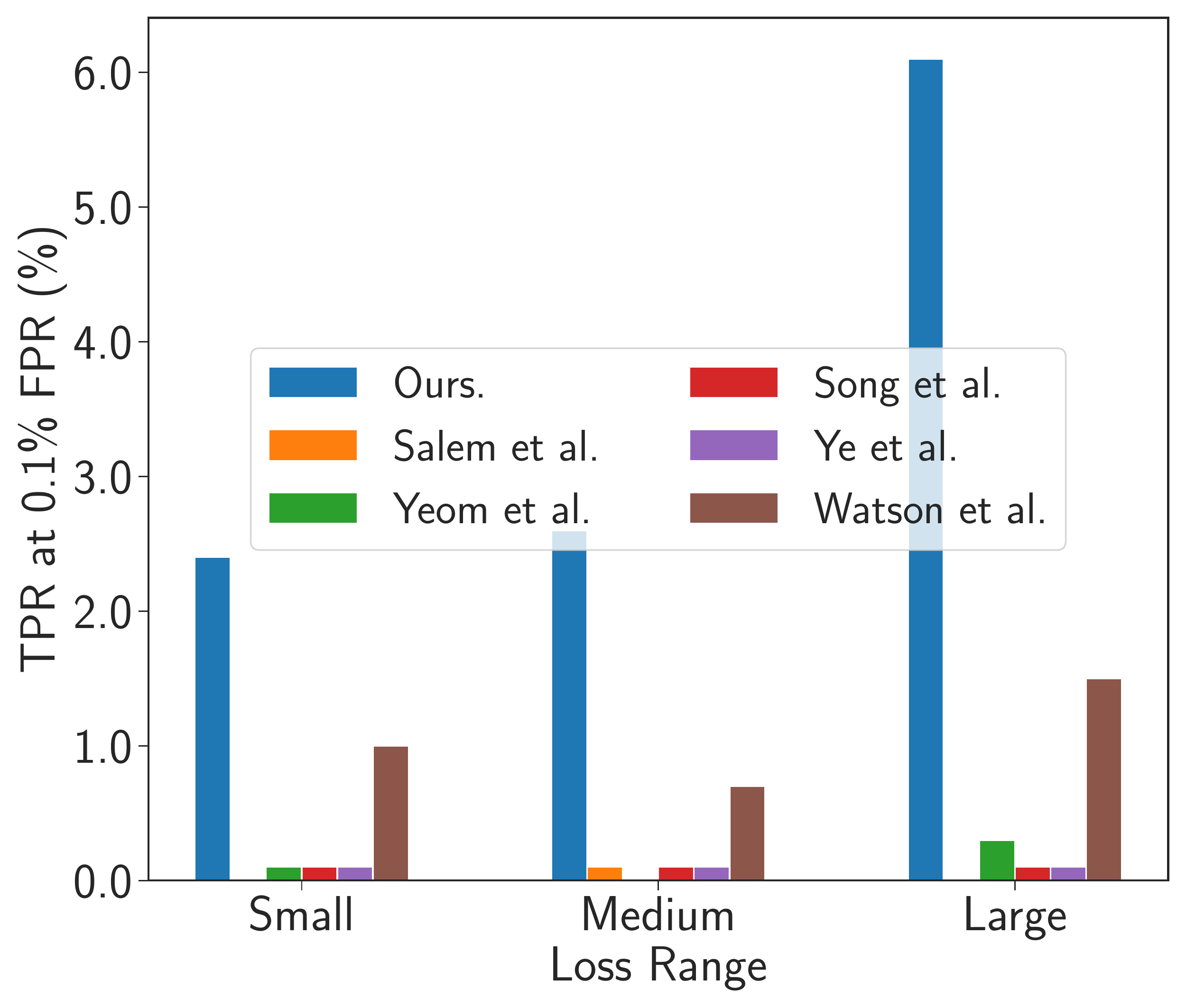}\label{cifar10_tpr_at_each_range_resnet}}
\subfloat[CINIC-10]{\includegraphics[width=0.250\linewidth]{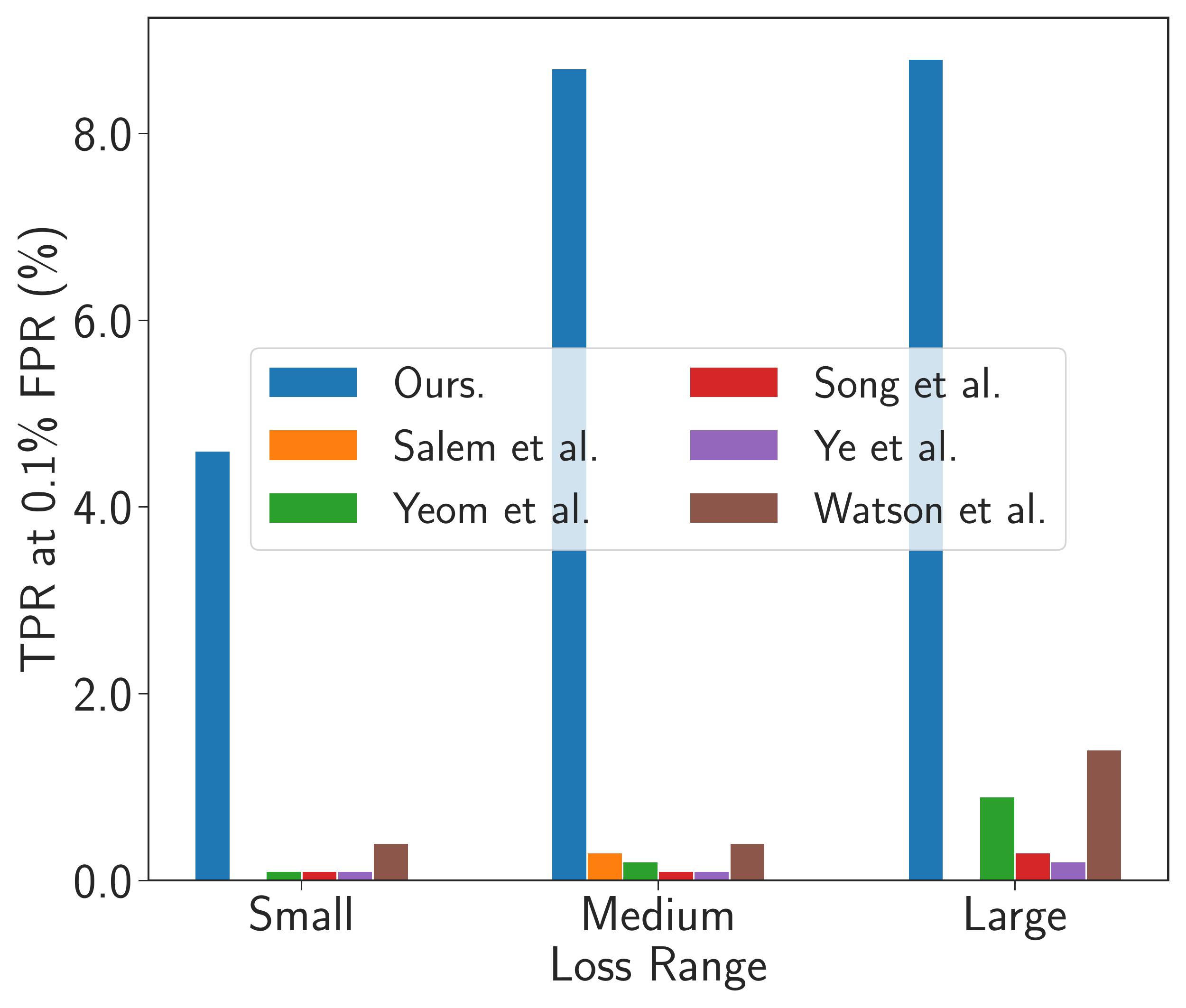}\label{cinic10_tpr_at_each_range_resnet}}
\subfloat[CIFAR-100]{\includegraphics[width=0.250\linewidth]{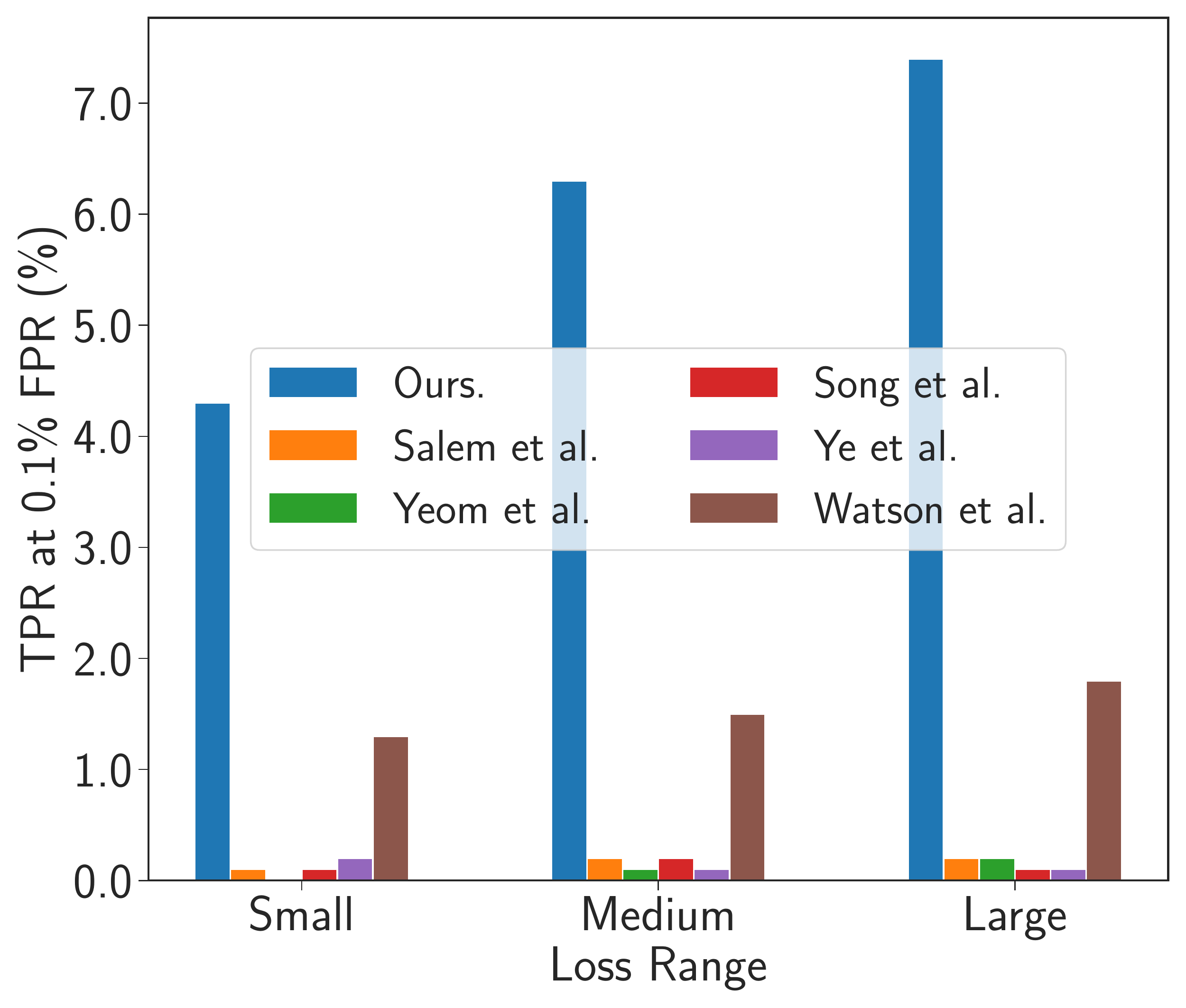}\label{cifar100_at_each_range_resnet}}
\subfloat[GTSRB]{\includegraphics[width=0.250\linewidth]{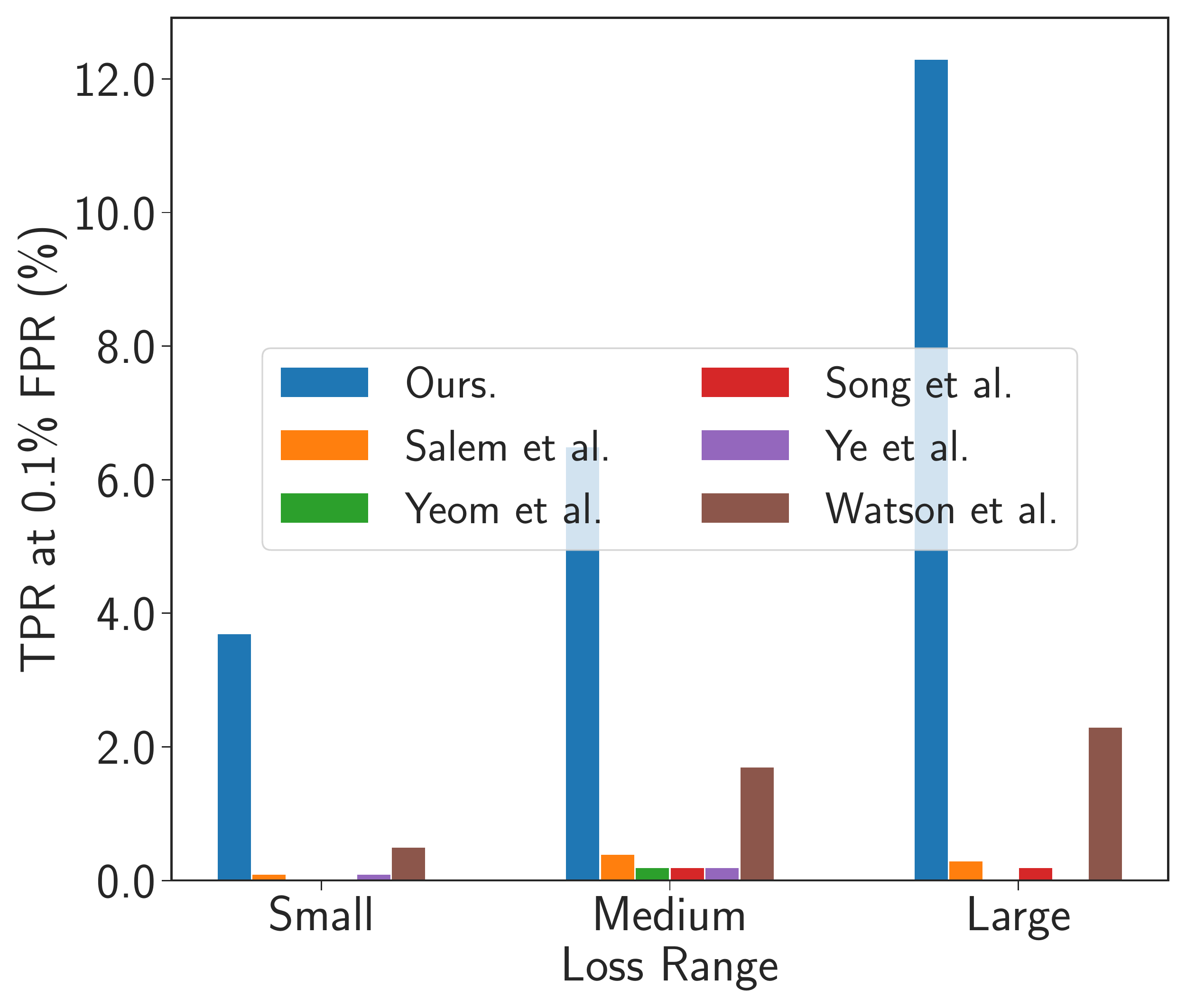}\label{gtsrb_at_each_range_resnet}}
\caption{TPR at 0.1\% FPR of different attacks for ResNet-56 trained on four datasets for samples with different ranges of losses obtained from the target model. 
Here we consider three loss ranges, 'small': [0.0,0.02), 'medium': [0.02,0.2), and 'large': [0.2,$+\infty$]. 
Additional results for the other model architectures with a similar pattern can be found in \autoref{app:model}.}
\label{fig:resnet_tpr_at_each_range}
\end{figure*} 

We also show the effectiveness of our approach on the following three datasets beyond the image domain:
\begin{itemize}
\item \textbf{Purchase.} The Purchase is a simplified dataset based on Kaggle's ``acquire valued shoppers'' dataset (with 197324 records), where each record has 600 binary features. 
Following Shokri et al.~\cite{SSSS17}, we use it to conduct a 100-classes classification task.
\item \textbf{Location~\cite{YZQ16}.} The Location is a ``check-in'' dataset in the Foursquare social network. 
We use the same method in ~\cite{SSSS17} to filter out the whole dataset and get 5010 records with 30 classes in total, and each record has 446 binary features.
\item \textbf{News.} The News (20 Newsgroup) datast is a commonly used dataset for text classification. 
The dataset consists of 20000 newsgroup documents categorized into 20 classes. We follow ~\cite{SZHBFB19} to preprocess the dataset in our experiments.
\end{itemize}

For each dataset, the number of samples for constructing the training and testing sets of target and shadow models ($\mathcal{D}^t_{train}$, $\mathcal{D}^t_{test}$, $\mathcal{D}^s_{train}$, and $\mathcal{D}^s_{test}$, respectively) are the same and the remaining data samples are used for knowledge distillation dataset $\mathcal{D}^k$.
The details of data splitting for different datasets can be found in \autoref{app:split}.

\mypara{Models.} 
For image datasets, we consider four popular neural network architectures including ResNet-56~\cite{HZRS16}, MobileNetV2~\cite{SHZZC18}, VGG-16~\cite{SZ15}, and WideResNet-32~\cite{ZK16} for the target, shadow, and distilled models. 
And for the other three datasets, we apply a 2-layer MLP.
We use SGD with a learning rate of 0.1, Nesterov momentum of 0.9 and a cosine learning rate schedule for optimization. 
We also adopt data augmentations for images in order to make the models more generalized.
All the models are trained from 20 to 150 epochs in terms of the model size and dataset complexity.
For the attack model, we train a 4-layer MLP.

\mypara{Metrics.}
We consider the following evaluation metrics:
\begin{itemize}
\item \textbf{Full Log-scale ROC.} It is the commonly-used Receiver Operating Characteristic (ROC) curve comparing the ratio of true-positives to false-positives, but reported in logarithmic scale for emphasizing the low FPR regime~\cite{CCNSTT21} 
\item \textbf{TPR at Low FPR.} It summarizes the attack performance at a single low false-positive rate for quick evaluation~\cite{CCNSTT21}. 
We also take a step further to apply this metric to separate groups of samples that have different levels of loss obtained from the target model.
\item \textbf{Balanced Accuracy and AUC.} They are two widely used average-case metrics to measure the performance for binary classification tasks, including most previous MIAs~\cite{MSCS19,SZHBFB19,WGCS21,YMMS21}. 
Here the ``Balanced'' means that the number of the member and non-member samples is the same. 
Since they are not the most suitable metrics for evaluating MIAs, we adopt them here just for completeness.
\end{itemize}

\mypara{Attack Baselines.}
We mainly compare our new MIA method with five representative existing methods~\cite{SZHBFB19,YGFJ18,SM21,YMMS21,WGCS21} as the baselines. 
Among them, Salem et al.~\cite{SZHBFB19} utilize posteriors to conduct the attack while Yeom et al.~\cite{YGFJ18} leverage the loss from target model; Song et al.~\cite{SM21} propose a metric-based method without the use of attack model; Watson et al.~\cite{WGCS21} and Ye et al.~\cite{YMMS21} both consider sample hardness where the former use reference models and the latter leverage distilled models.
To ensure a fair comparison, those methods that involve model training have access to the same auxiliary dataset, and use a single shadow/reference model like ours.
Please refer to \autoref{sec:mia} for more descriptions of these five methods.

\subsection{Experimental Results}
\label{sec:result}

Here, we show the detailed attack results in the black-box setting with a comparison to 5 advanced baseline methods; among them, we compare our attack to a state-of-the-art method LiRA as well. 
Besides, we apply our attack in the label-only scenario and also attack the model defended with DP-SGD. \autoref{table:train_test_acc} reports the performance of the models we attack.

\mypara{Black-box Attacks.}
We first evaluate different attacks in the black-box scenario.
As can be seen from \autoref{fig:log_auc}, our method consistently achieves the best performance on the low-FPR regime.
The TPR at 0.1\% FPR further confirms this conclusion, as shown in \autoref{table:attack_performance_on_resnet}.
Regarding the two aggregate metrics, balanced accuracy and AUC, we can also observe that our method strictly dominates all the baselines.

We then evaluate different attacks at a more fine-grained level by looking at the TPR at 0.1\% FPR on separate groups of samples that have different levels of loss values from the target model.
Here we consider three levels of loss values, i.e., 'small': [0.0,0.02), 'medium': [0.02,0.2), and 'large': [0.2,+$\infty$).
For the GTSRB dataset, we run the experiments for five times and take the average because some loss ranges may suffer from unstable results due to the limited number of data samples.
As demonstrated in \autoref{fig:resnet_tpr_at_each_range}, our attack outperforms all other baselines across all different loss ranges.
The above experimental observations also hold for the other three model architectures (MobileNetV2, VGG-16, and WideResNet-32) and three datasets (Purchase, Location, and News), for which the results can be found in \autoref{app:NLP}.

\begin{table}[!t]
\definecolor{mygray}{gray}{1}
\newcolumntype{a}{>{\columncolor{mygray}}c}
\centering
\caption{Attack performance of our attack and LiRA (online version)~\cite{CCNSTT21} for ResNet-56 trained on CINIC-10.}
\label{table:result_of_online_offline_attack}
\setlength{\tabcolsep}{3.0pt}
\scalebox{0.75}{
\begin{tabular}{l|ca}
\rowcolor{white}
\toprule
Metric& LiRA~\cite{CCNSTT21}&Ours \\
\midrule
TPR at 0.1\% FPR& 5.0\%& \textbf{5.3\%}\\
Balanced accuracy & 0.713& \textbf{0.730}\\
AUC& 0.793& \textbf{0.819}\\
\bottomrule
\end{tabular}
}
\end{table}

\begin{figure}[t]
\centering
\includegraphics[width=0.6\linewidth]{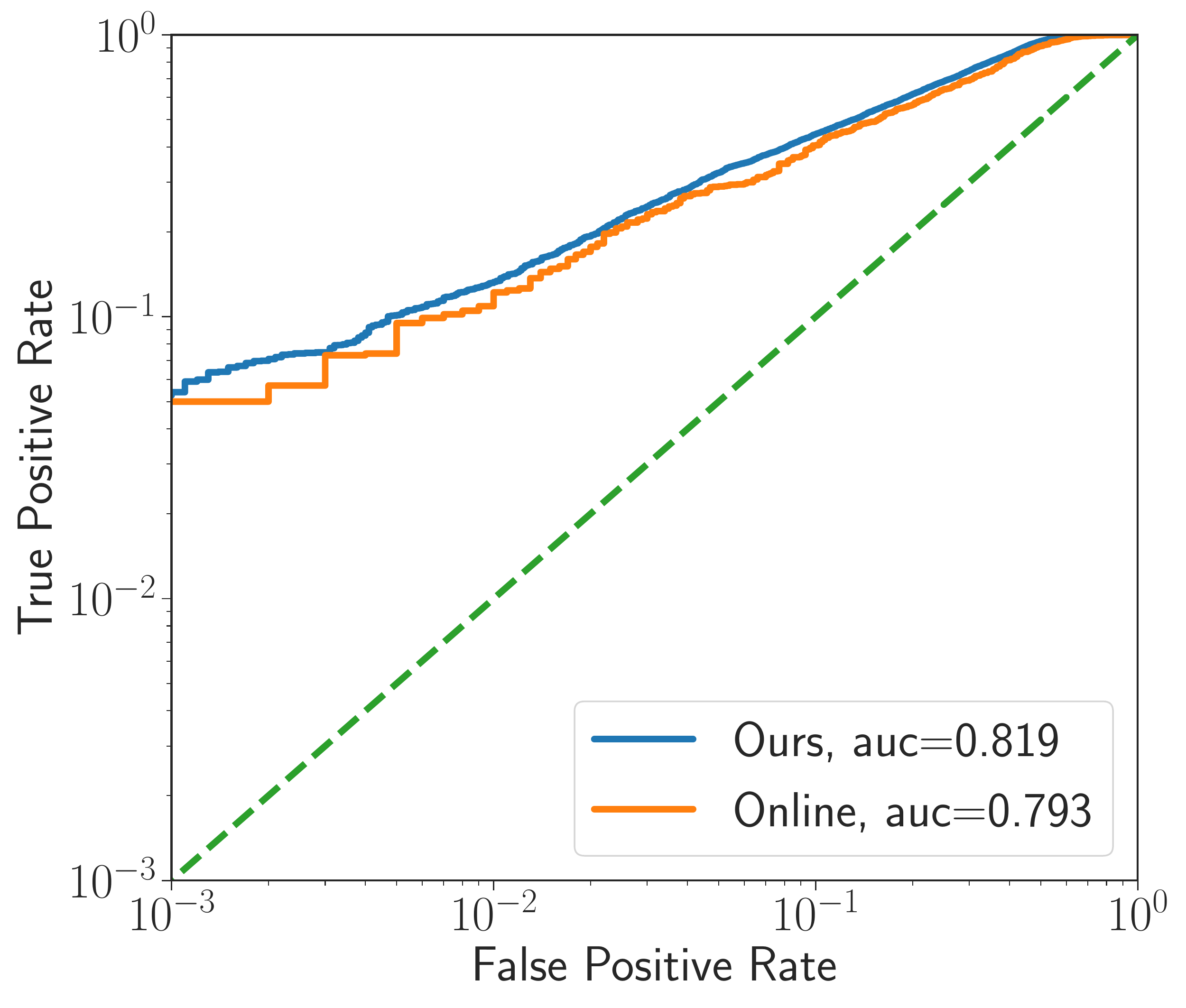}
\caption{ROC curves of our attack compared with LiRA (online version)~\cite{CCNSTT21} for ResNet-56 trained on CINIC-10.}
\label{fig:attack_online_offline}
\end{figure}

\mypara{Comparison with LiRA.}
Besides the evaluation recommendation on the low-FPR regime, Carlini et al.~\cite{CCNSTT21} also introduce a new attack method called Likelihood Ratio Attack (LiRA), which achieves the state-of-the-art attack performance.
For LiRA, the adversary trains N shadow models, of which half are IN models (trained with target sample ($\DataPoint,y$) and the other half are OUT models (trained without $(\DataPoint,y)$). 
Then Gaussians are fitted to the confidences of the IN and OUT models on $(\DataPoint,y)$. 
Finally, the confidence of $(\DataPoint,y)$ from the target model will be used to conduct a parametric Likelihood-ratio test.

\begin{table}[!t]
\definecolor{mygray}{gray}{1}
\newcolumntype{a}{>{\columncolor{mygray}}c}
\centering
\caption{TPR at 0.1\% FPR of different attacks on CINIC-10 in the label-only setting. }
\label{table:cinic10_label_only_attack_tpr_at_low_fpr}
\setlength{\tabcolsep}{2pt}
\scalebox{0.75}{
\begin{tabular}{l|cca}
\toprule
\rowcolor{white}
Model Architecture& Li \& Zhang~\cite{LZ21}&  Watson et al.~\cite{WGCS21}& Ours\\
\midrule
MobileNetV2&  {0.2\%}& 0.2\%&  \textbf{0.8\%}\\
VGG-16&  {0.2\%}& 0.1\%&  \textbf{0.3\%}\\
ResNet-56&  \textbf{0.3\%}& 0.2\%&  \textbf{0.3\%}\\
WideResNet-32&  {0.0\%}& 0.0\%&  \textbf{0.1\%}\\
\bottomrule
\end{tabular}
}
\end{table}

\begin{table}[!t]
\definecolor{mygray}{gray}{1}
\newcolumntype{a}{>{\columncolor{mygray}}c}
\centering
\caption{Balanced acc of different attacks on CINIC-10 in the label-only setting.}
\label{table:cinic_label_only_attack_acc}
\setlength{\tabcolsep}{2pt}
\scalebox{0.75}{
\begin{tabular}
{l|cca}\toprule
\rowcolor{white}
Model Architecture& Li \& Zhang~\cite{LZ21}&  Watson et al.~\cite{WGCS21}& Ours\\
\midrule
MobileNetV2&  {0.782}& 0.761&  \textbf{0.828}\\
VGG-16&  {0.735}& 0.728&  \textbf{0.806}\\
ResNet-56&  {0.713}& 0.676&  \textbf{0.733}\\
WideResNet-32&  {0.615}& 0.595&  \textbf{0.667}\\
\bottomrule
\end{tabular}
}
\end{table}

Following the original work of LiRA, 256 shadow models (128 IN models and 128 OUT models) are trained.
We find that using fewer shadow models results in worse attack performance, and using only (128) out models also leads to a low performance, 2.1\%.
To ensure a fair comparison, LiRA also queries the target model once for each sample instead of multiple times, the same as in our attack.
It can be seen from \autoref{table:result_of_online_offline_attack} that LiRA achieves comparable (but a bit lower) attack performance to ours, and the detailed comparison in terms of ROC curve in \autoref{fig:attack_online_offline} further confirms this. 
However, LiRA is not practically feasible due to the necessity of training N shadow models for each given target sample at inference time. 
In contrast, our attack requires no inference-time model training but only queries to obtain the corresponding loss trajectory.

\begin{table}[!t]
\newcommand{\tabincell}[2]{\begin{tabular}{@{}#1@{}}#2\end{tabular}}
\centering
\caption{Attack performance of our attack against DP-SGD for ResNet-56 trained on CINIC-10.}
\label{table:result_of_model_trained_with_dp}
\setlength{\tabcolsep}{2.5pt}
\scalebox{0.75}{
\begin{tabular}{l|cccc}
\toprule
\rowcolor{white}
Noise& \multicolumn{4}{c}{C = 10}\\
\rowcolor{white}
Multiplier ($\sigma$)& $\epsilon$& Model acc& Attack acc& TPR at 0.1\% FPR\\
\midrule
0.0& $\infty$& 0.520& 0.560& 0.2\%\\
0.2& > 1000& 0.485& 0.544& 0.2\%\\
0.5& > 100& 0.438& 0.528& 0.1\%\\
1.0& 8& 0.332& 0.512& 0.1\%\\
\bottomrule
\end{tabular}
}
\end{table}

\begin{figure}[t]
\centering
\includegraphics[width=0.6\linewidth]{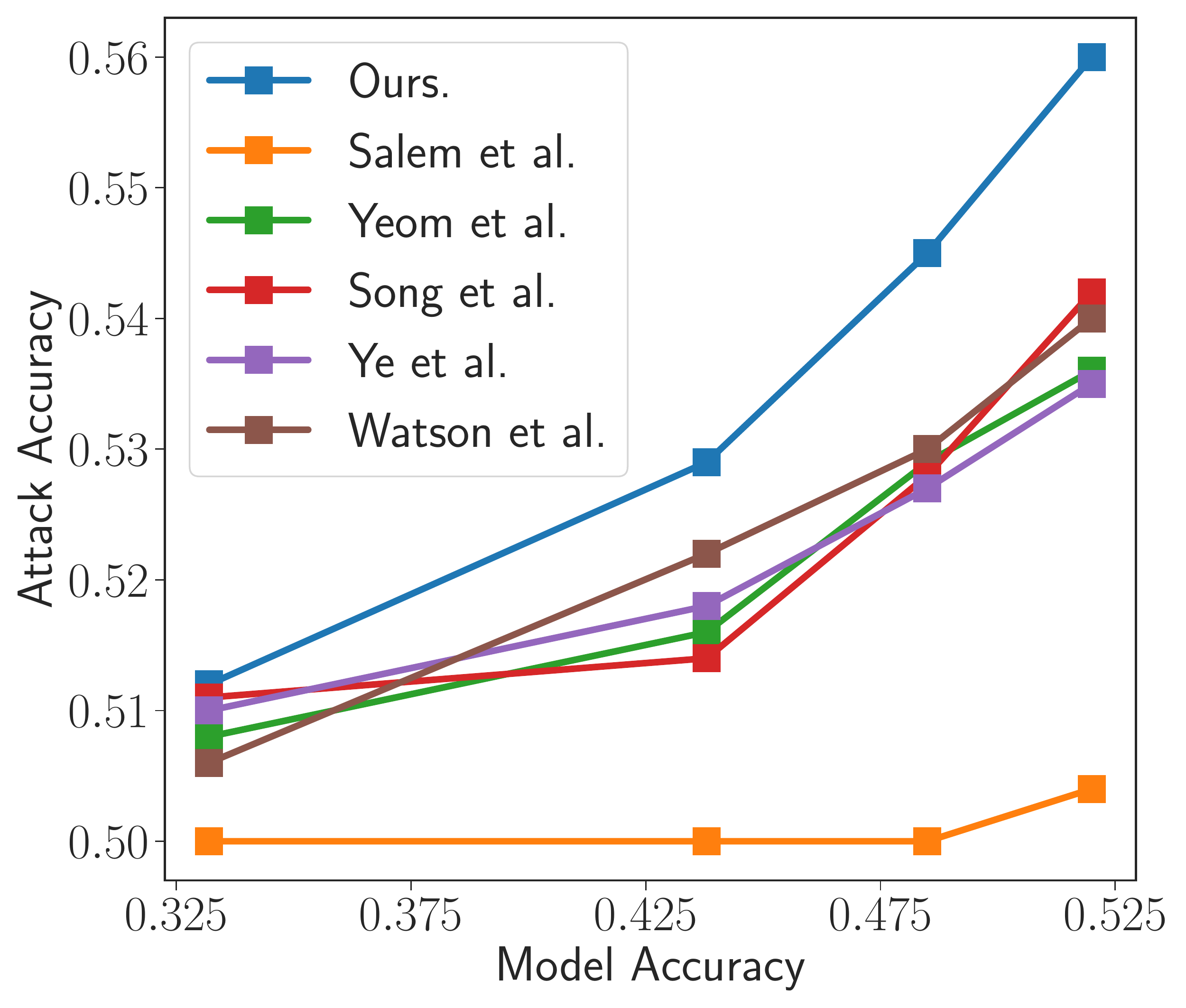}
\caption{The trade-off between model accuracy and attack accuracy of different attacks against DP-SGD with different noise multiplier for ResNet-56 trained on CINIC-10.}
\label{fig:attack_dp_sgd}
\end{figure}

\mypara{Label-only Attacks.}
When the target model only returns a hard prediction, that is a single label, rather than a continuous-valued output like posteriors, membership inference can still be conducted by label-only attacks. 
Here we use the CINIC-10 dataset and ResNet-56 to evaluate our attack in the label-only scenario. 
We compare the result to the original boundary-attack in~\cite{LZ21} and label-only attacks with calibration in~\cite{WGCS21} as other baselines do not take this scenario into consideration. 

\autoref{table:cinic10_label_only_attack_tpr_at_low_fpr} and \autoref{table:cinic_label_only_attack_acc} report the attack performance between \system and other baselines. 
Similar to the results in \autoref{table:attack_performance_on_resnet}, label-only attacks can also benefit from loss trajectory with increased TPR at 0.1\% FPR and balanced attack accuracy. 
However, the performance increase is observably smaller compared to the results in score-based attacks. 
One possible reason is that solely relying on the hard predicted labels from the target model, the \emph{distilled loss trajectory} contains much less information than using the posteriors; besides, due to the same reason, the loss from the original model is replaced by the corresponding HopSkipJump boundary distance, which is not as reliable as using the loss to infer the membership.

\mypara{Attacking DP-SGD.}
Differential Privacy (DP)~\cite{DMNS06} is a widely used mechanism to defend machine learning models against different privacy attacks~\cite{LAGHJ19,KTPPI20,SS19}. 
Essentially, it provides a bound on the ability to distinguish two neighboring datasets that differ in the presence of one data sample, which has a close connection to our problem of membership inference.

Here we adopt the popular mechanism DP-SGD~\cite{ACGMMTZ16} to evaluate our attack under the DP-based defense. 
We fix the clip bound C to 10 and change the noise multiplier from $0.0$ to $1.0$ to control the privacy budget $\epsilon$.
As can be seen from \autoref{table:result_of_model_trained_with_dp} and \autoref{fig:attack_dp_sgd}, applying DP in the training process achieves strong defense effects against all attacks, although our attack still outperforms others across all privacy budgets. 
However, DP also reduces the classification accuracy heavily even when $C=10$, $\sigma=0.0$, and $\epsilon=\infty$. 
This trade-off between defense strength and classification accuracy makes DP not feasible in practical scenarios. 

\begin{table}[!t]
\newcommand{\tabincell}[2]{\begin{tabular}{@{}#1@{}}#2\end{tabular}}
\definecolor{mygray}{gray}{1}
\newcolumntype{a}{>{\columncolor{mygray}}c}
\centering
\caption{The impact of the knowledge distillation set size for ResNet-56 trained on CINIC-10. The accuracy of the target model is 60.7 \%.}
\label{table:test_acc_with_different_model_stealing_size}
\setlength{\tabcolsep}{5pt}
\scalebox{0.75}{
\begin{tabular}{l|ccc}
\toprule
\rowcolor{white}
\tabincell{l}{Knowledge distillation\\set size} & \tabincell{c}{Distilled\\model acc} &\tabincell{c}{Attack\\acc}&\tabincell{c}{Attack\\TPR at 0.1\% FPR}  \\
\midrule
20000& 63.5\%&0.713&1.8\%\\
70000& 63.4\%&0.720&2.2\%\\
120000& 63.5\%&0.729&3.5\%\\
170000& 63.9\%&0.725&2.8\%\\
220000& 63.6\%&0.730&5.3\%\\
\bottomrule
\end{tabular}
}
\end{table}

\begin{figure}[t]
\centering
\includegraphics[width=0.6\linewidth]{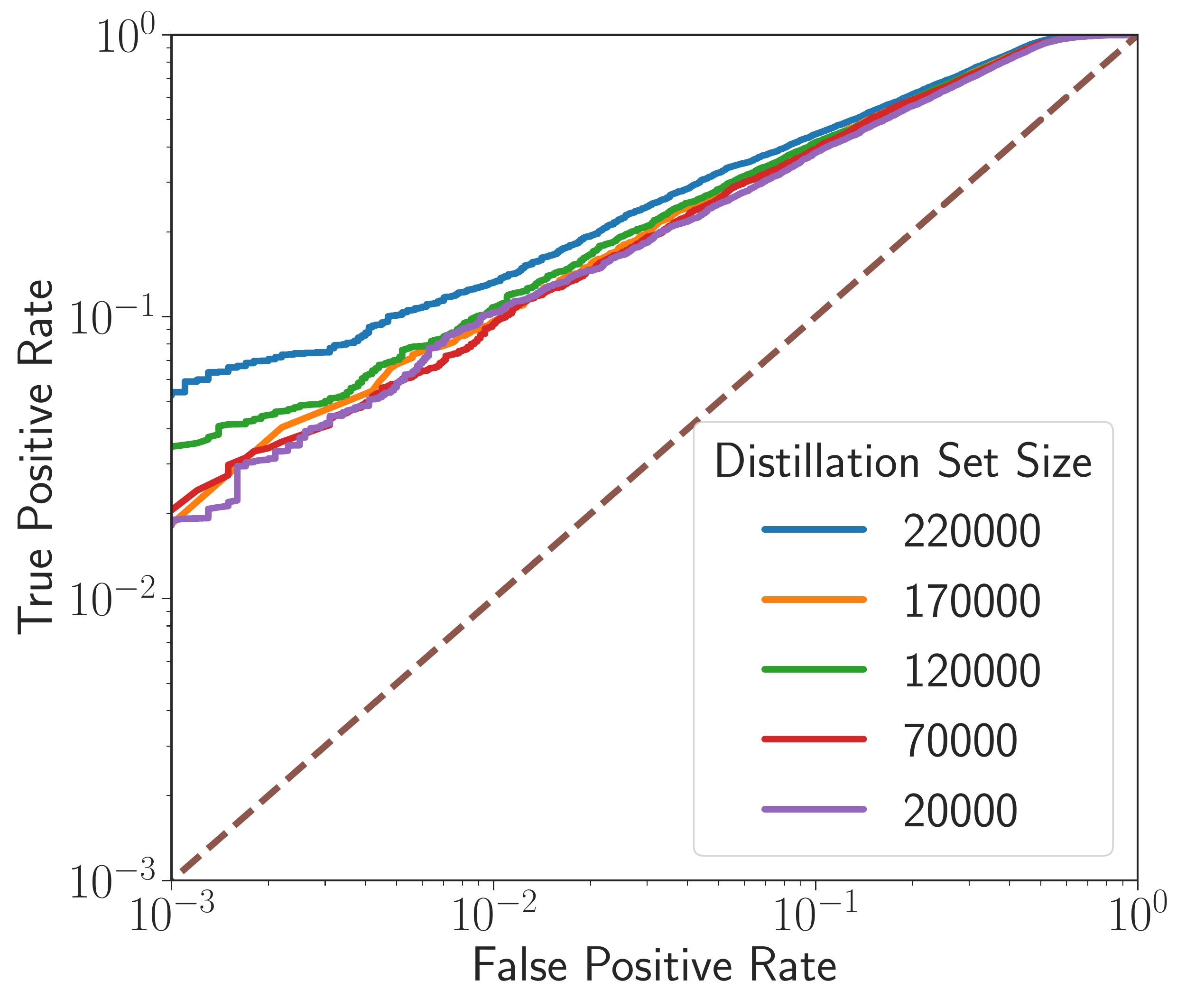}
\caption{ROC curves of our attack with different knowledge distillation set size for ResNet-56 trained on CINIC-10.}
\label{fig:knowledge_distillation_set_size_attack_performance}
\end{figure}

\section{Ablation Study}
\label{sec:Ablation}

In this section, we analyze the impact of several important factors on attack performance. 
We first discuss the impact of the size of the knowledge distillation dataset as well as the number of epochs used in distillation for getting the distilled loss trajectory. 
We then explore the impact of the overfitting level of the target model by changing the size of the target model training set. 
Finally, we relax the two major assumptions about the adversary, namely the data distributions of the auxiliary dataset $\mathcal{D}^a$ and the architectures of the target model.

\subsection{Knowledge Distillation Set Size}

\begin{figure*}[t]
\centering
{\includegraphics[width=0.245\linewidth]{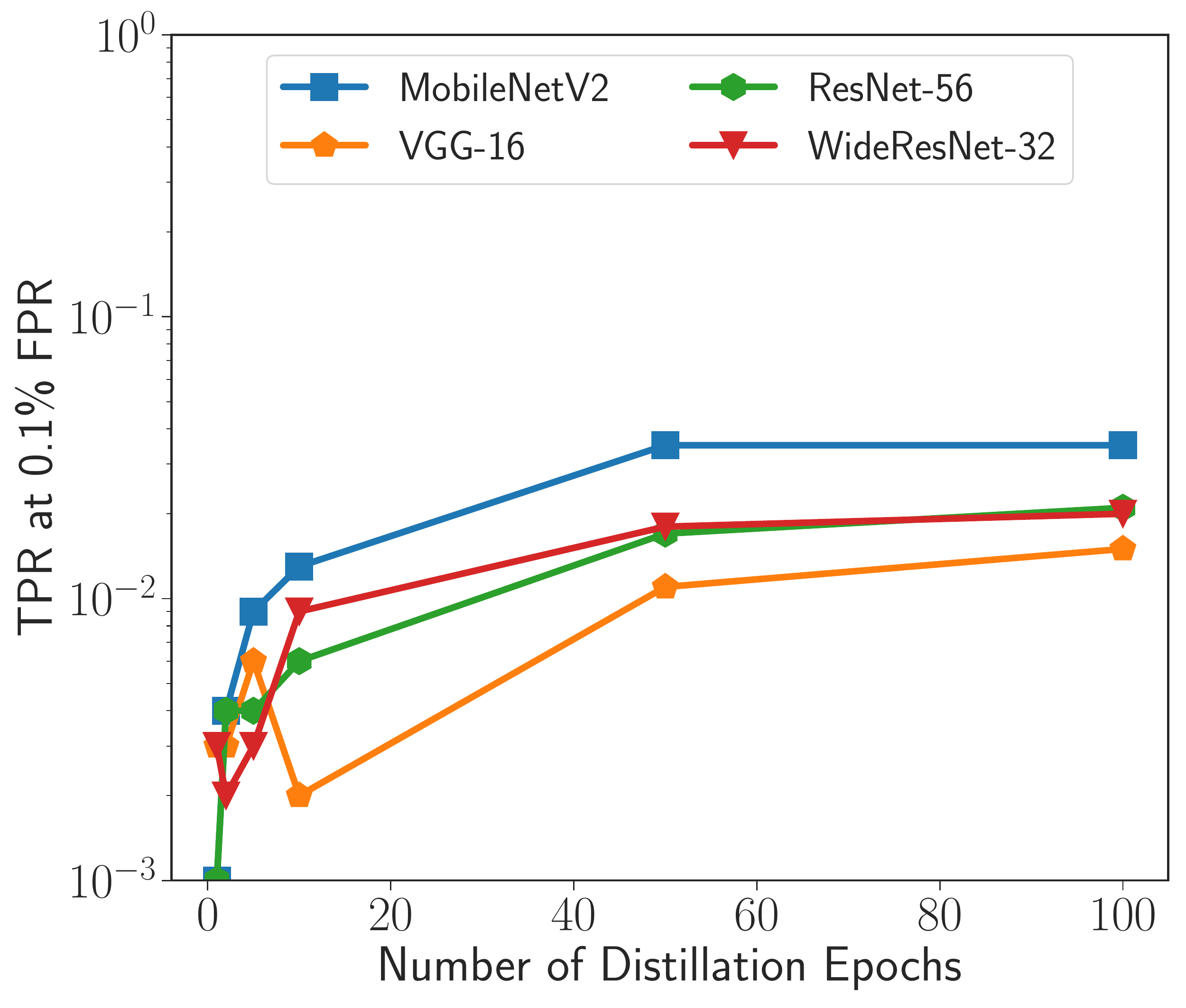}\label{cifar10_resnet_model_stealing_epochs_attack_auc}
\includegraphics[width=0.245\linewidth]{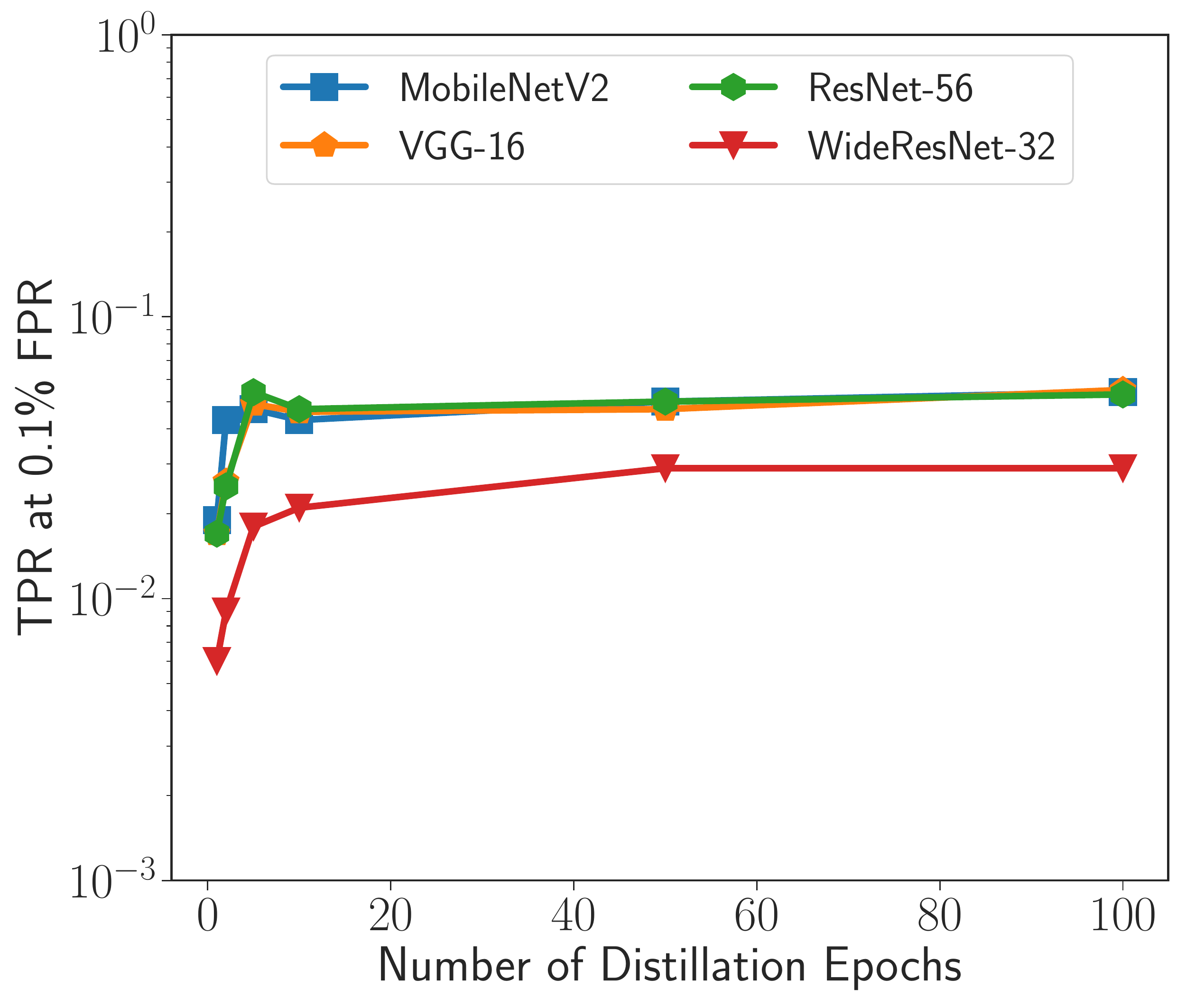}\label{cinic10_resnet_model_stealing_epochs_attack_auc}
\includegraphics[width=0.245\linewidth]{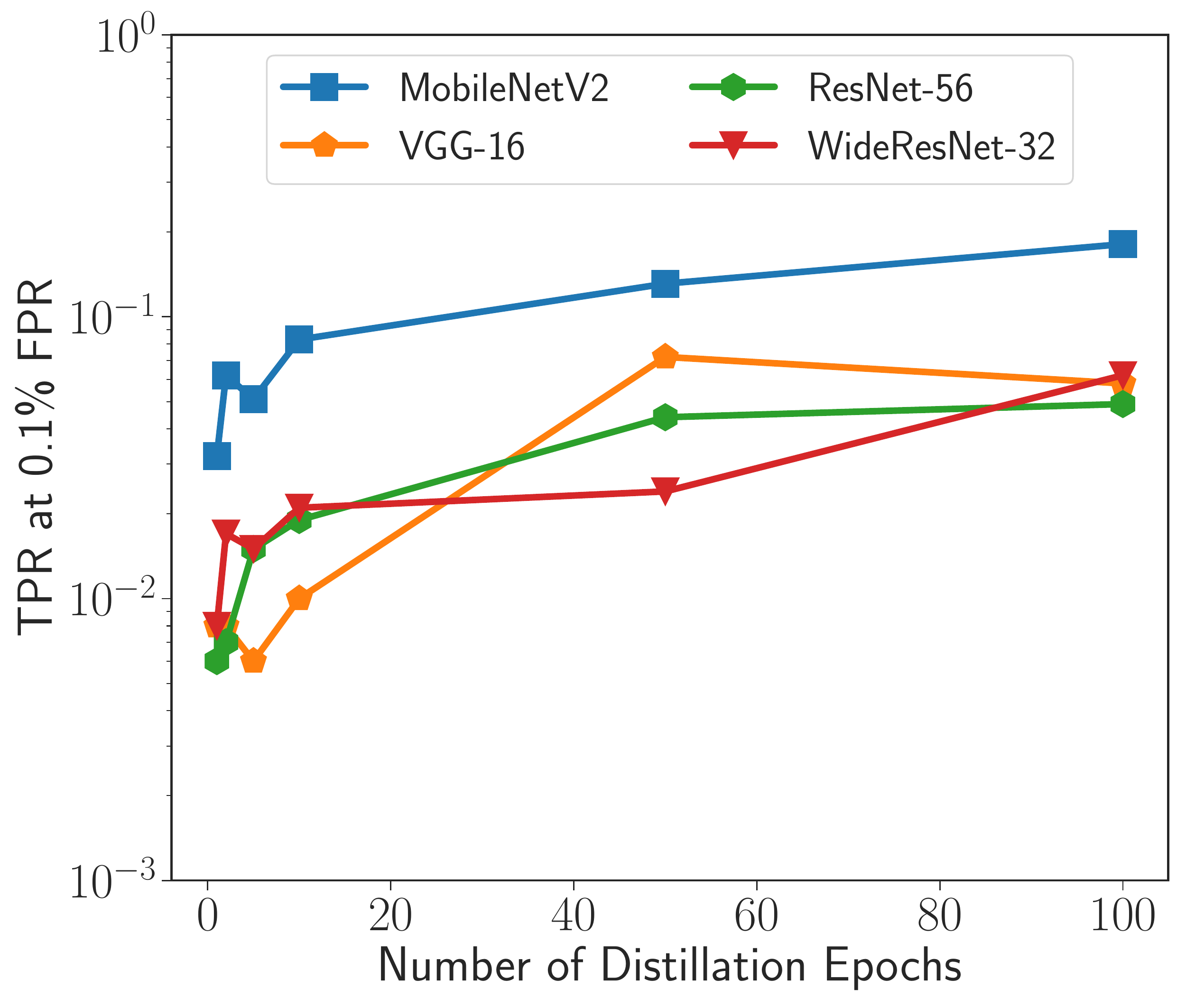}\label{cifar100_esnet_model_stealing_epochs_attack_auc}
\includegraphics[width=0.245\linewidth]{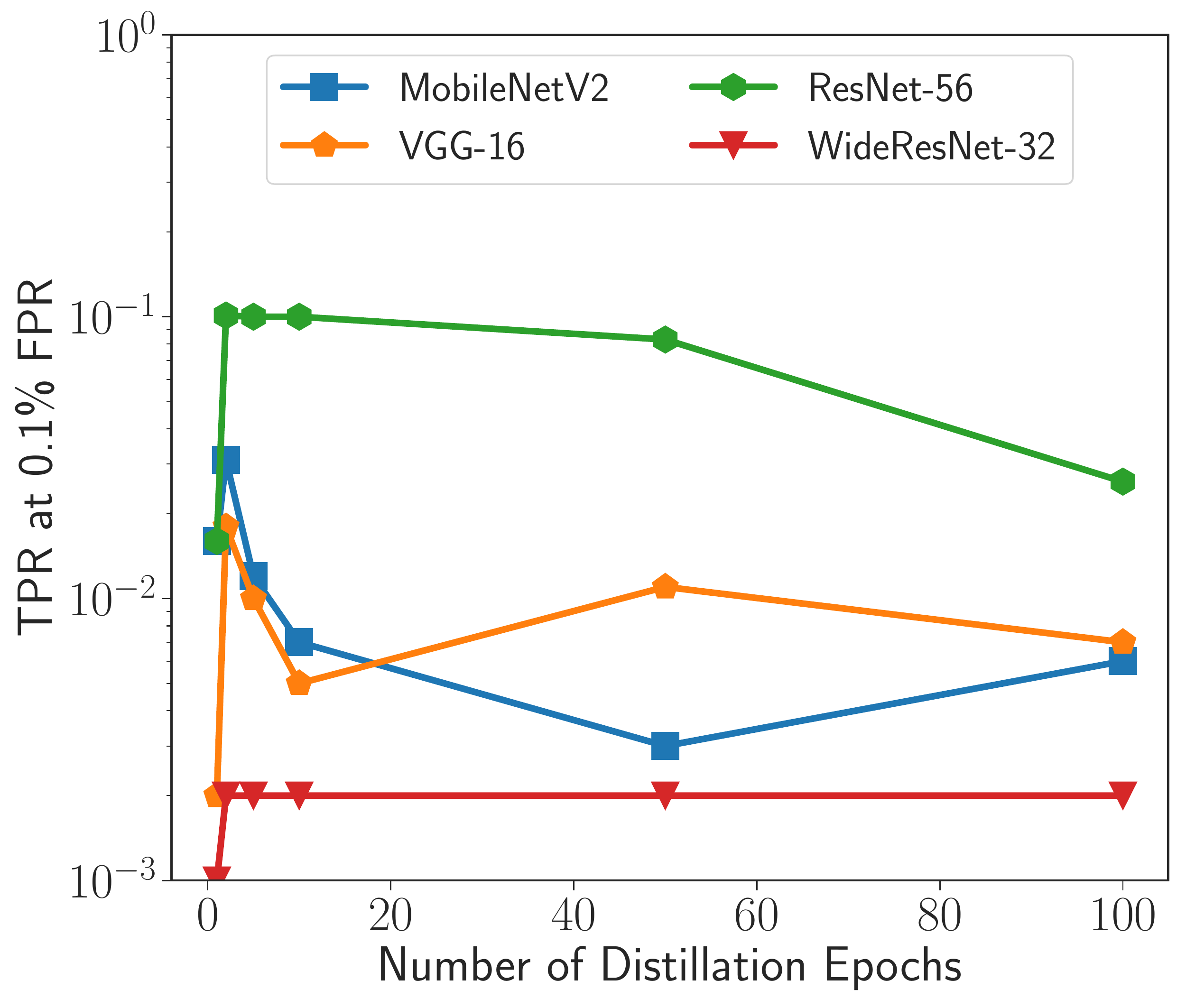}\label{gtsrb_resnet_model_stealing_epochs_attack_auc}}
\subfloat[CIFAR-10]{\includegraphics[width=0.25\linewidth]{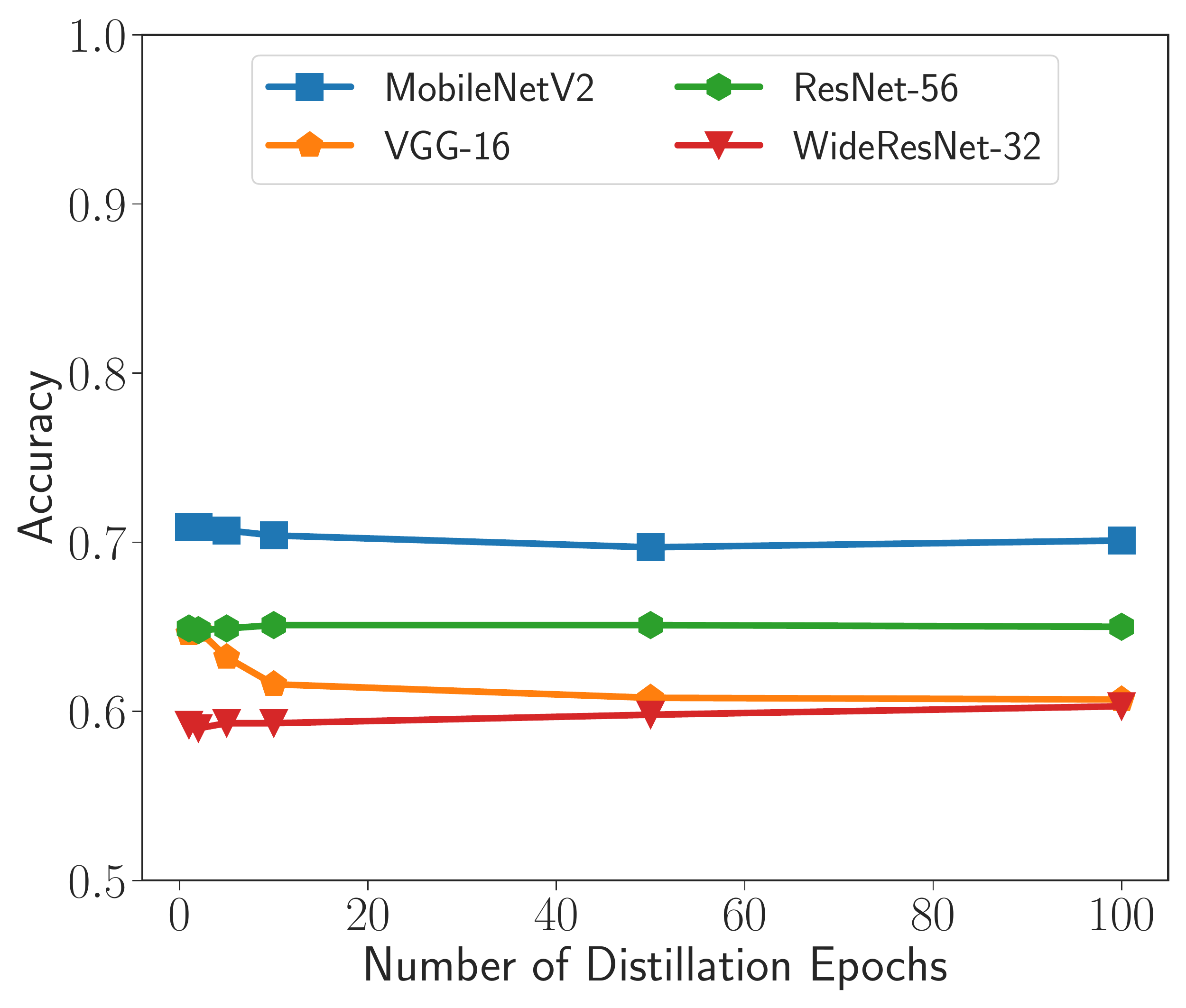}\label{cifar10_resnet_model_stealing_epochs_attack_acc}}
\subfloat[CINIC-10]{\includegraphics[width=0.25\linewidth]{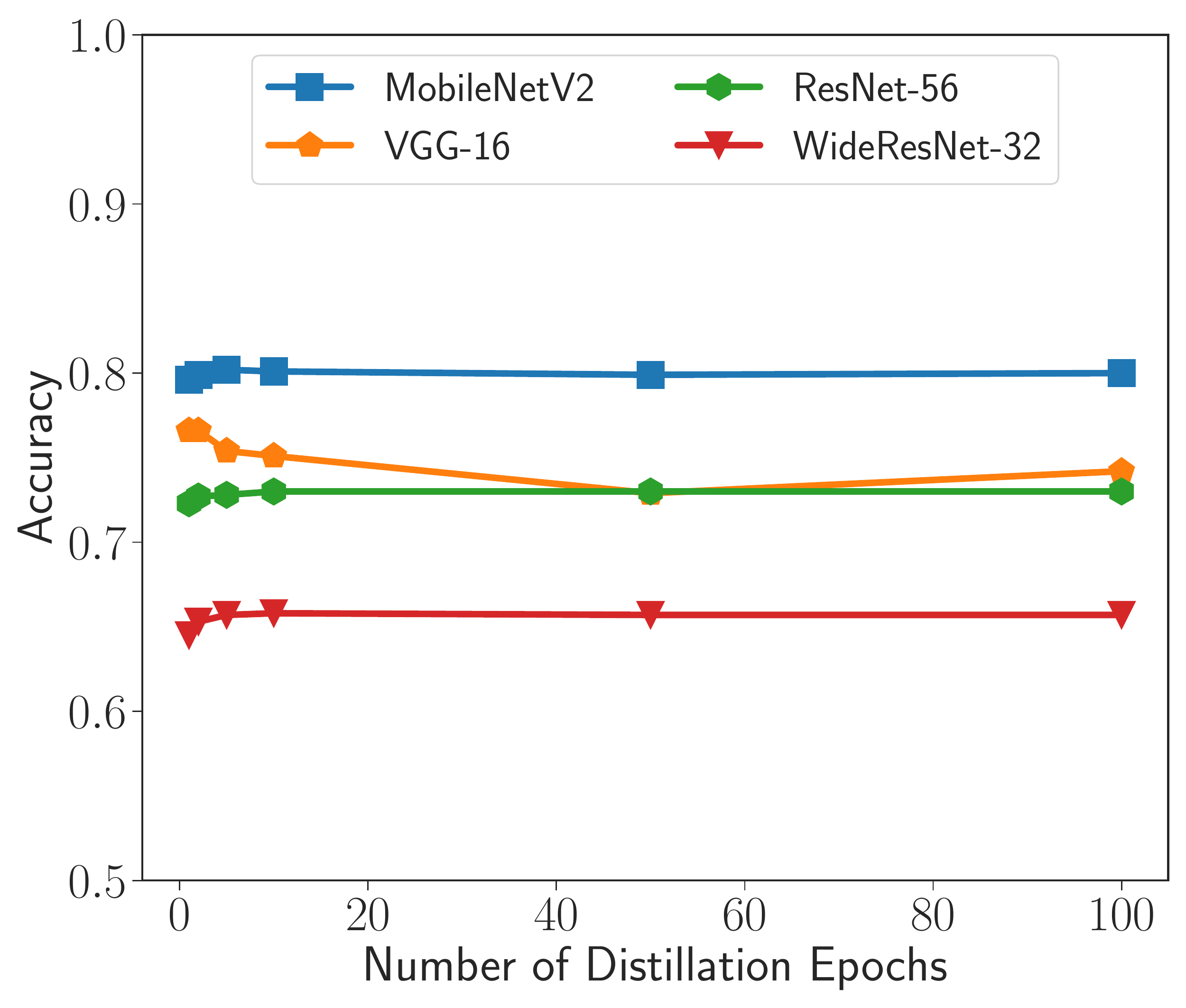}\label{cinic10_resnet_model_stealing_epochs_attack_acc}}
\subfloat[CIFAR-100]{\includegraphics[width=0.25\linewidth]{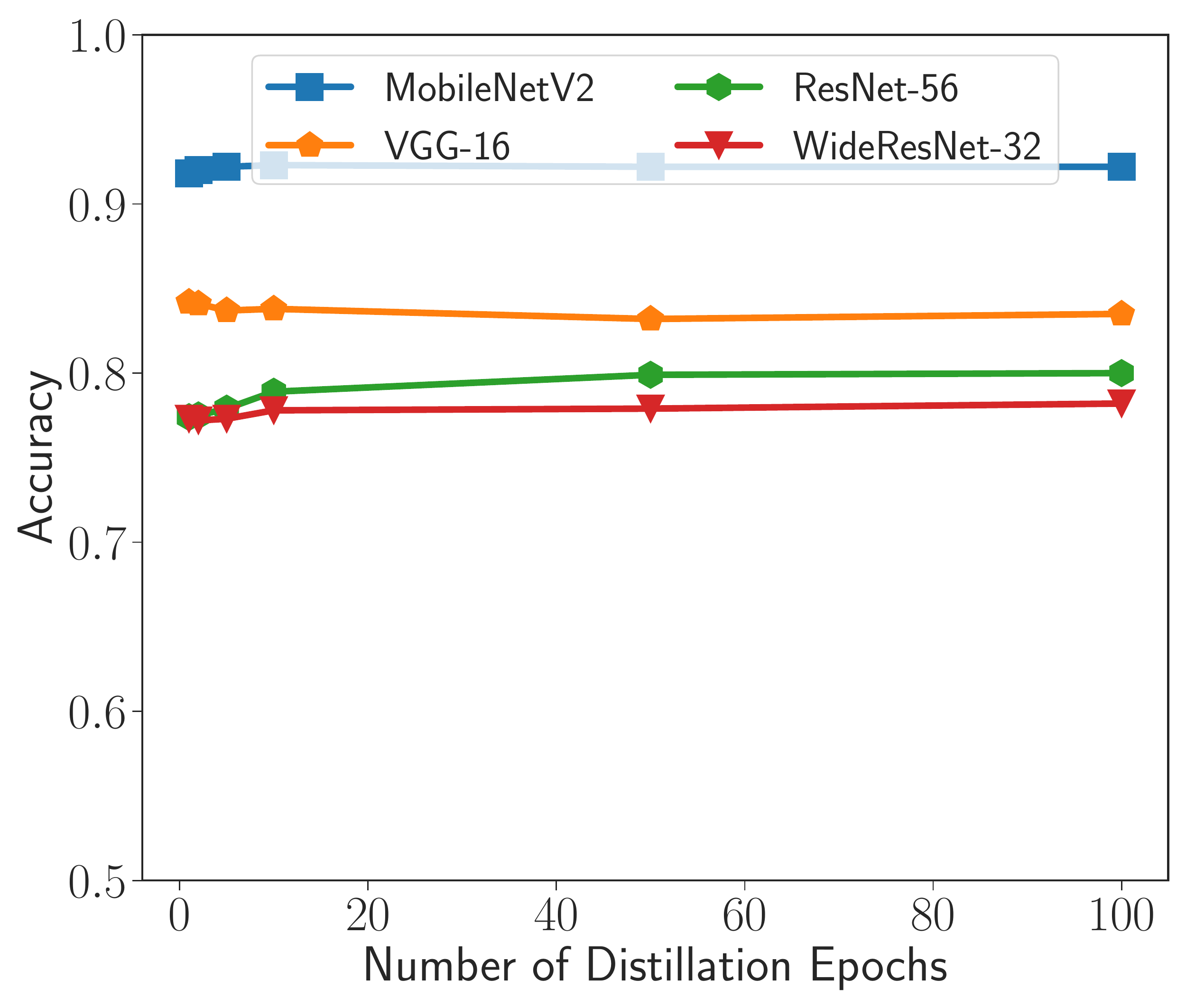}\label{cifar100_esnet_model_stealing_epochs_attack_acc}}
\subfloat[GTSRB]{\includegraphics[width=0.25\linewidth]{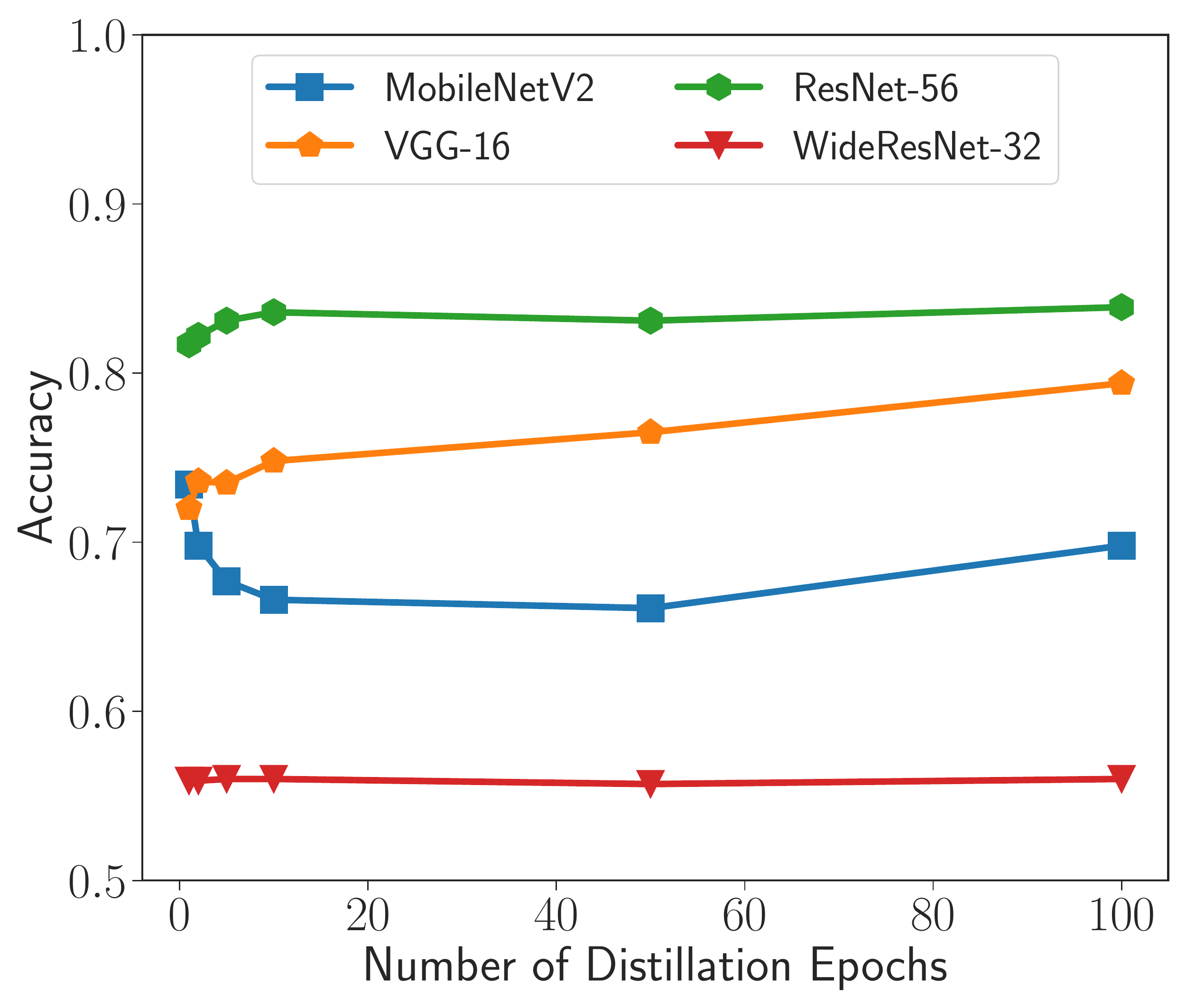}\label{gtsrb_resnet_model_stealing_epochs_attack_acc}}
\caption{The impact of number of knowledge distillation epochs on TPR at 0.1\% FPR (top) and balanced accuracy (bottom).}
\label{fig:resnet_model_stealing_epochs_attack_performance}
\end{figure*}

For knowledge distillation, a generally important factor that influences the distillation performance is the size of the distillation dataset.
Here we explore the impact of this factor on our attack performance.
We follow the same settings in \autoref{sec:result} for training the target model and the distilled models, except that the size of the knowledge distillation dataset $\mathcal{D}^k$ is varied from 20000 to 220000.

\autoref{table:test_acc_with_different_model_stealing_size} and \autoref{fig:knowledge_distillation_set_size_attack_performance} summarize the results. 
As expected, when the adversary is able to acquire a larger set of auxiliary data, the attack performance could be improved both in TPR at 0.1\% FPR and balanced accuracy. 
Differently, the classification accuracy of different distilled models remains very similar. 
This difference indicates that the membership information distilled from the training process does not have a direct connection to the functionality (as represented by the classification accuracy) of the distilled model, which we will discuss more in \autoref{sec:discuss}.

\subsection{Number of Knowledge Distillation Epochs}
\label{sec:epoch}

The number of epochs for knowledge distillation will influence both the computational cost in the distillation process and the input dimension to the attack model. 
Thus, it is crucial to figure out how many epochs are necessary in the distillation process to obtain the \emph{distilled loss trajectory}.

We can see from \autoref{fig:resnet_model_stealing_epochs_attack_performance} that distilling the target model for more epochs indeed increases the attack TPR at low FPR whereas has little impact on the attack accuracy across all different datasets and model architectures.
However, although more distillation epochs can give us a better attack performance, the larger the number of epochs is, the lower the marginal benefits will be, thus we need to find a trade-off between them. 
Take CIFAR-10 and CINIC-10 for example, the number of epochs for the distilled models trained on these two datasets to reach a similar classification accuracy as the target model are around fifty and five respectively, and interestingly, just using the same epochs for distillation can also make the attack performance near the best. 
Another extreme case is the GTSRB dataset where if we continue the distillation for too many epochs after the distilled model already reaches the time point to have comparable functionality as the target model, it even degrades the attack performance. 
Thus, this time point can be used as a strong signal for us to stop the distillation so as to save the computational cost and keep a considerable attack performance at the same time.

\begin{figure}[t]
\centering
\subfloat[TPR at 0.1\% FPR]{\includegraphics[width=0.5\linewidth]{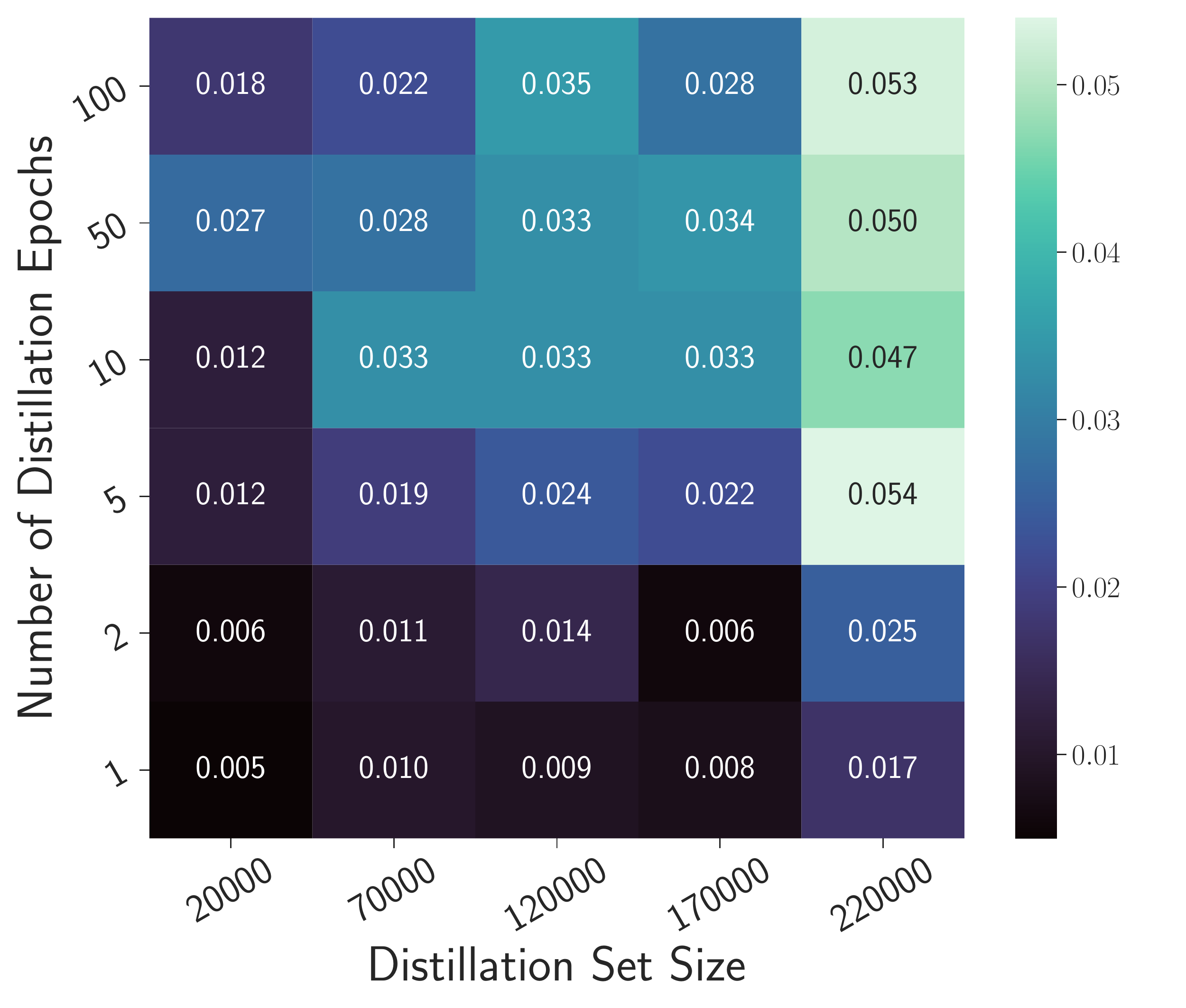}\label{cinic10_resnet_distill_size_distill_epoch_attack_tpr}}
\subfloat[Accuracy]{\includegraphics[width=0.5\linewidth]{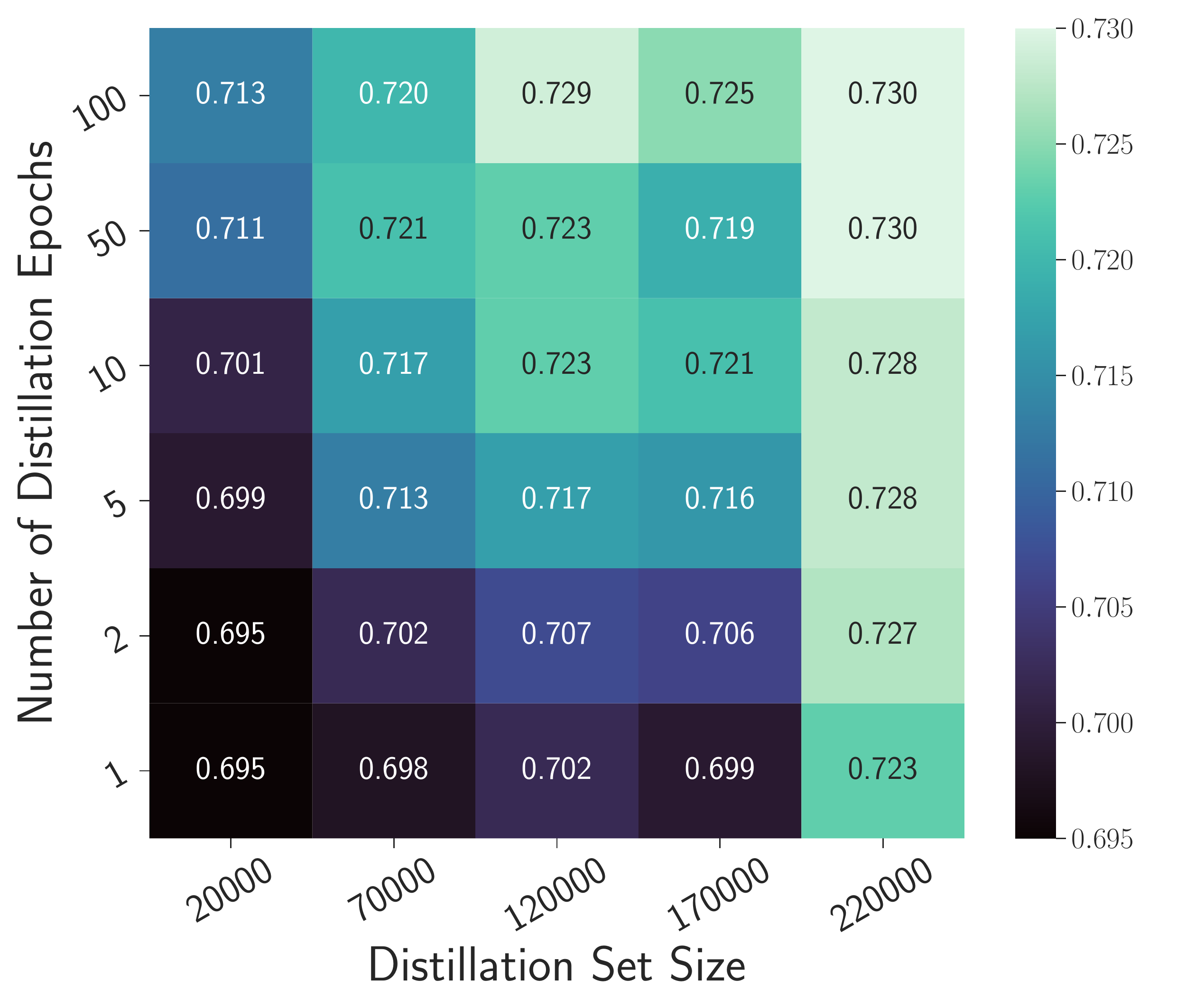}\label{cinic10_resnet_distill_size_distill_epoch_attack_acc}}
\caption{The impact of both knowledge distillation set size and number of distillation epochs on attack success rate for ResNet-56 trained on CINIC-10.}
\label{fig:cinic10_resnet_distill_size_distill_epoch_attack_performance}
\end{figure}

Actually, there is an interrelationship between the size of the knowledge distillation set and the epochs needed in the distillation process, where more distillation data samples mean the model can be distilled quickly. 
Thus, we go a step further to investigate the connection between each other. 
As demonstrated in \autoref{fig:cinic10_resnet_distill_size_distill_epoch_attack_performance}, first, both larger distillation set size ($\mathcal{D}^k$) and more distillation epochs can improve the attack performance. 
More interestingly, regarding the TPR at 0.1\% FPR, using 220000 distillation data samples with only 1 epoch can have comparable performance as using 20000 samples in distillation with 100 epochs, although the number of distillation epochs in the latter setting is 100 times larger than the former one. 
This indicates that the impact of distillation dataset size seems more significant. 
This is reasonable as a larger size of the distillation set can help the distilled model to be more representative and the \emph{distilled loss trajectory} can obtain more information from the original one while more epochs contribute a little especially after the distilled model already has a similar classification accuracy as the original model.

\begin{figure}[t]
\centering
\includegraphics[width=0.6\linewidth]{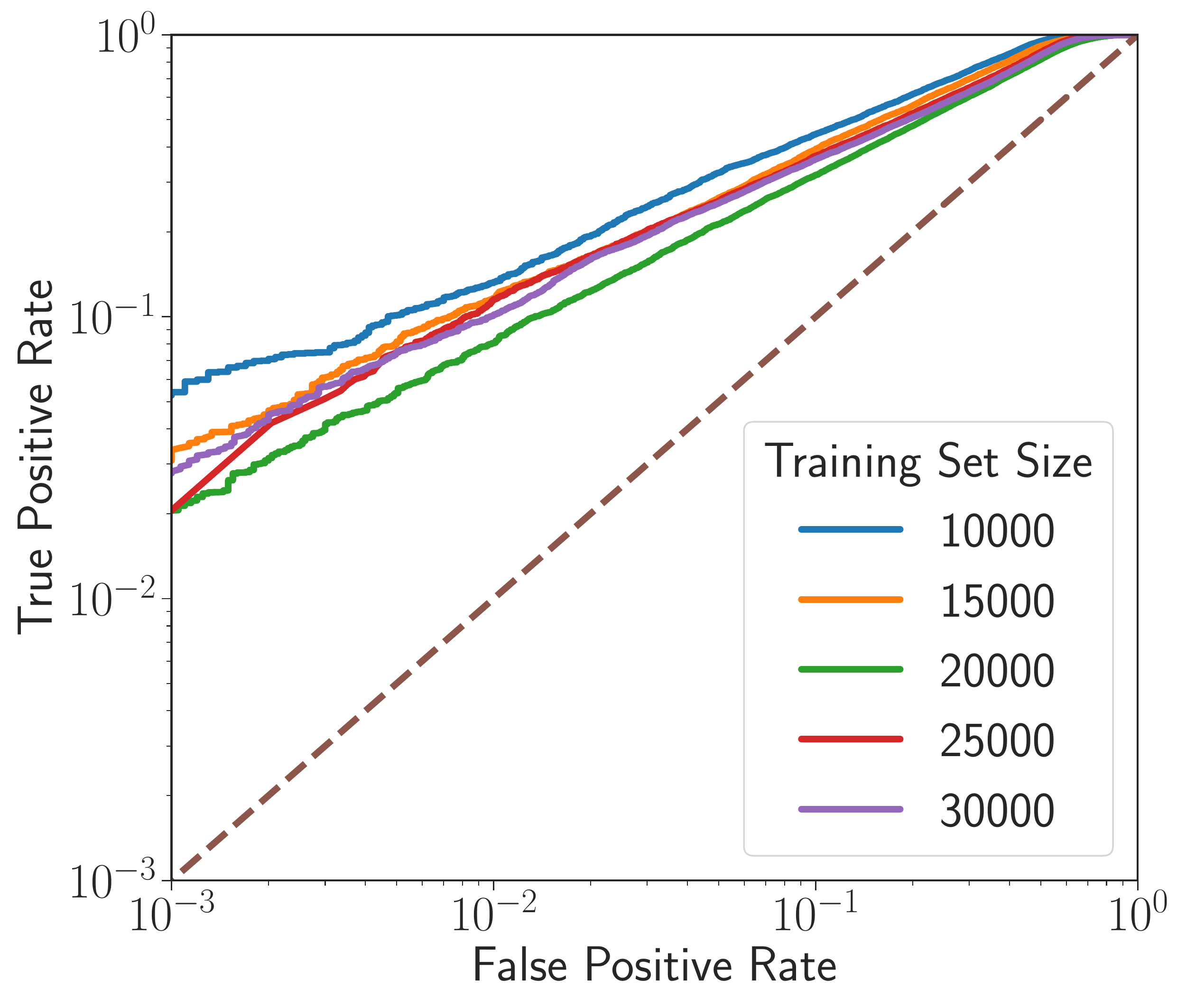}
\caption{ROC curves for ResNet-56 trained on CINIC-10 with different training set size.}
\label{fig:training_set_size_attack_performance}
\end{figure}

\subsection{Overfitting Level of the Target Model}

There is a general consensus that the success of MIAs is relevant to the overfitting level of the target model~\cite{SZHBFB19,SSSS17}. 
Here we represent the overfitting level by the training and testing accuracy gap and control it by varying the training set size.
Specifically, we vary the size of $\mathcal{D}^t_{train}$, $\mathcal{D}^t_{test}$, $\mathcal{D}^s_{train}$ and $\mathcal{D}^s_{test}$ from 10000 to 30000, while fixing the size of the knowledge distillation dataset $\mathcal{D}^k$ as 150000.

As can be seen from \autoref{table:training_size}, as expected, a larger training set size incurs a lower training testing accuracy gap, indicating a lower overfitting level.
\autoref{fig:training_set_size_attack_performance} evaluates the corresponding attack performance, and we can observe that the overfitting problem does make the target model more vulnerable to our attack. 
However, even when the size of the training set is increased to 30000 and the model is well-generalized, our attack can still secure a good attack performance (2.8\% in terms of TPR at 0.1\% FPR), which remains much better than other attack baselines that are mounted on more overfitted target models with 10000 training samples (cf. \autoref{table:attack_performance_on_resnet}).

\begin{table}[!t]
\definecolor{mygray}{gray}{0.9}
\newcolumntype{a}{>{\columncolor{mygray}}c}
\centering
\caption{The impact of the overfitting level of target model for ResNet-56 trained on CINIC-10.}
\label{table:training_size}
\setlength{\tabcolsep}{4.0pt}
\scalebox{0.75}{
\begin{tabular}{l|c|cc}
\toprule
\rowcolor{white}
Training& Training testing& \multicolumn{2}{c}{Attack}\\
\cmidrule{3-4}
set size& acc gap& Acc& TPR at 0.1\% FPR\\
\midrule
10000&  0.339&0.730& 5.3\%\\
15000&  0.300&0.708& 3.2\%\\
20000&  0.249&0.659& 2.3\%\\
25000&  0.234&0.685& 2.1\%\\
30000&  0.214&0.674& 2.8\%\\
\bottomrule
\end{tabular}
}
\end{table}

\begin{figure}[t]
\centering
\subfloat[TPR at 0.1\% FPR]{\includegraphics[width=0.5\linewidth]{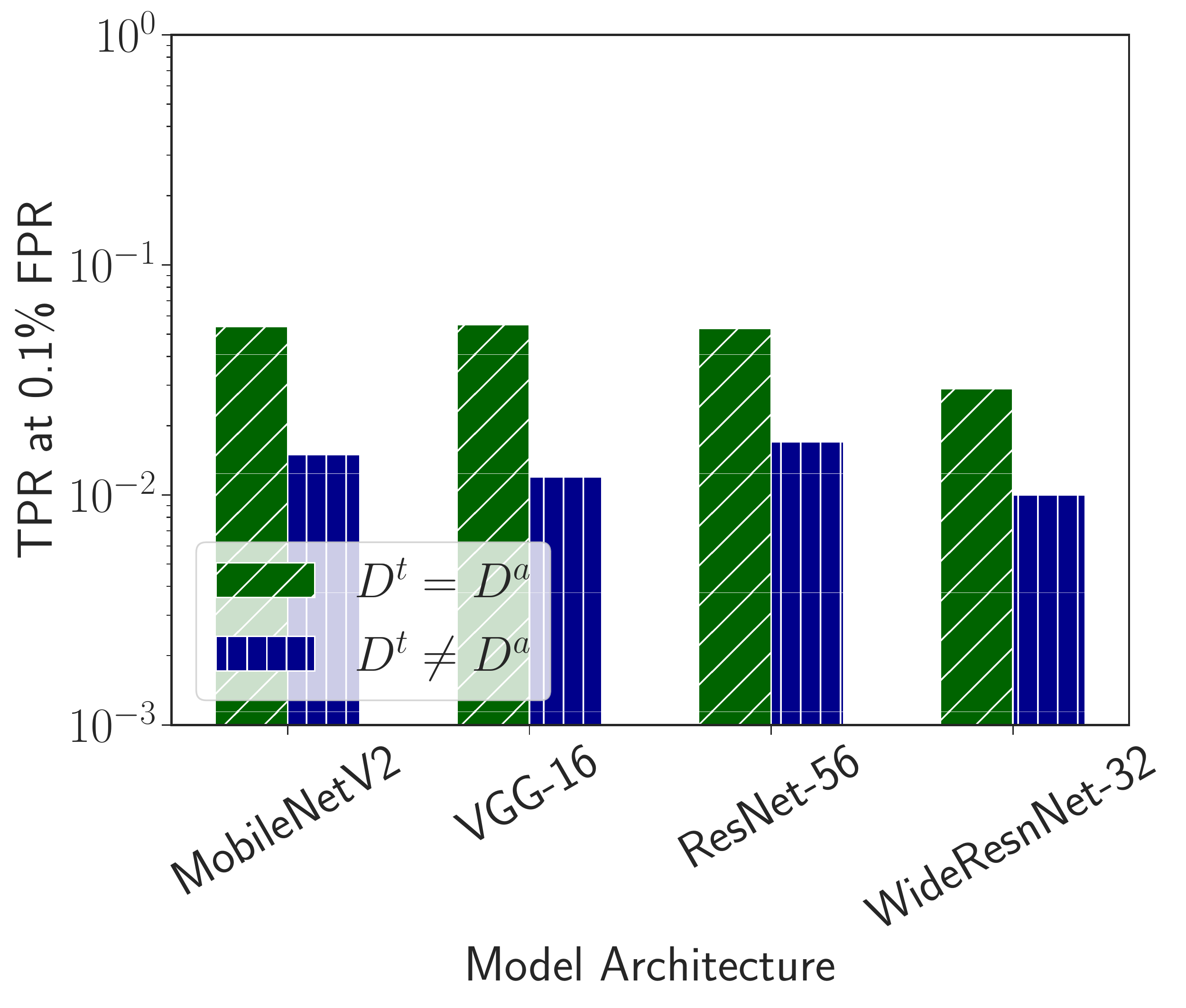}\label{cinic10_resnet_disjoint_data_attack_auc}}
\subfloat[Accuracy]{\includegraphics[width=0.5\linewidth]{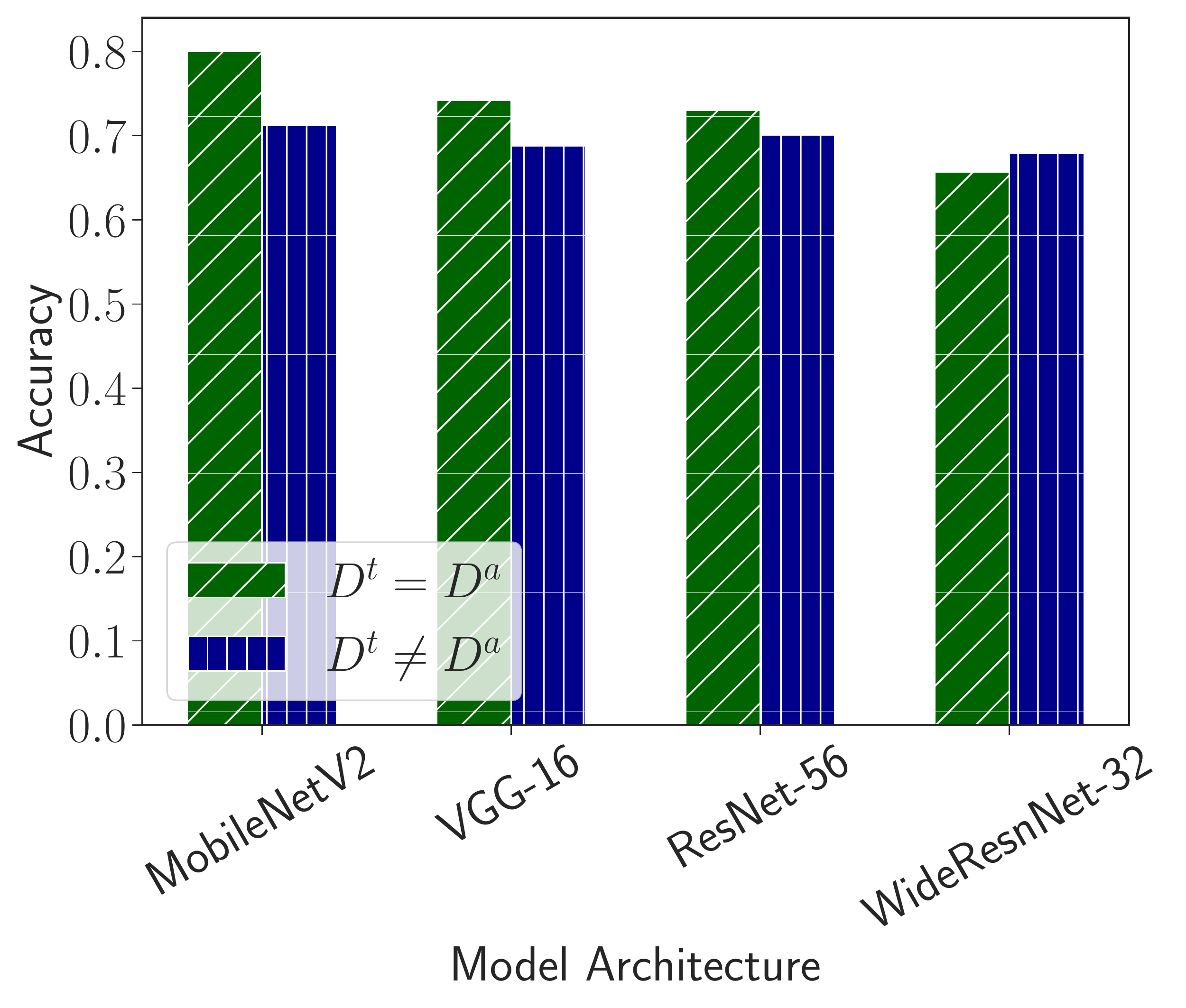}\label{cinic10_resnet_disjoint_data_attack_acc}}
\caption{The impact of data distribution shift between the target model training set and auxiliary dataset for different models trained on CINIC-10.}
\label{fig:cinic10_resnet_disjoint_data_attack_performance}
\end{figure}

\subsection{Disjoint Datasets}

In this section, we will relax the assumption that the adversary has the auxiliary dataset from the same distribution as the training dataset of the target model.
We compare two distribution settings:
\begin{itemize}
\item $\mathbb{D}^t = \mathbb{D}^a$, which means the dataset for training the target model, shadow model, and distilled models (target and shadow) are from the same distribution, i.e., the CINIC-10 dataset.
\item $\mathbb{D}^t\neq\mathbb{D}^a$, which means the dataset for training the target model dataset and the auxiliary dataset are from different distributions. 
Specifically, the target model is trained on the CIFAR-10 portion of CINIC-10 but the adversary only accesses the ImageNet portion as the auxiliary dataset. 
\end{itemize}

\autoref{fig:cinic10_resnet_disjoint_data_attack_performance} shows that a distribution shift between the training dataset of the target model and the auxiliary dataset will indeed decrease the attack performance in most cases. This can be explained from two aspects.
On the one hand, the shadow dataset is different from the target dataset, which may lead to different functionality of the shadow model and target model.
On the other hand, the knowledge distillation dataset is also different from the target dataset, which may cause the distilled loss trajectory to behave more different from the actual one. As a result, less membership information can be extracted to train the attack model.

\subsection{Model Architecture}

\begin{figure}[t]
\centering
\subfloat[TPR at 0.1\% FPR]{\includegraphics[width=0.5\linewidth]{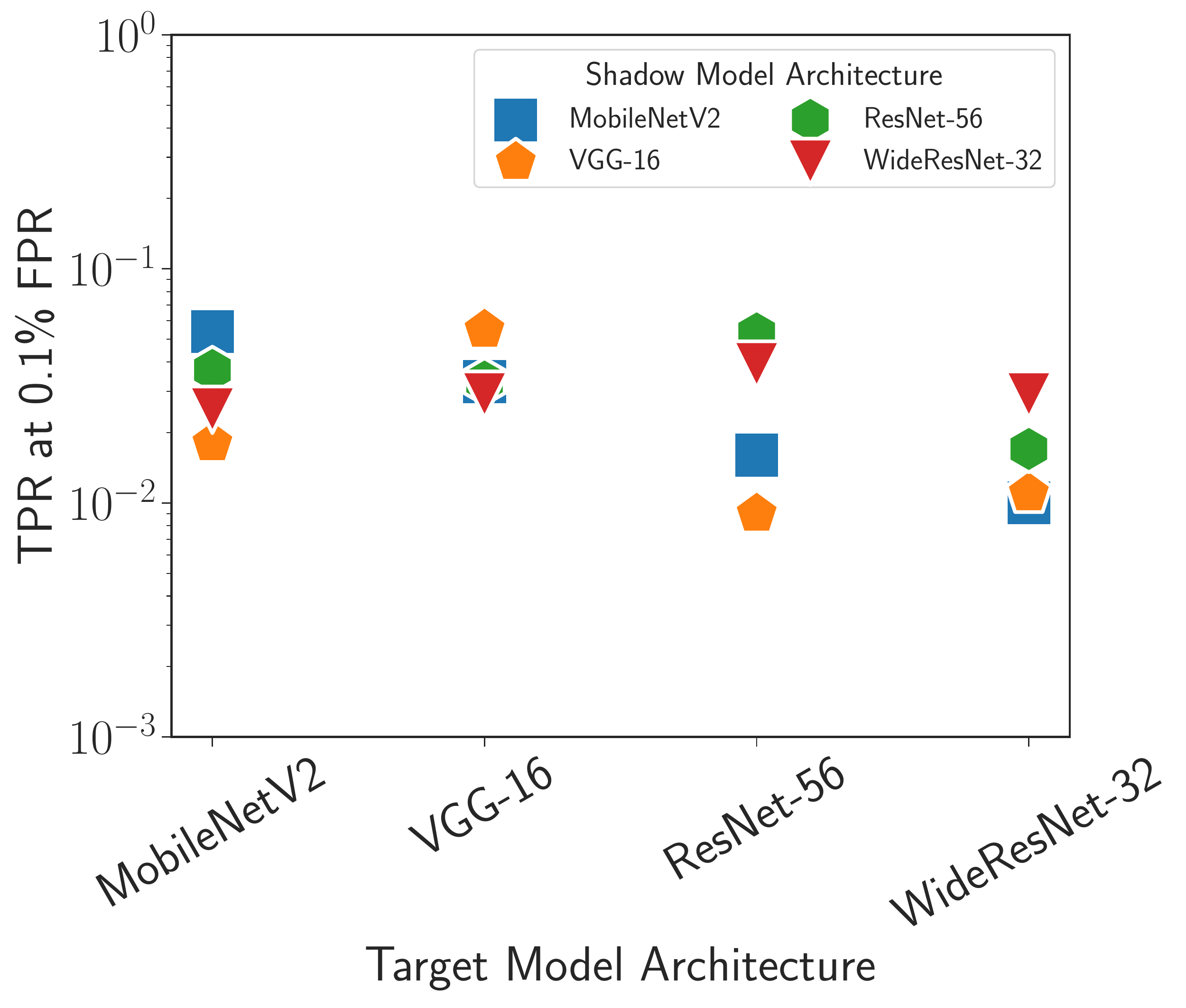}\label{cinic10_different_architectures_attack_auc}}
\subfloat[Accuracy]{\includegraphics[width=0.5\linewidth]{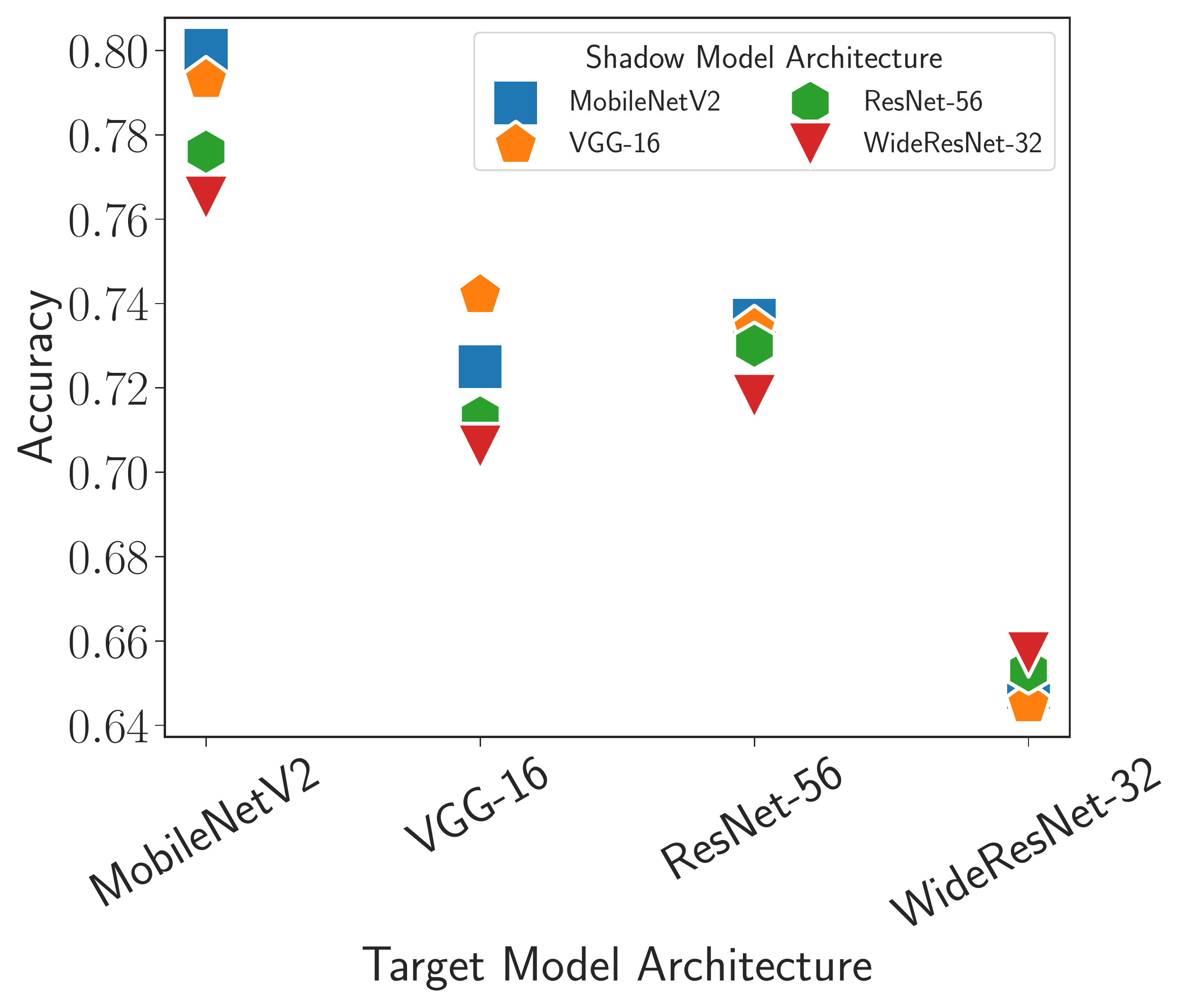}\label{cinic10_different_architectures_attack_acc}}
\caption{The impact of the architecture differences between the target model and local (shadow and distilled) models trained on CINIC-10.}
\label{fig:cinic10_different_architectures_attack_performance}
\end{figure}

After validating the impact of dataset distribution shift, here we focus on another assumption on the knowledge of the adversary about the architecture of the target model.
We vary the architectures of the target model, shadow model, and distilled models while keeping the architectures of the shadow model and distilled models the same since both of them are under the full control of the adversary locally. As can be seen from \autoref{fig:cinic10_different_architectures_attack_performance}, our attack performs the best when the shadow model and distilled models have the same architecture as the target model. 
In addition, using models from the kindred family of model architectures (e.g., ResNet-56 and WideResNet-32) will lead to a similar attack performance compared to using exactly the same architecture. 
When the architectures are totally different, although the attack performance will decrease, the result is still better than those achieved by other baselines using the same architecture for the target model and shadow model (or reference model), as reported in \autoref{sec:Eval}.
One possible reason could be that although the target model uses a different architecture, the same architecture of the distilled target and shadow models still achieve a close enough loss trajectory.
Overall, the performance of our attack in the above harder settings with relaxed assumptions remains better than that of other attacks achieved in easier settings (as shown in \autoref{table:attack_performance_on_resnet}).
 
\section{Discussion}
\label{sec:discuss}

In this section, we discuss in more detail the characteristics of our attack.
We first analyze the different roles of the distilled loss trajectory and the loss from the target model in our attack.
We then provide evidence that it is indeed better to use the loss trajectory from the distilled shadow models rather than directly using that from the actual training process of the shadow model.

\begin{figure}[t]
\centering
\includegraphics[width=0.6\linewidth]{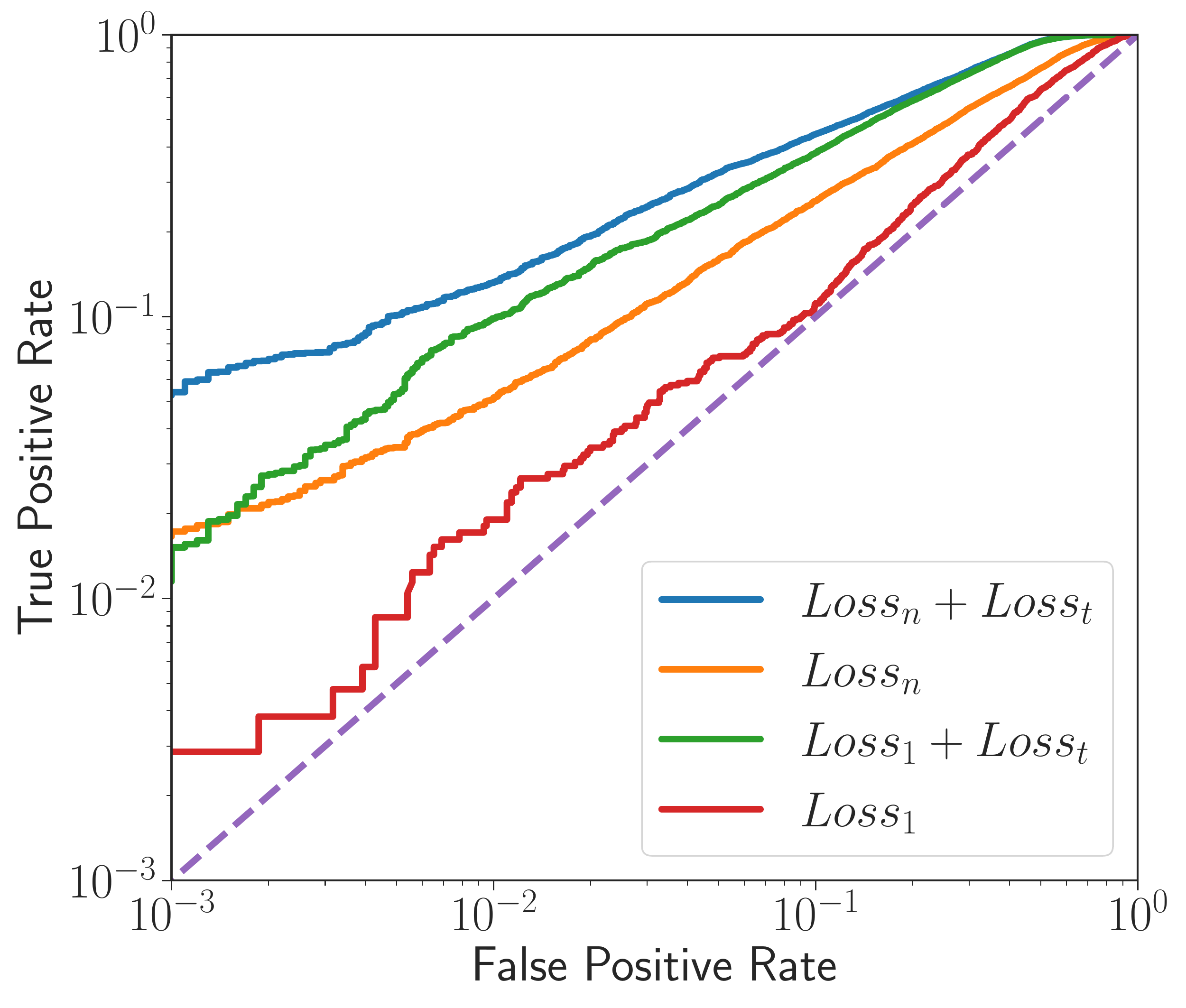}
\caption{The attack performance showing the effect of each component in our attack, the name of the four methods are the same as in \autoref{table:effect_of_loss_trajectory_and_loss_from_target_model}.
}
\label{fig:effect_of_loss_and_model_stealing}
\end{figure}

\mypara{Distilled Loss Trajectory.}
Here we show the impact of the distilled loss trajectory to the attack performance. Comparing the result of $Loss_n$ and $Loss_1$ in \autoref{table:effect_of_loss_trajectory_and_loss_from_target_model}, using the distilled loss trajectory can achieve more than 10$\times$ TPR at 0.1\% FPR than solely using the loss from the last distilled model. 
In addition, our original attack can also have a considerable improvement over the improved variant of the attack in~\cite{YMMS21}, where we concatenate the loss from the original model and the last distilled model together.
A more comprehensive picture of this observation is shown by the ROC curves in \autoref{fig:effect_of_loss_and_model_stealing}, where our attack achieves consistently best TPR across the whole range of FPR.

\mypara{Loss from Original Models.}
Apart from the \emph{distilled loss trajectory}, our attack also concatenates the loss from the original models (the target model and the trained shadow model).
We argue that it is due to the difference between the functionality distillation and membership information distillation.
More concretely, although the classification accuracy of the last distilled model can be similar to or even better than the original model as illustrated in \autoref{table:test_acc_with_different_model_stealing_size}, it omits some information about membership, thus there is still membership information that can be extracted from the loss of the original models.
As can be seen from \autoref{table:effect_of_loss_trajectory_and_loss_from_target_model}, when concatenating the loss from the original model, our attack is substantially improved, e.g., TPR at 0.1\% FPR increases from 1.7\% to 5.3\%.
Similarly, if we leave out the loss from the original model for the last distilled loss used in~\cite{YMMS21}, the attack is substantially degraded, e.g., TPR at 0.1\% FPR decreases from 1.2\% to 0.1\%.

\begin{table}[!t]
\definecolor{mygray}{gray}{1}
\newcolumntype{a}{>{\columncolor{mygray}}c}
\centering
\caption{The impact of incorporating the loss from the target model ($Loss_t$) into the loss trajectory ($Loss_{n}$ for using losses from all distilled models, and $Loss_{1}$ for using only the last one).
The model is ResNet-56 trained on CINIC-10.}
\label{table:effect_of_loss_trajectory_and_loss_from_target_model}
\setlength{\tabcolsep}{2.5pt}
\scalebox{0.75}{
\begin{tabular}{l|ccca}
\toprule
\rowcolor{white}
Metric& $Loss_{1}$~\cite{YMMS21}& $Loss_{1}$+$Loss_t$& $Loss_{n}$& $Loss_{n}$+$Loss_t$ (ours)\\
\midrule
TPR at 0.1\% FPR&  {0.1\%}& 1.2\%&  {1.7\%}& \textbf{5.3\%}\\
Accuracy &  {0.603}& 0.726&  {0.633}& \textbf{0.730}\\
\bottomrule
\end{tabular}
}
\end{table}

\begin{figure}[t]
\centering
\includegraphics[width=0.6\linewidth]{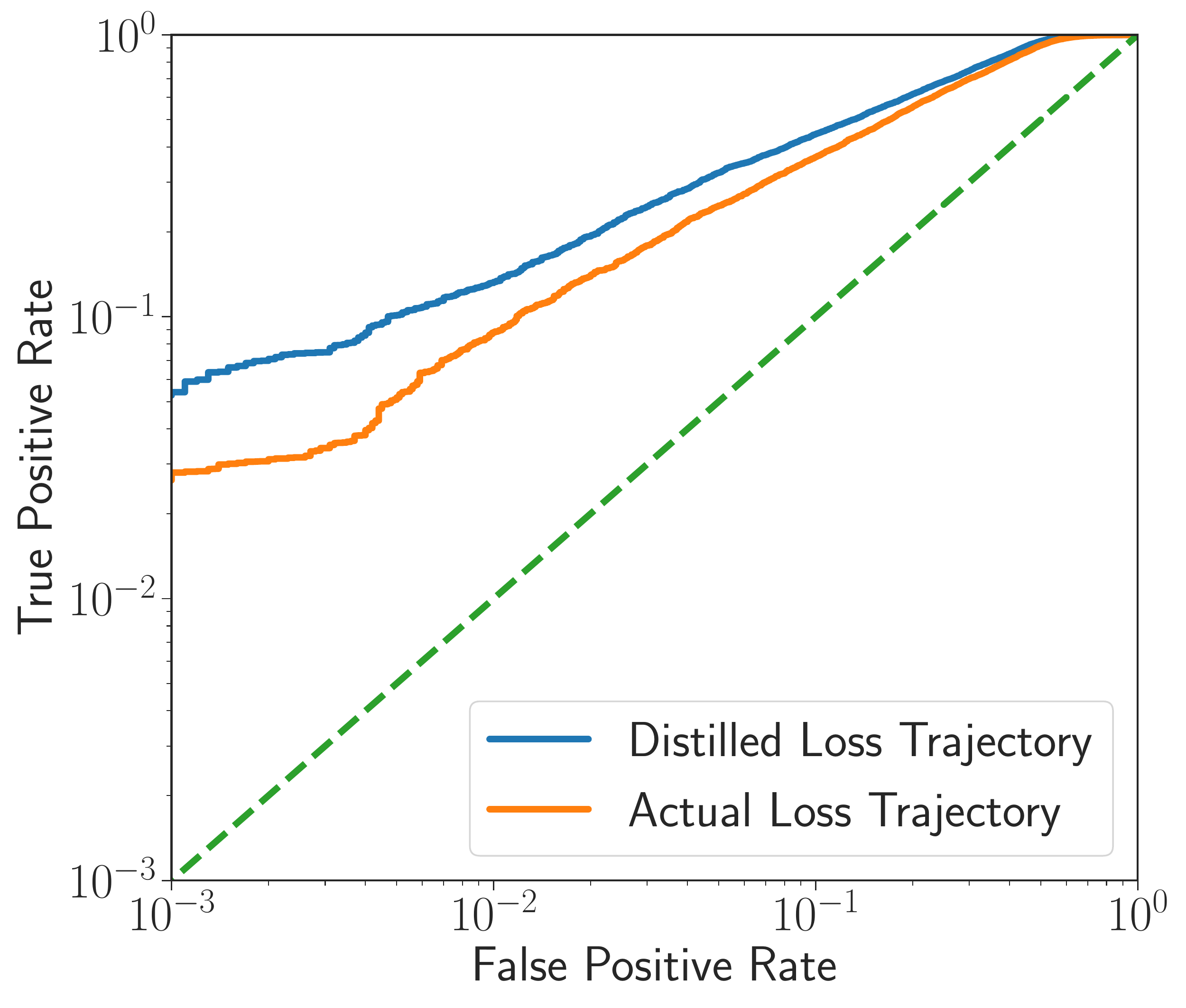}
\caption{ROC curves of our attack with distilled loss trajectory vs. actual loss trajectory for the shadow model.}
\label{fig:result_of_actual_loss_trajectory_shadow_model_attack}
\end{figure}

\mypara{Distilled Shadow Models.}
As mentioned before, we are able to obtain the actual loss trajectory from the shadow model as it is trained locally, but we still use the distilled one. 
\autoref{table:result_of_actual_loss_trajectory_shadow_model_attack} and \autoref{fig:result_of_actual_loss_trajectory_shadow_model_attack} support this specific design by showing that using the distilled loss trajectory can indeed lead to better attack performance than directly using the actual loss trajectory. 
For example, the distilled loss trajectory yields about 2$\times$ higher TPR at 0.1\% FPR than the actual loss trajectory. 

This observation is understandable because directly using the actual loss trajectory makes the loss trajectory not well aligned with the distilled loss trajectory from the target model. 
This worse alignment is due to the fact that the \emph{distilled loss trajectory} could be similar to the actual one yet not the same.
For instance, when the distillation dataset is larger than the shadow dataset, the distilled model needs fewer epochs to reach a similar classification accuracy as the original shadow model. 
Thus if we use the actual loss trajectory to train the attack model, the different patterns in the \emph{distilled loss trajectory} from the target model will incur a dissatisfied membership prediction.

\section{Related Work}
\label{sec:related}

\subsection{Membership Inference Attacks}
\label{sec:mia}

Currently, membership inference attack (MIA) has gaining attentions for quantifying the privacy risks of machine learning models~\cite{LLHYBZ222,HLXCZ22,SSSS17,YGFJ18,HMDC19,SZHBFB19,NSH19,SM21,LF20,HWWBSZ21}. 
Shokri et al.~\cite{SSSS17} propose the first membership inference attack in black-box settings. 
They train multiple shadow models to mimic the behavior of the target model and use the posteriors obtained from these shadow models to train multiple attack models. 
Salem et al.~\cite{SZHBFB19} later relax the assumptions made in~\cite{SSSS17}, and only train one shadow model without the knowledge of architecture and data distribution used in target model. 
Song et al.~\cite{SM21} introduce a metric-based attack without training any attack model. Similarly, 
Yeom et al.~\cite{YGFJ18} assume that the adversary knows the target model's training dataset distribution and size, and conducts membership inference by relying on the samples' loss. 

\begin{table}[!t]
\definecolor{mygray}{gray}{1}
\newcolumntype{a}{>{\columncolor{mygray}}c}
\centering
\caption{Attack performance of our method with distilled loss trajectory vs. actual loss trajectory for the shadow model.}
\label{table:result_of_actual_loss_trajectory_shadow_model_attack}
\setlength{\tabcolsep}{3.0pt}
\scalebox{0.75}{
\begin{tabular}{l|ca}
\toprule
\rowcolor{white}
Metric& Actual & Distilled \\
\midrule
TPR at 0.1\% FPR& 2.8\%& \textbf{5.3\%}\\
Acc& 0.693& \textbf{0.730}\\
AUC& 0.794& \textbf{0.819}\\
\bottomrule
\end{tabular}
}
\end{table}

Recently, more studies begin to focus on addressing one major problem shared by most MIAs, which is the high false-positive rate. 
Ye et al.~\cite{YMMS21} design a model-dependent and sample-dependent MIA via knowledge distillation, which trains multiple distilled models to approximate the samples from the retrained model distribution. 
Sablayrolles et al.~\cite{SDSOJ19} introduce a calibration term, computed by the loss from both the models trained with and without the target sample, in order to calibrate the loss from the target model. 
Similarly, Watson et al.~~\cite{WGCS21} view the loss from reference models (trained without the target sample) as the sample's hardness, and a smaller difference between the loss from the target model and the sample's hardness implies that the target sample is more likely to be a non-member. 
Carlini et al.~\cite{CCNSTT21} go a step further from~\cite{SDSOJ19} to develop a Likelihood Ratio Attack. 
By taking advantage of the logit scaling, Gaussian likelihood, and multiple queries, this attack can achieve high TPR at low FPR.

\subsection{Knowledge Distillation}

The notion of transferring knowledge from larger models (teacher models) to smaller ones (student models) emerges quite early. 
After forming a framework under the name of knowledge distillation (KD), this line of work is extended by either finding new approaches for KD or applying KD in different domains. 
For the former direction, Romero et al.~\cite{RBKCGB15} use additional linear projection to train a thinner and deeper student model. 
Zagoruykoet et al.~\cite{ZK17} adopt attention mechanism to advance the performance of knowledge distillation. 
Yim et al.~\cite{YJBK17} propose a new method that adds additional losses to enhance the performance but also speeds up the optimization process. 
Xu et al.~\cite{XHH18} utilize conditional adversarial networks to learn a loss function for KD. 
Regarding the applications of knowledge distillation, many studies have explored KD in other domains, such as semi-supervised learning~\cite{TV17}, sequence modeling~\cite{KR16} and multi-modal learning~\cite{GHM16}.

Our work is more related to a specific direction of knowledge distillation, that is self-distillation~\cite{FLTIA18}. 
In this direction, the teacher model and student model have identical model architectures, and the distillation is used to improve the performance of the student over the teacher. 
However, there is still a core difference between this direction and our work, that is we adopt the knowledge distillation to extract the membership information represented by the loss trajectory, but care less about the general performance of the model.

\section{Conclusion}
\label{sec:con}

In this paper, we take the first step to exploit the information from the training process of the target model to conduct membership inference. 
We demonstrate that the \emph{loss trajectory}, i.e., losses of the sample evaluated on the target model at its different training epochs, can be used to represent such membership information. 
Specifically, we propose a new attack method, call \system, that leverages knowledge distillation to extract the loss trajectory information of the target model with only black-box access.
Our extensive experiments demonstrate the state-of-the-art performance of \system, especially on a practically meaningful metric that measures the true-positive rate at a low false-positive rate. 
We also show the general advantage of our attack over existing methods for separate groups of target samples that have different loss values. 
Additional analyses provide evidence on the importance of each component of our \system for its attack success. 
For future work, one promising direction could be exploring more fine-grained modeling of loss trajectories beyond simply using the whole trajectory as the input feature.

\section*{Acknowledgements}
This work is partially funded by the Helmholtz Association within the project ``Trustworthy Federated Data Analytics'' (TFDA) (funding number ZT-I-OO1 4).

\begin{small}
\bibliographystyle{plain}
\bibliography{normal_generated_py3}
\end{small}

\clearpage
\newpage
\appendix

\section{Appendix}

\subsection{Data Splits on Different Datasets}
\label{app:split}

\begin{table}[htbp]
\begin{minipage}{\textwidth}
\captionof{table}{Data splits on different datasets in our evaluation.}
\label{table:data_split_for_score_based}
\centering
\scalebox{0.75}{
\begin{tabular}{l|ccccc}
\toprule
Dataset& $\mathcal{D}^t_{train}$&  $\mathcal{D}^t_{test}$& $\mathcal{D}^s_{train}$& $\mathcal{D}^s_{test}$& $\mathcal{D}^k$\\
\midrule
CIFAR-10&  {10000}& 10000&  {10000}& 10000& 20000\\
CINIC-10&  {10000}& 10000&  {10000}& 10000& 220000\\
CIFAR-100&  {10000}& 10000&  {10000}& 10000& 20000\\
GTSRB&  {1500}& 1500&  {1500}& 1500& 45839\\
Purchase& {20000}& 20000&  {20000}& 20000& 110000\\
Location& {800}& 800&  {800}& 800& 1400\\
News& {3000}& 3000&  {3000}& 3000& 6000\\
\bottomrule
\end{tabular}
}
\end{minipage}

\subsection{Additional Results on Different Models}
\label{app:model}

\begin{minipage}{1.0\textwidth}
\vspace{5mm}
\captionof{table}{Attack performance of different attacks on four datasets with MobileNetV2 model.}
\centering
\setlength{\tabcolsep}{2.5pt}
\scalebox{0.75}{
\begin{tabular}{l|cccccccccccc}
\toprule
\rowcolor{white}
Attack & \multicolumn{4}{c}{TPR at 0.1\% FPR}&  \multicolumn{4}{c}{Balanced accuracy}& \multicolumn{4}{c}{AUC}\\
\cmidrule(l{5pt}r{5pt}){2-5}\cmidrule(l{5pt}r{5pt}){6-9}\cmidrule(l{5pt}r{0pt}){10-13}
method& CIFAR-10& CINIC-10& CIFAR-100& GTSRB&CIFAR-10& CINIC-10& CIFAR-100& GTSRB&CIFAR-10& CINIC-10& CIFAR-100& GTSRB\\
\midrule
Salem et al.~\cite{SZHBFB19}&0.2\%&0.3\%&0.4\%&0.1\%&0.652&0.696&0.709&0.616&0.694&0.757&0.839&0.616\\
Yeom et al.~\cite{YGFJ18}&0.1\%&0.2\%&0.2\%&0.2\%&0.686&0.778&0.919&0.729&0.726&0.826&0.942&0.786\\
Song et al.~\cite{SM21}&0.0\%&0.1\%&0.1\%&0.1\%&0.700&0.788&0.916&\textbf{0.772}&0.721&0.826&0.941&0.784\\
Ye et al.~\cite{YMMS21}&0.1\%&0.3\%&0.4\%&0.0\%&0.545&0.665&0.680&0.624&0.556&0.706&0.728&0.610\\
Watson et al.~\cite{WGCS21}&0.7\%&0.4\%&2.2\%&0.2\%&0.679&0.758&0.787&0.719&0.730&0.797&0.845&\textbf{0.812}\\
\midrule
Ours&\textbf{3.5\%}&\textbf{5.4\%}&\textbf{18.1\%}&\textbf{0.6\%}&\textbf{0.701}&\textbf{0.800}&\textbf{0.922}&0.698&\textbf{0.779}&\textbf{0.882}&\textbf{0.976}&0.765\\
\bottomrule
\end{tabular}
}
\end{minipage}

\vspace{5mm}
\begin{minipage}{1\textwidth}
\captionof{table}{Attack performance of different attacks on four datasets with VGG-16 model.}
\centering
\setlength{\tabcolsep}{2.5pt}
\scalebox{0.75}{
\begin{tabular}{l|cccccccccccc}
\toprule
\rowcolor{white}
Attack & \multicolumn{4}{c}{TPR at 0.1\% FPR}&  \multicolumn{4}{c}{Balanced accuracy}& \multicolumn{4}{c}{AUC}\\
\cmidrule(l{5pt}r{5pt}){2-5}\cmidrule(l{5pt}r{5pt}){6-9}\cmidrule(l{5pt}r{0pt}){10-13}
method& CIFAR-10& CINIC-10& CIFAR-100& GTSRB&CIFAR-10& CINIC-10& CIFAR-100& GTSRB&CIFAR-10& CINIC-10& CIFAR-100& GTSRB\\
\midrule
Salem et al.~\cite{SZHBFB19}&0.2\%&0.0\%&0.6\%&0.1\%&0.626&0.682&0.597&0.643&0.664&0.745&0.757&0.704\\
Yeom et al.~\cite{YGFJ18}&0.1\%&0.0\%&0.1\%&0.0\%&0.615&0.747&0.831&0.693&0.658&0.767&0.872&0.734\\
Song et al.~\cite{SM21}&0.0\%&0.0\%&0.2\%&0.0\%&\textbf{0.659}&\textbf{0.764}&\textbf{0.851}&0.696&0.657&0.766&0.874&0.738\\
Ye et al.~\cite{YMMS21}&0.1\%&0.2\%&0.1\%&0.0\%&0.538&0.649&0.601&0.636&0.554&0.697&0.629&0.633\\
Watson et al.~\cite{WGCS21}&0.8\%&0.5\%&1.1\%&0.0\%&0.613&0.729&0.761&0.694&\textbf{0.690}&0.772&0.815&0.735\\
\midrule
Ours&\textbf{1.5\%}&\textbf{5.5\%}&\textbf{7.3\%}&\textbf{0.7\%}&0.607&0.742&0.835&\textbf{0.794}&0.676&\textbf{0.821}&\textbf{0.911}&\textbf{0.865}\\
\bottomrule
\end{tabular}
}
\end{minipage}

\vspace{5mm}
\begin{minipage}{1\textwidth}
\captionof{table}{Attack performance of different attacks on four datasets with WideResNet-32 model.}
\centering
\setlength{\tabcolsep}{2.5pt}
\scalebox{0.75}{
\begin{tabular}{l|cccccccccccc}
\toprule
\rowcolor{white}
Attack & \multicolumn{4}{c}{TPR at 0.1\% FPR}&  \multicolumn{4}{c}{Balanced accuracy}& \multicolumn{4}{c}{AUC}\\
\cmidrule(l{5pt}r{5pt}){2-5}\cmidrule(l{5pt}r{5pt}){6-9}\cmidrule(l{5pt}r{0pt}){10-13}
method& CIFAR-10& CINIC-10& CIFAR-100& GTSRB&CIFAR-10& CINIC-10& CIFAR-100& GTSRB&CIFAR-10& CINIC-10& CIFAR-100& GTSRB\\
\midrule
Salem et al.~\cite{SZHBFB19}&0.0\%&0.0\%&0.2\%&0.0\%&0.553&0.541&0.555&0.510&0.570&0.562&0.670&0.508\\
Yeom et al.~\cite{YGFJ18}&0.0\%&0.0\%&0.2\%&0.0\%&0.590&0.620&0.766&0.543&0.605&0.654&0.813&0.558\\
Song et al.~\cite{SM21}&0.1\%&0.1\%&0.1\%&0.0\%&0.600&0.633&0.759&0.548&0.600&0.650&0.812&0.559\\
Ye et al.~\cite{YMMS21}&0.0\%&0.1\%&0.1\%&0.0\%&0.536&0.580&0.589&0.537&0.533&0.597&0.612&0.551\\
Watson et al.~\cite{WGCS21}&0.2\%&0.5\%&1.3\%&0.0\%&0.584&0.625&0.736&0.543&0.622&0.666&0.792&0.558\\
\midrule
Ours&\textbf{2.0\%}&\textbf{2.9\%}&\textbf{6.2\%}&\textbf{0.2\%}&\textbf{0.603}&\textbf{0.657}&\textbf{0.782}&\textbf{0.560}&\textbf{0.650}&\textbf{0.723}&\textbf{0.870}&\textbf{0.579}\\
\bottomrule
\end{tabular}
}
\end{minipage}
\end{table}

\begin{figure*}[t]
\centering
\subfloat[CIFAR-10]{\includegraphics[width=0.250\linewidth]{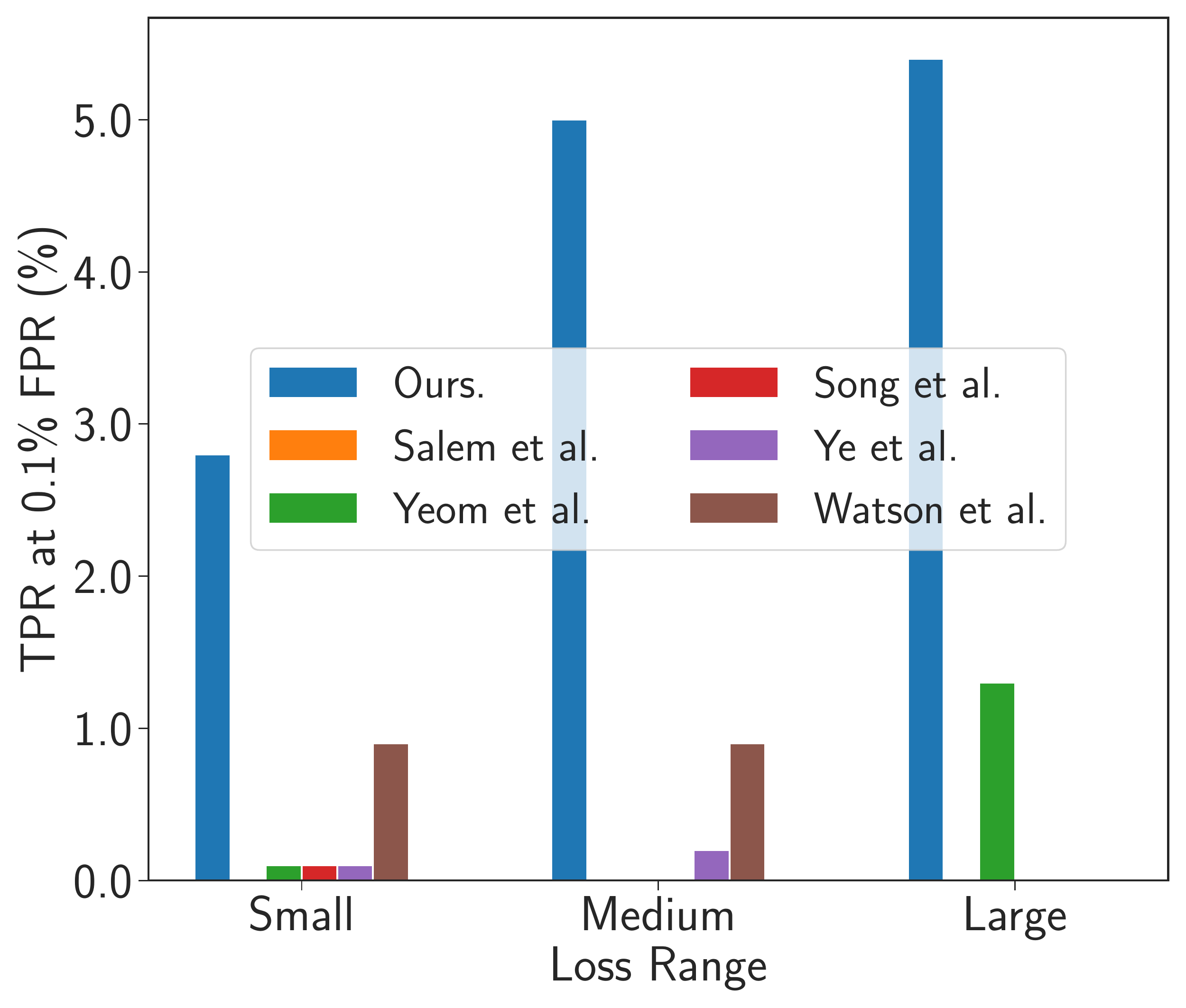}\label{cifar10_tpr_at_each_range_mobilenet}}
\subfloat[CINIC-10]{\includegraphics[width=0.250\linewidth]{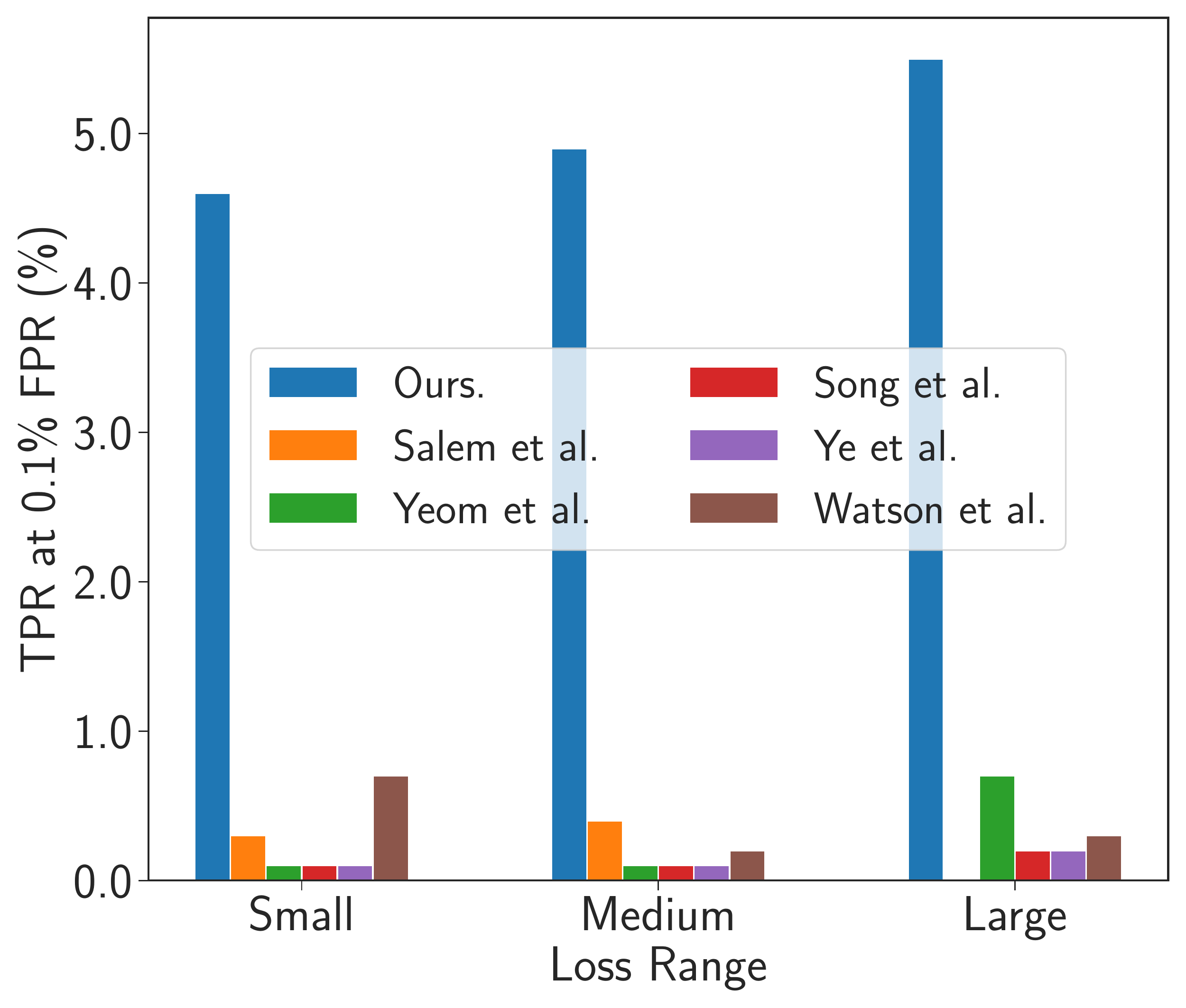}\label{cinic10_tpr_at_each_range_mobilenet}}
\subfloat[CIFAR-100]{\includegraphics[width=0.250\linewidth]{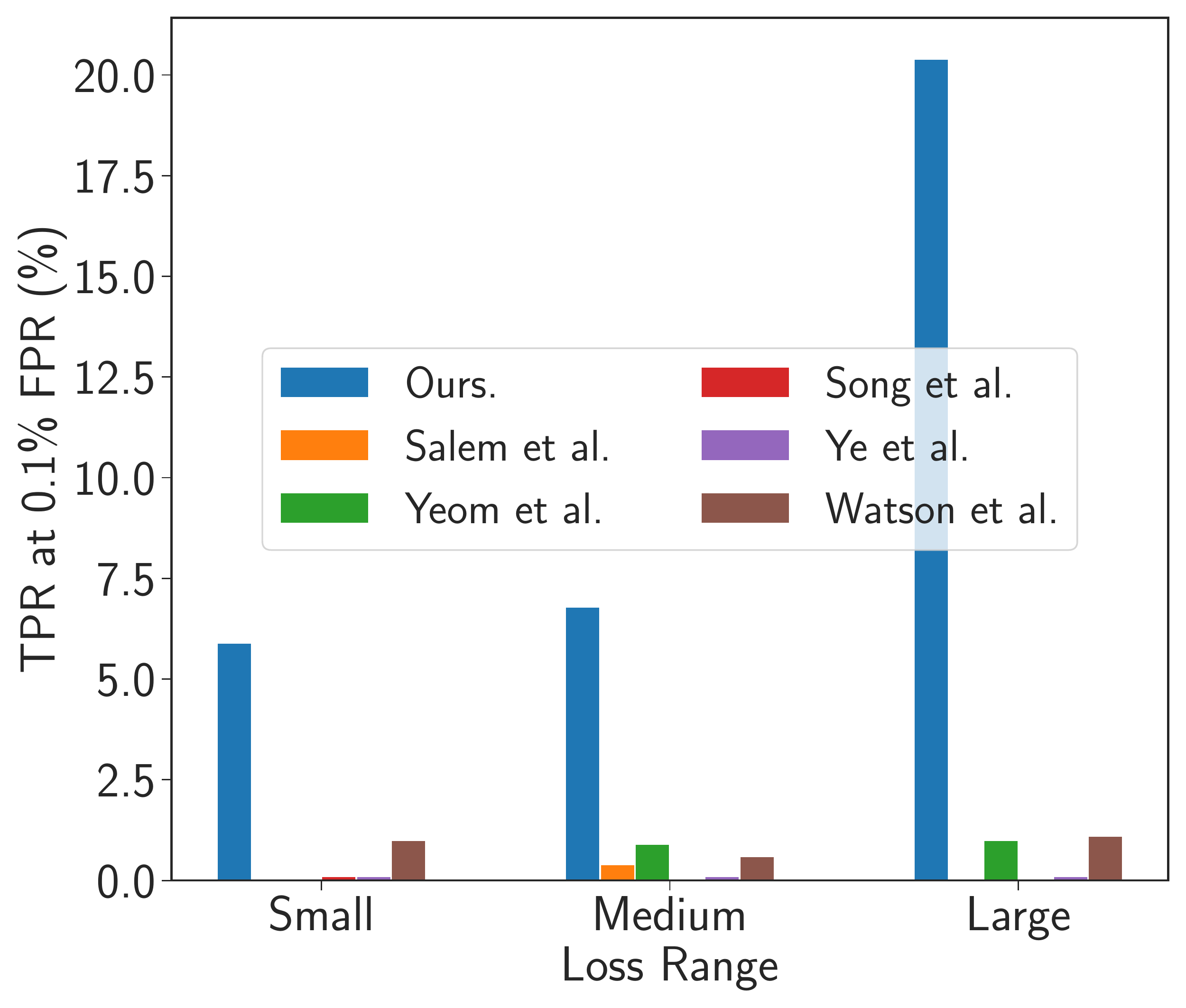}\label{cifar100_at_each_range_mobilenet}}
\subfloat[GTSRB]{\includegraphics[width=0.250\linewidth]{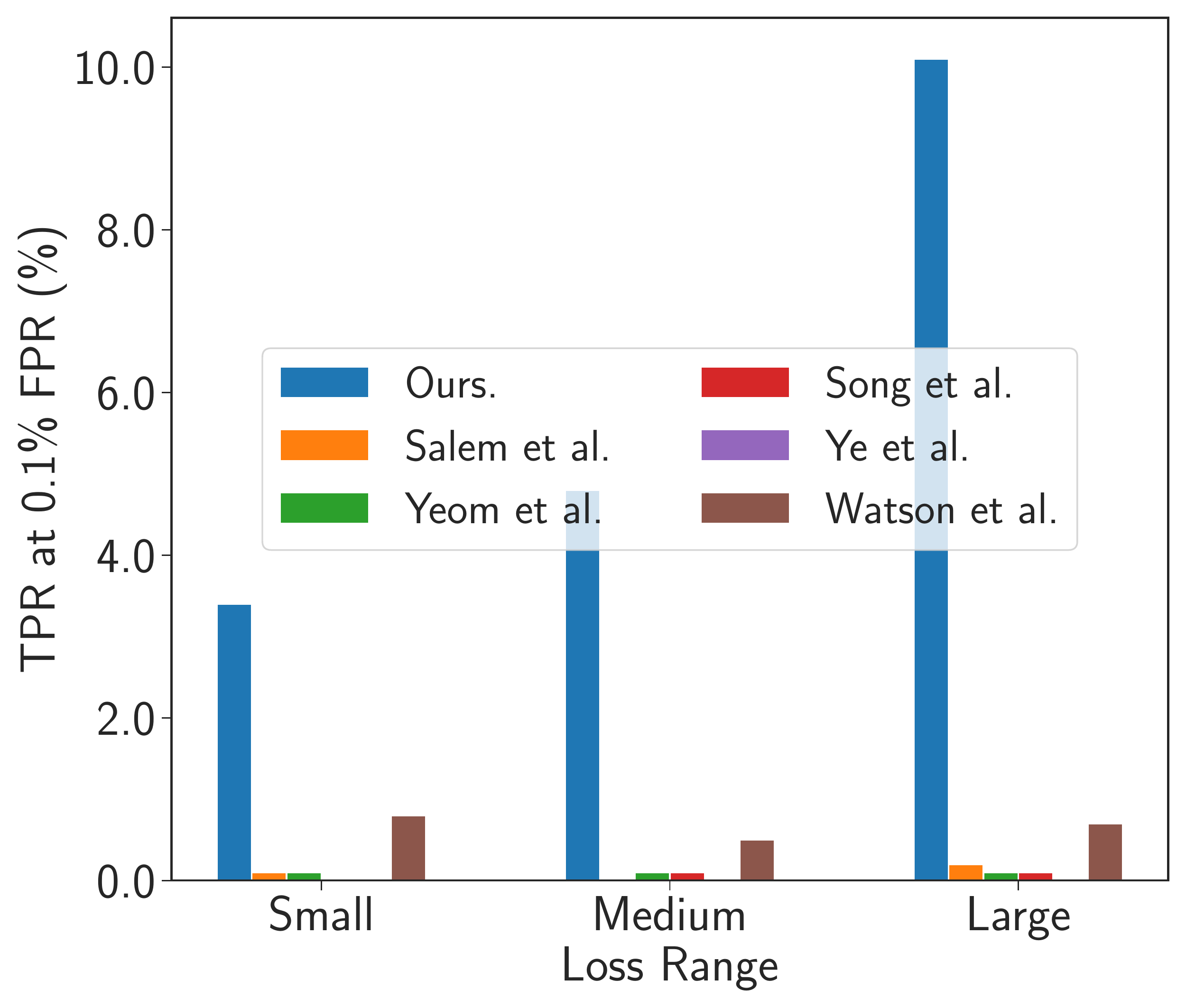}\label{gtsrb_at_each_range_mobilenet}}
\caption{TPR at 0.1\% FPR of different attacks for MobileNetV2 trained on four datasets for samples with different losses obtained from the target model. 
Here we consider three loss ranges: 'small' [0.0,0.02), 'medium' [0.02,0.2), and 'large' [0.2,$+\infty$].}
\label{fig:mobilenet_tpr_at_each_range}
\end{figure*}

\begin{figure*}[t]
\centering
\subfloat[CIFAR-10]{\includegraphics[width=0.250\linewidth]{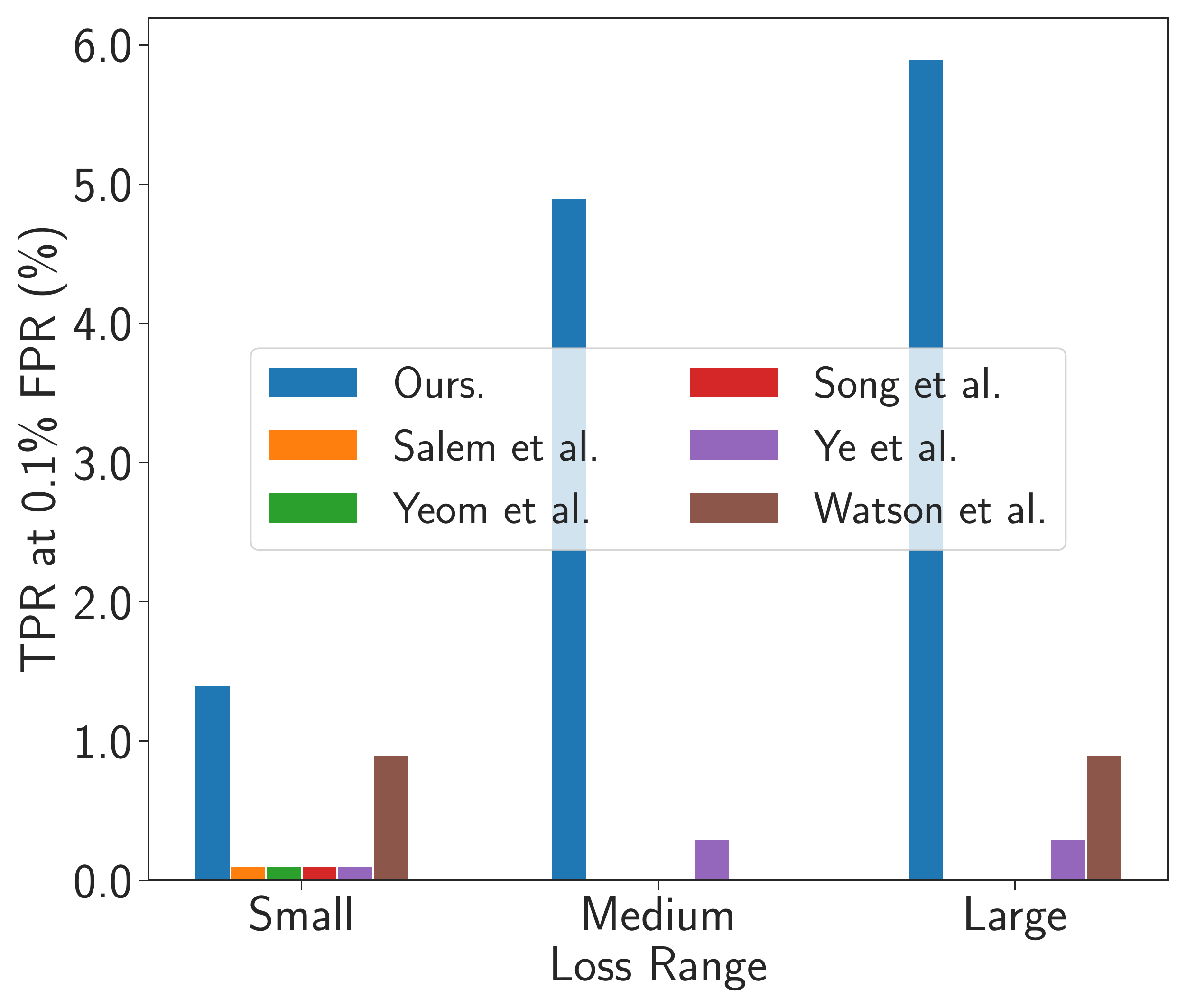}\label{cifar10_tpr_at_each_range_vgg}}
\subfloat[CINIC-10]{\includegraphics[width=0.250\linewidth]{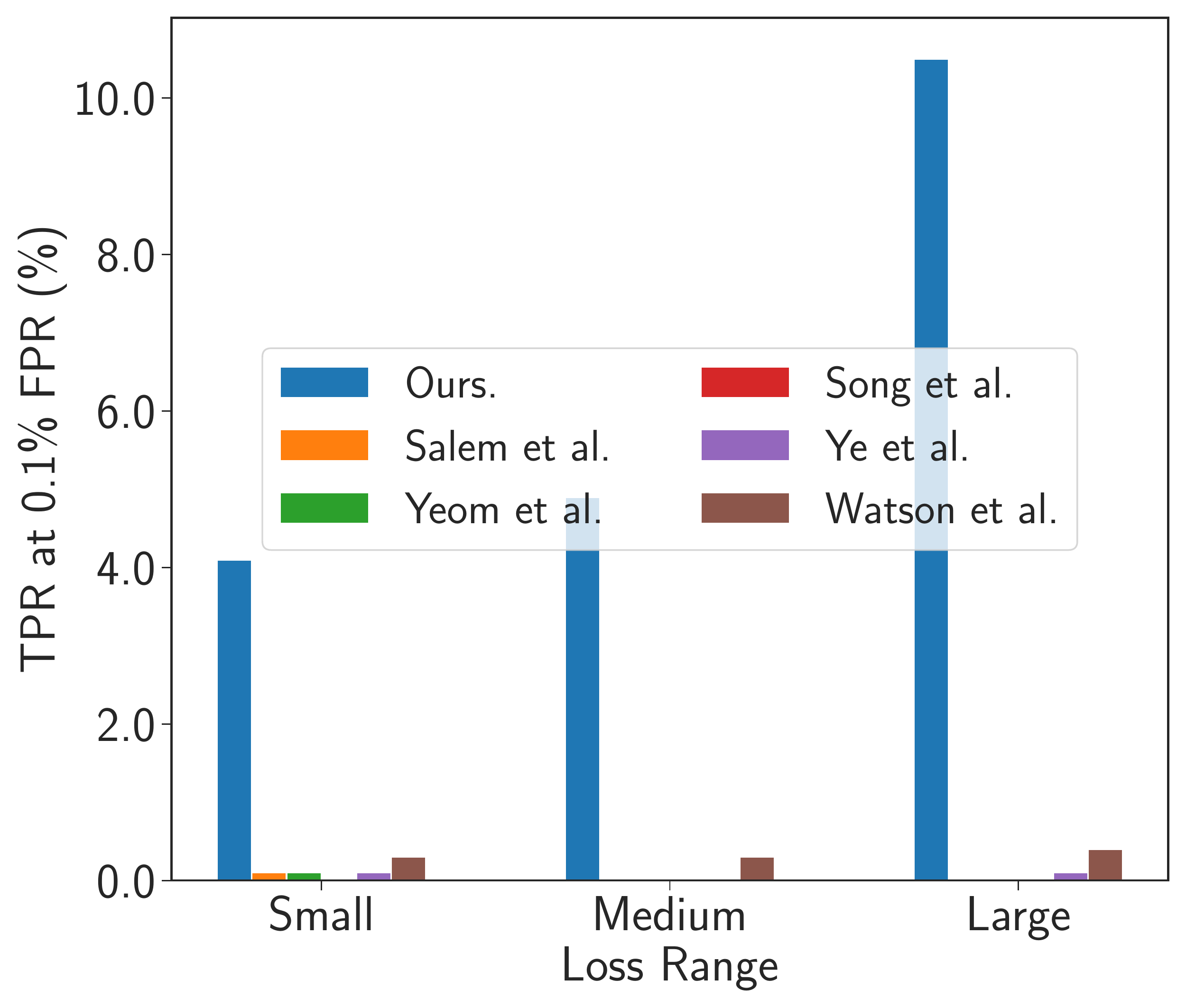}\label{cinic10_tpr_at_each_range_vgg}}
\subfloat[CIFAR-100]{\includegraphics[width=0.250\linewidth]{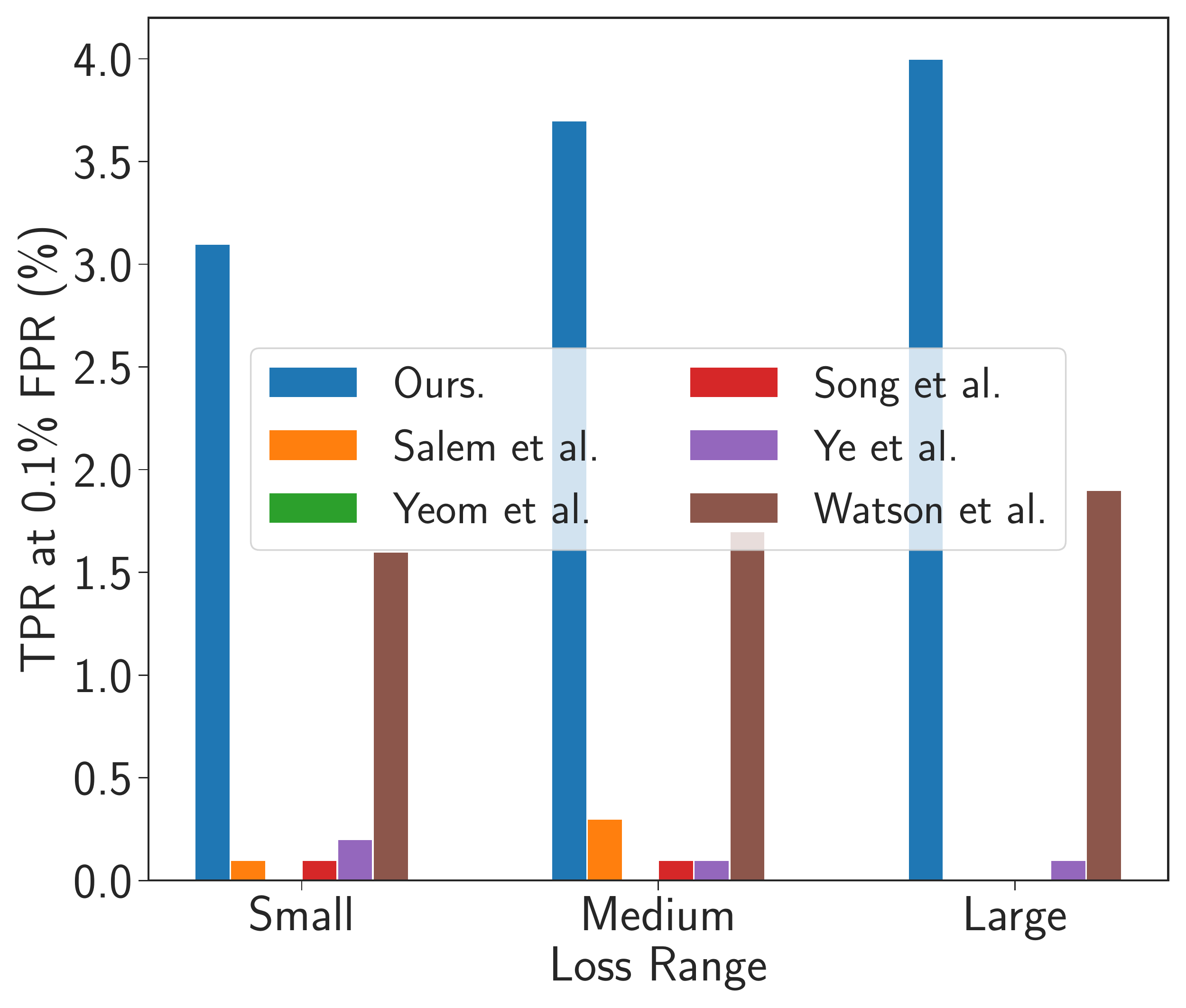}\label{cifar100_at_each_range_vgg}}
\subfloat[GTSRB]{\includegraphics[width=0.250\linewidth]{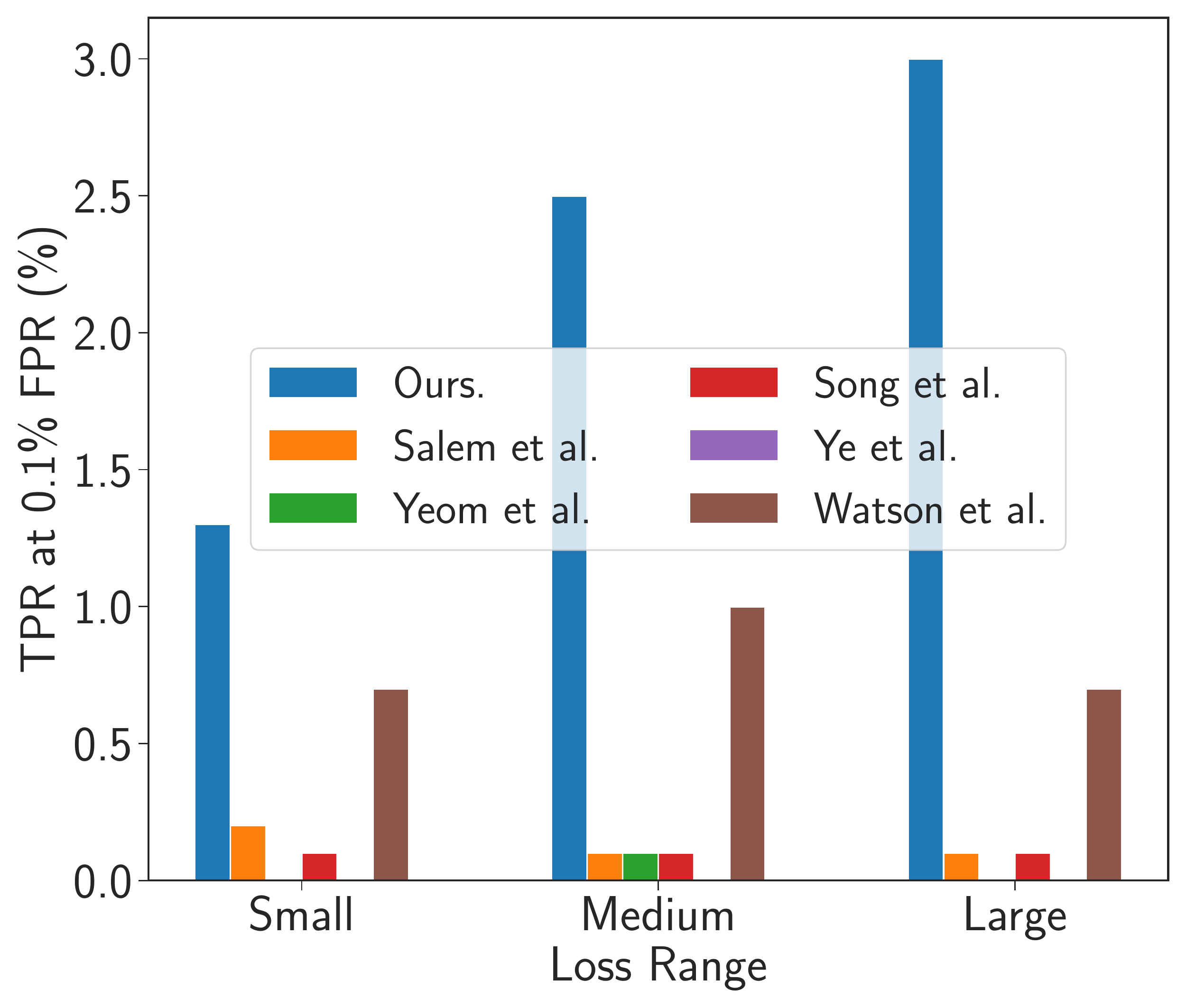}\label{gtsrb_at_each_range_vgg}}
\caption{TPR at 0.1\% FPR of different attacks for VGG-16 trained on four datasets for samples with different losses obtained from the target model. Here we consider three loss ranges: 'small' [0.0,0.02), 'medium' [0.02,0.2), and 'large' [0.2,$+\infty$].}
\label{fig:vgg_tpr_at_each_range}
\end{figure*}

\begin{figure*}[t]
\centering
\subfloat[CIFAR-10]{\includegraphics[width=0.250\linewidth]{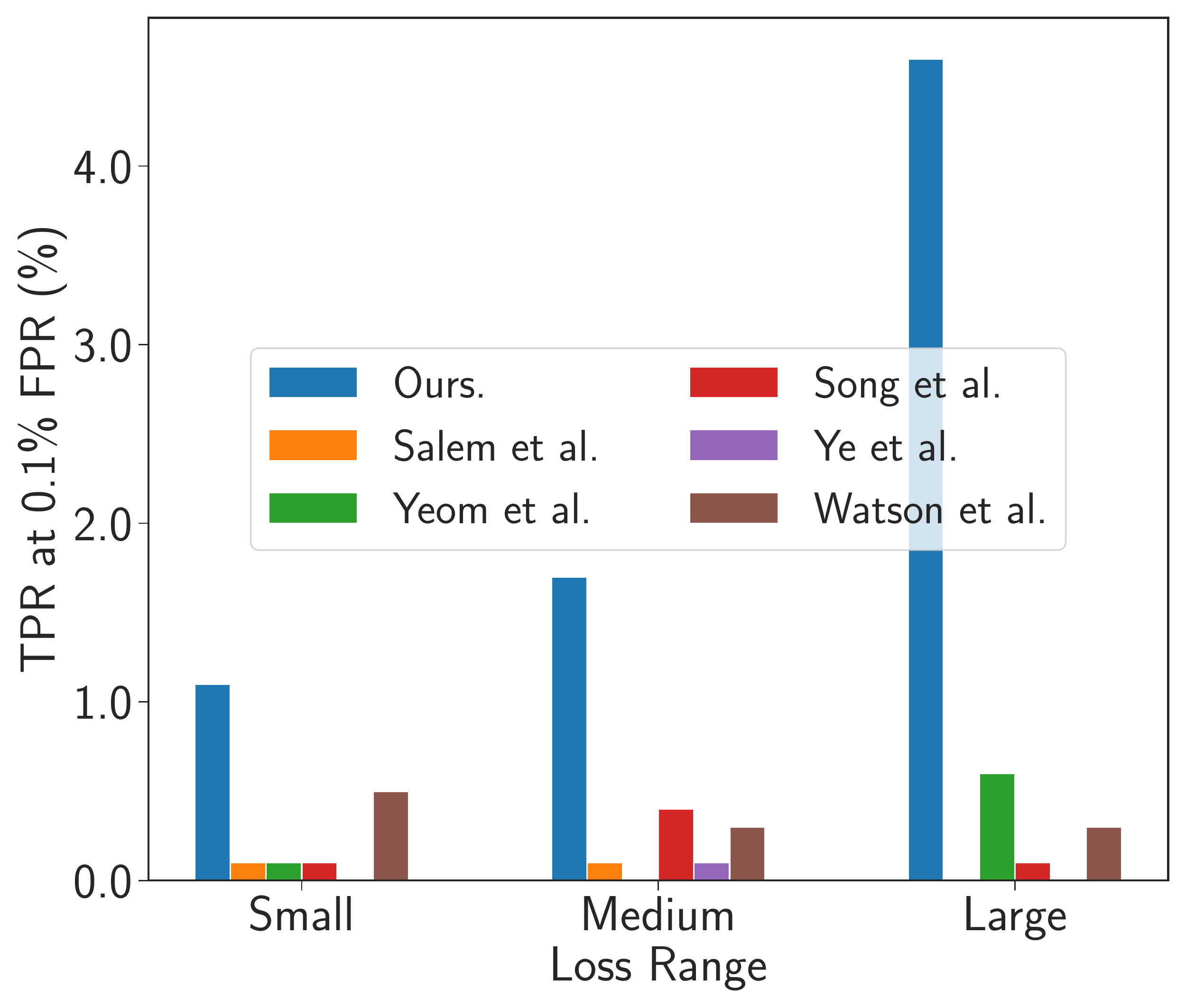}\label{cifar10_tpr_at_each_range_wideresnet}}
\subfloat[CINIC-10]{\includegraphics[width=0.250\linewidth]{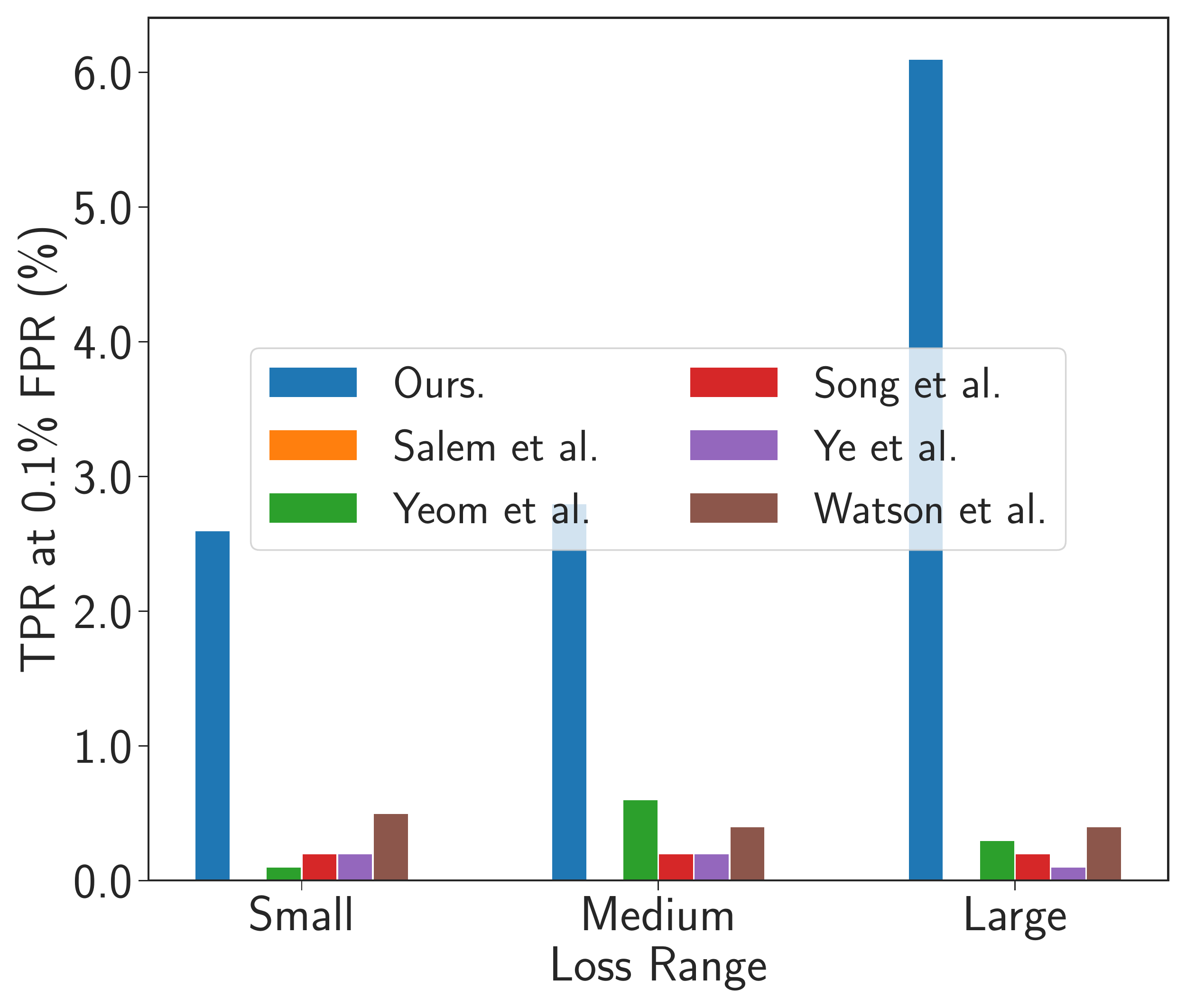}\label{cinic10_tpr_at_each_range_wideresnet}}
\subfloat[CIFAR-100]{\includegraphics[width=0.250\linewidth]{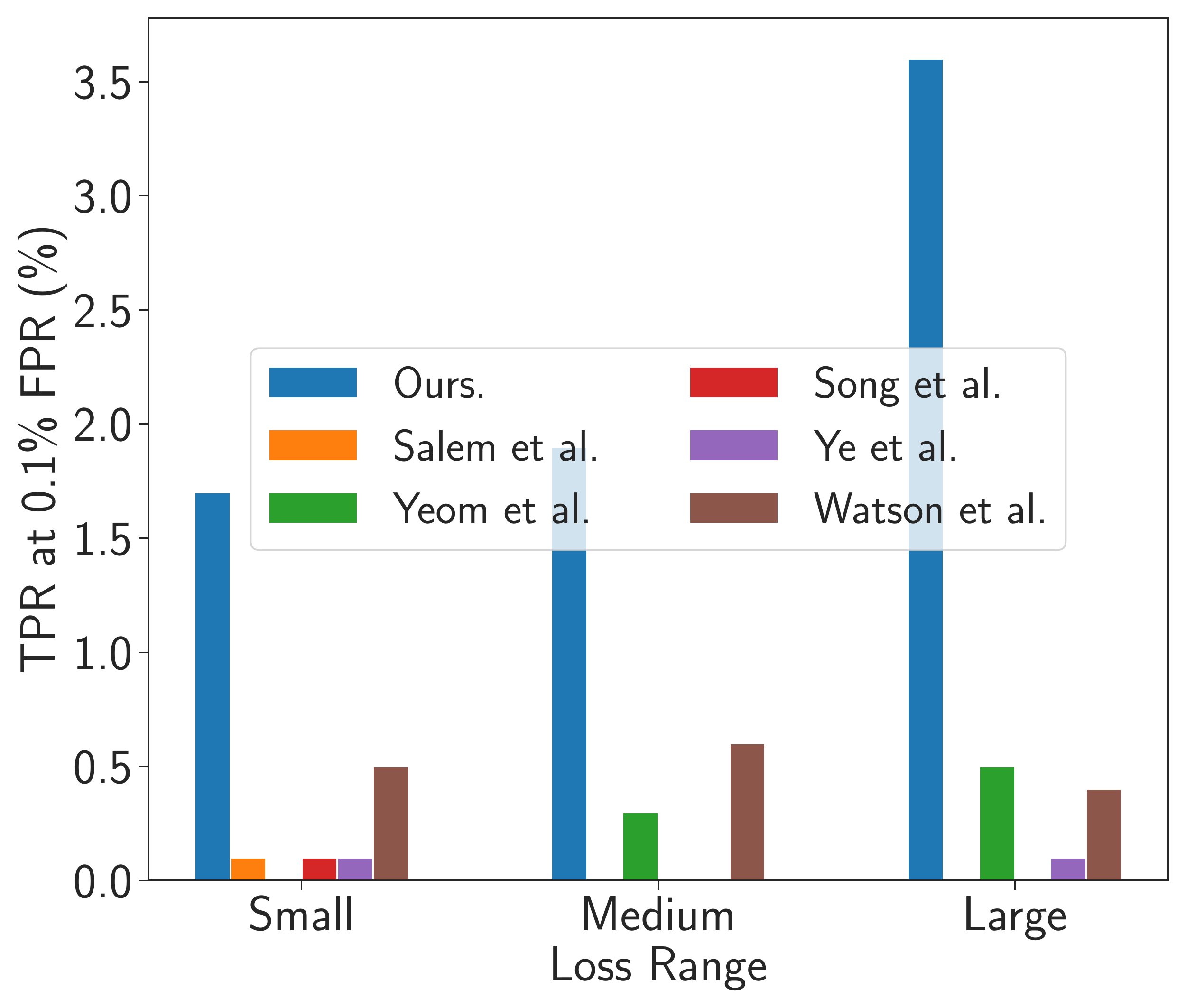}\label{cifar100_at_each_range_wideresnet}}
\subfloat[GTSRB]{\includegraphics[width=0.250\linewidth]{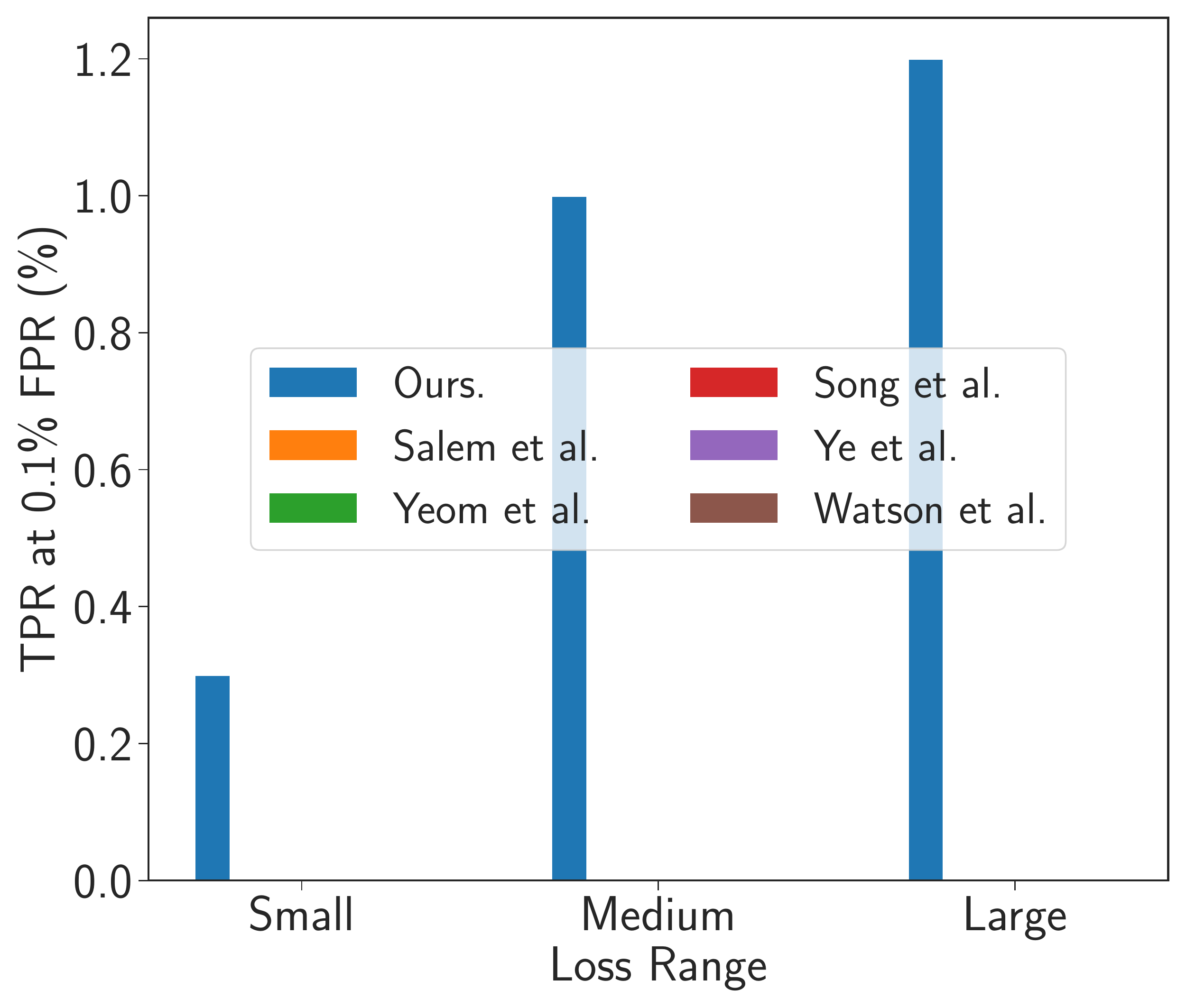}\label{gtsrb_at_each_range_wideresnet}}
\caption{TPR at 0.1\% FPR of different attacks for WideResNet-32 trained on four datasets for samples with different losses obtained from the target model. 
Here we consider three loss ranges: 'small' [0.0,0.02), 'medium' [0.02,0.2), and 'large' [0.2,$+\infty$].}
\label{fig:wideresnet_tpr_at_each_range}
\end{figure*}

\clearpage

\begin{minipage}{1.0\textwidth}
\subsection{Attack Results on Purchase, Location, and News Datasets}
\label{app:NLP}
\centering
\captionof{table}{Attack performance of different attacks against MLP trained on Purchase, Location and News.}
\label{table:attack_performance_on_MLP_for_additional_dataset}
\setlength{\tabcolsep}{4.0pt}
\scalebox{0.75}{
\begin{tabular}{l|cccccccccccc}
\toprule
\rowcolor{white}
Attack & \multicolumn{3}{c}{TPR at 0.1\% FPR}&  \multicolumn{3}{c}{Balanced accuracy}& \multicolumn{3}{c}{AUC}\\
\cmidrule(l{5pt}r{5pt}){2-4}\cmidrule(l{5pt}r{5pt}){5-7}\cmidrule(l{5pt}r{0pt}){8-10}
method& Purchase& Location& News& Purchase& Location& News& Purchase& Location& News\\
\midrule
Salem et al.~\cite{SZHBFB19}&0.2\%&0.3\%&1.3\%&0.666&0.753&0.746&0.711&0.820&0.790\\
Yeom et al.~\cite{YGFJ18}&0.1\%&0.8\%&1.5\%&0.790&0.869&0.764&0.802&0.917&0.790\\
Song et al.~\cite{SM21}&0.1\%&0.1\%&0.3\%&0.785&0.849&0.759&0.804&0.918&0.791\\
Ye et al.~\cite{YMMS21}&0.1\%&0.3\%&0.2\%&0.594&0.596&0.551&0.615&0.606&0.565\\
Watson et al.~\cite{WGCS21}&0.3\%&3.2\%&1.0\%&0.764&0.778&0.743&0.797&0.853&0.780\\
\midrule
Ours&\textbf{3.4\%}&\textbf{4.9\%}&\textbf{2.5\%}&\textbf{0.803}&\textbf{0.902}&\textbf{0.824}&\textbf{0.881}&\textbf{0.956}&\textbf{0.899}\\
\bottomrule
\end{tabular}
}
\end{minipage}

\begin{figure*}[t]
\centering
\subfloat[Purchase]{\includegraphics[width=0.3\linewidth]{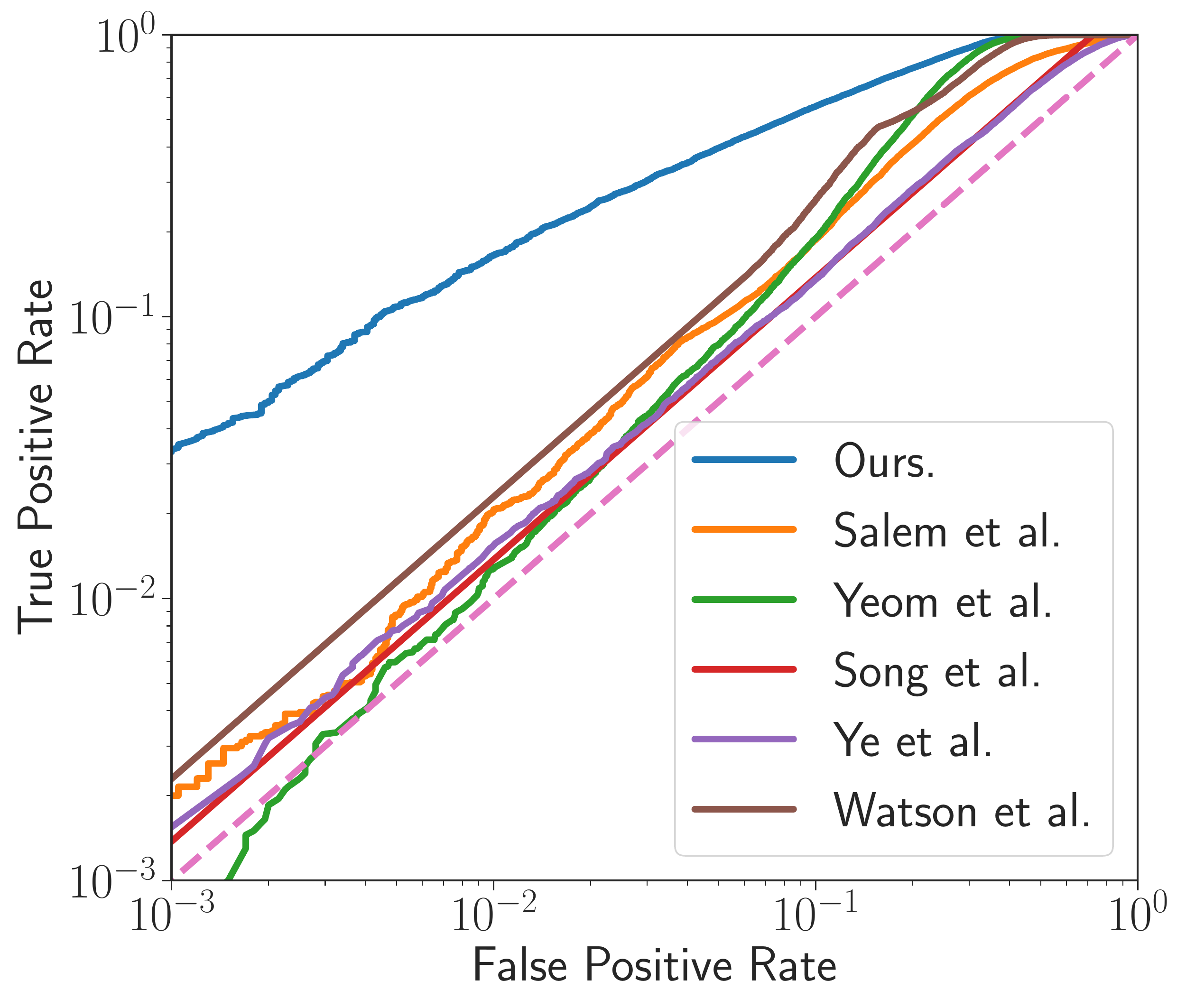}}
\subfloat[Location]{\includegraphics[width=0.3\linewidth]{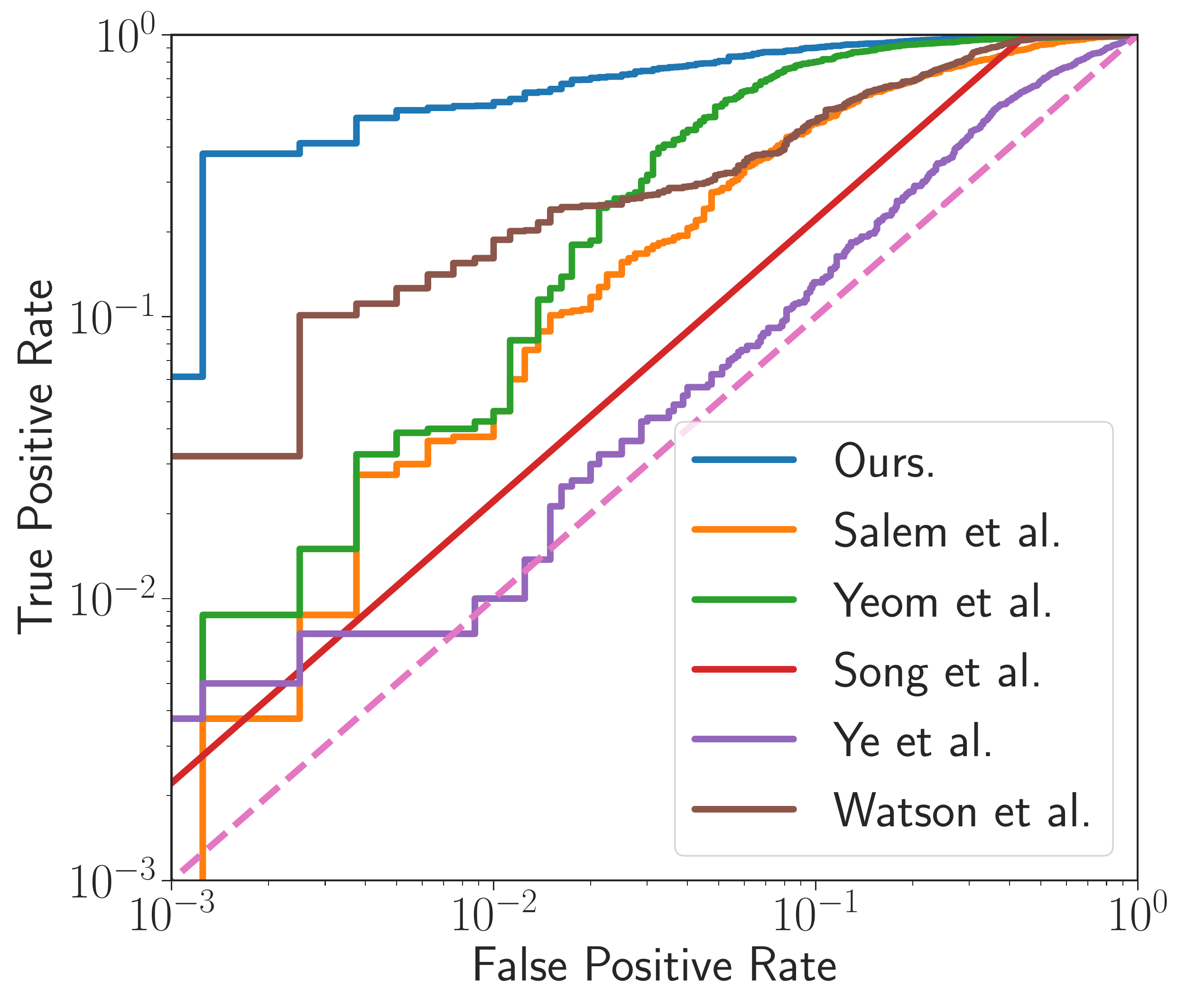}}
\subfloat[News]{\includegraphics[width=0.3\linewidth]{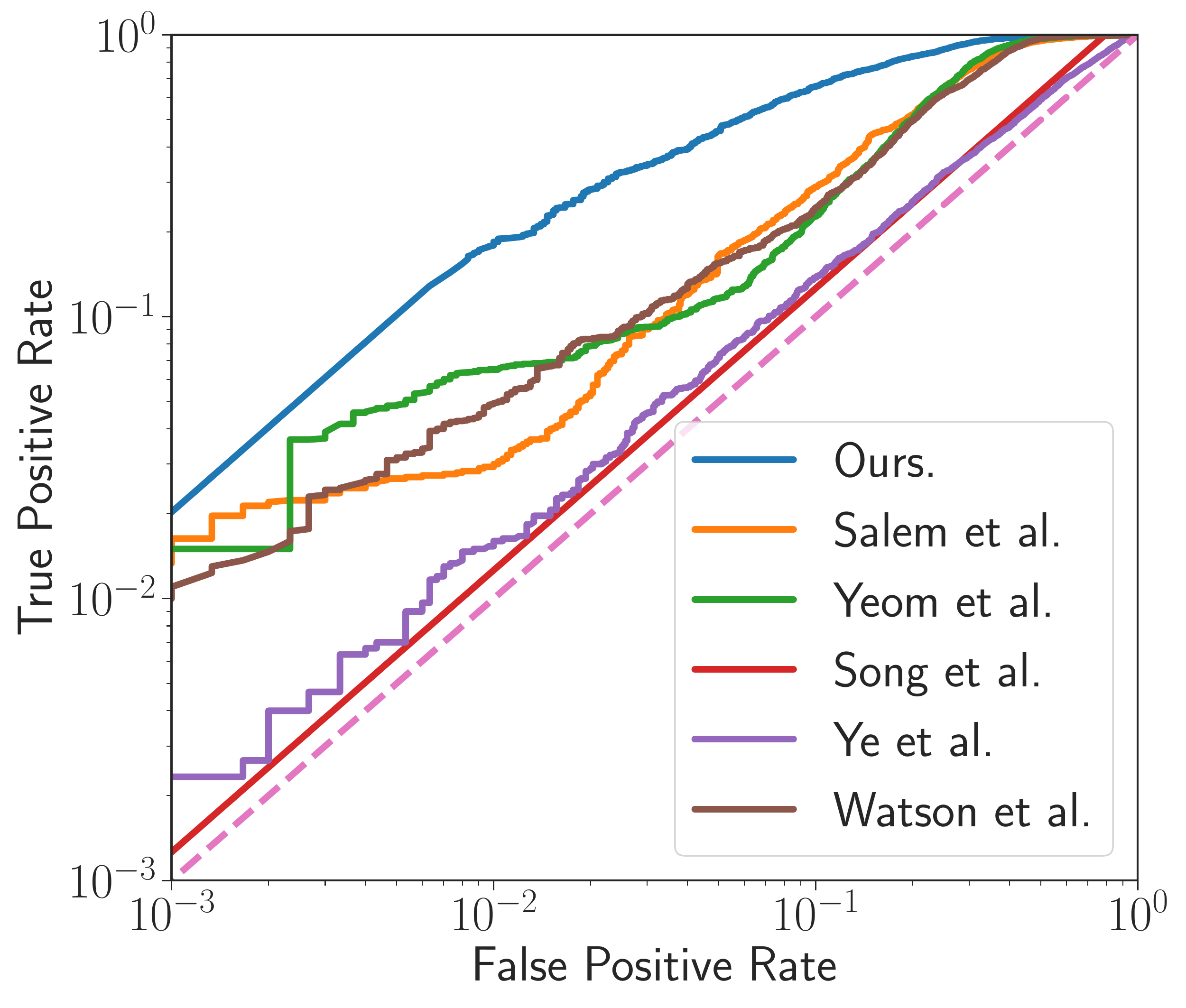}}
\caption{ROC curves for attacks against MLP on Purchase, Location and News.}
\label{fig:log_auc_mlp}
\end{figure*}

\begin{figure*}[t]
\centering
\subfloat[Purchase]{\includegraphics[width=0.3\linewidth]{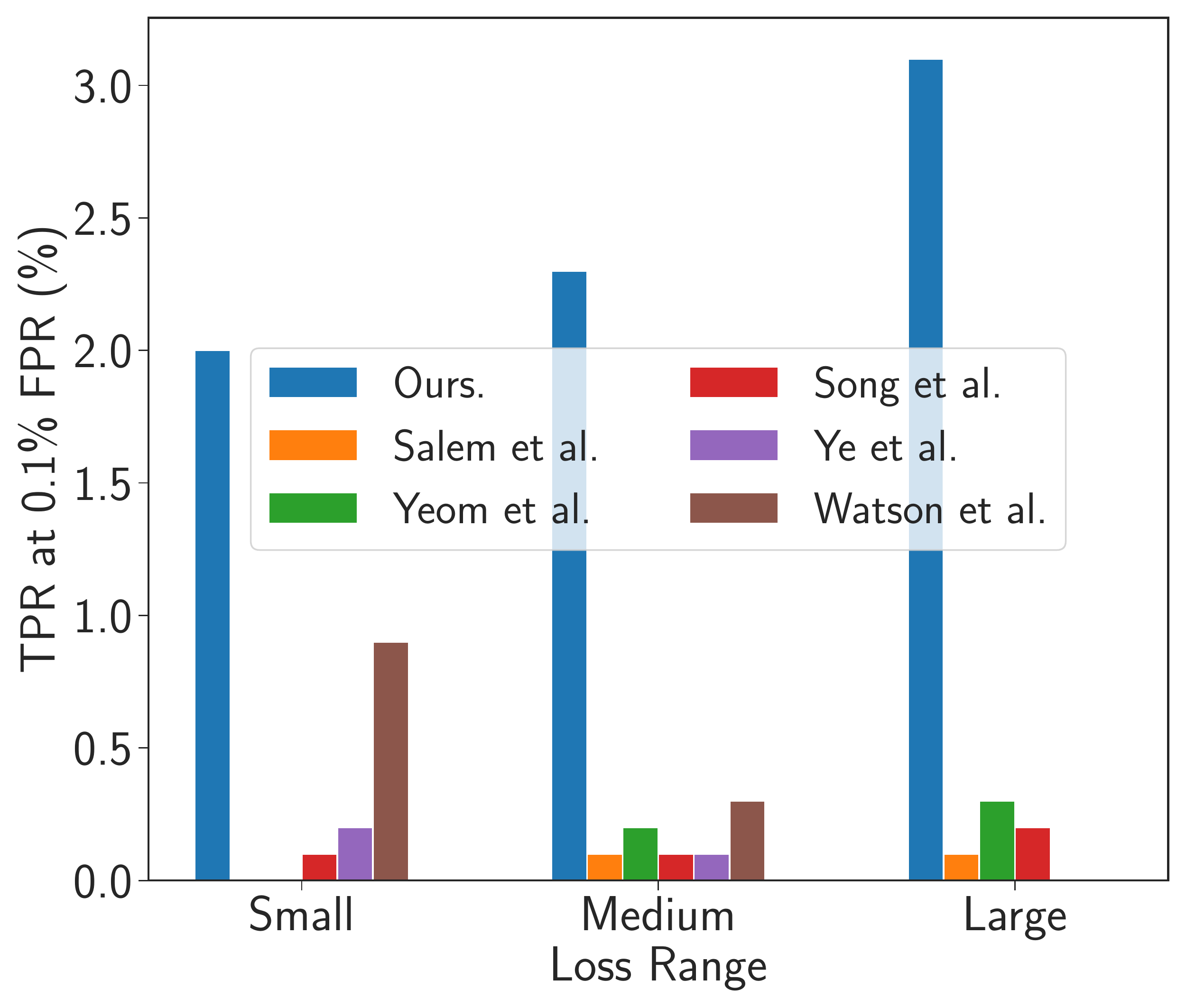}}
\subfloat[Location]{\includegraphics[width=0.3\linewidth]{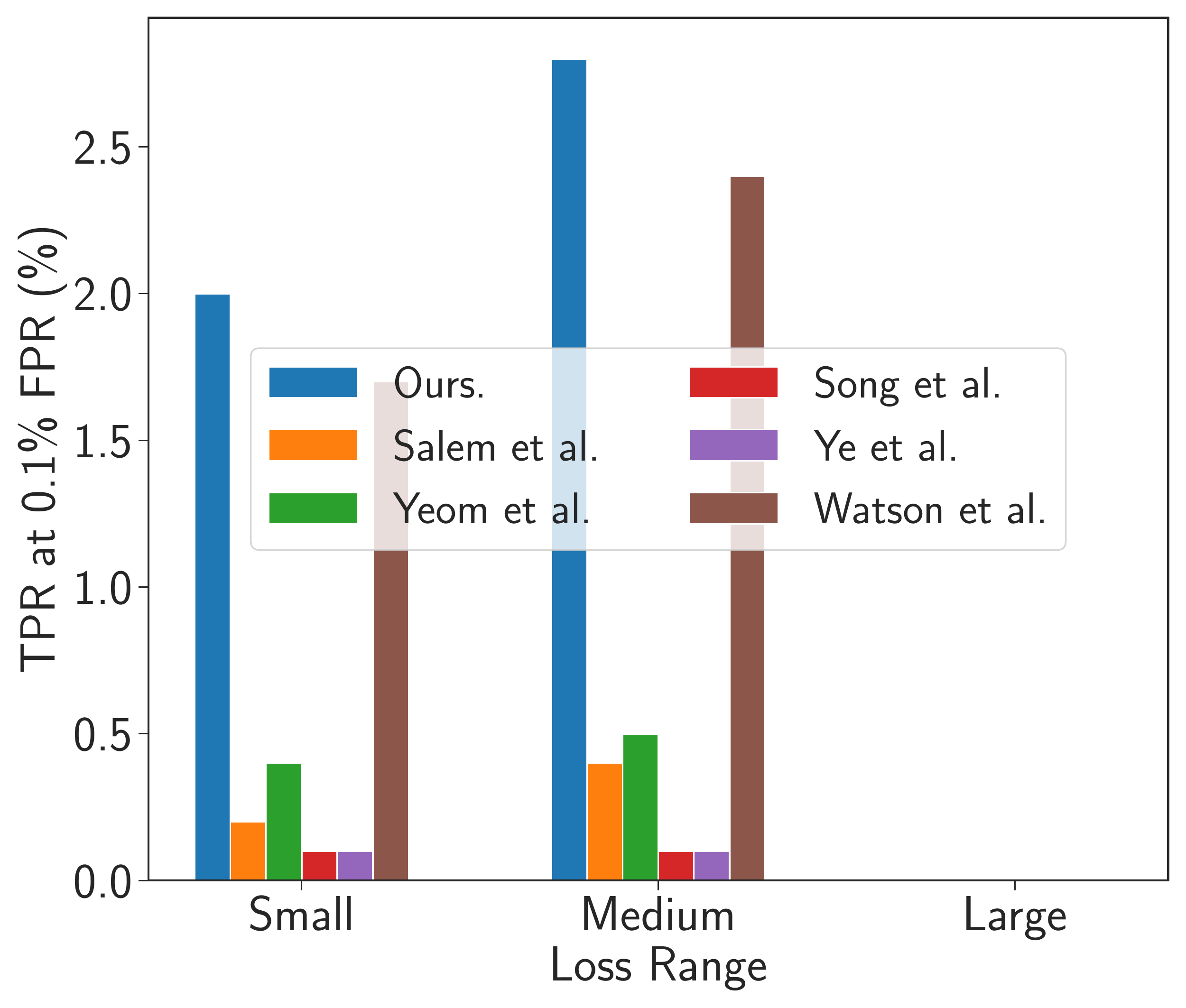}}
\subfloat[CIFAR-100]{\includegraphics[width=0.3\linewidth]{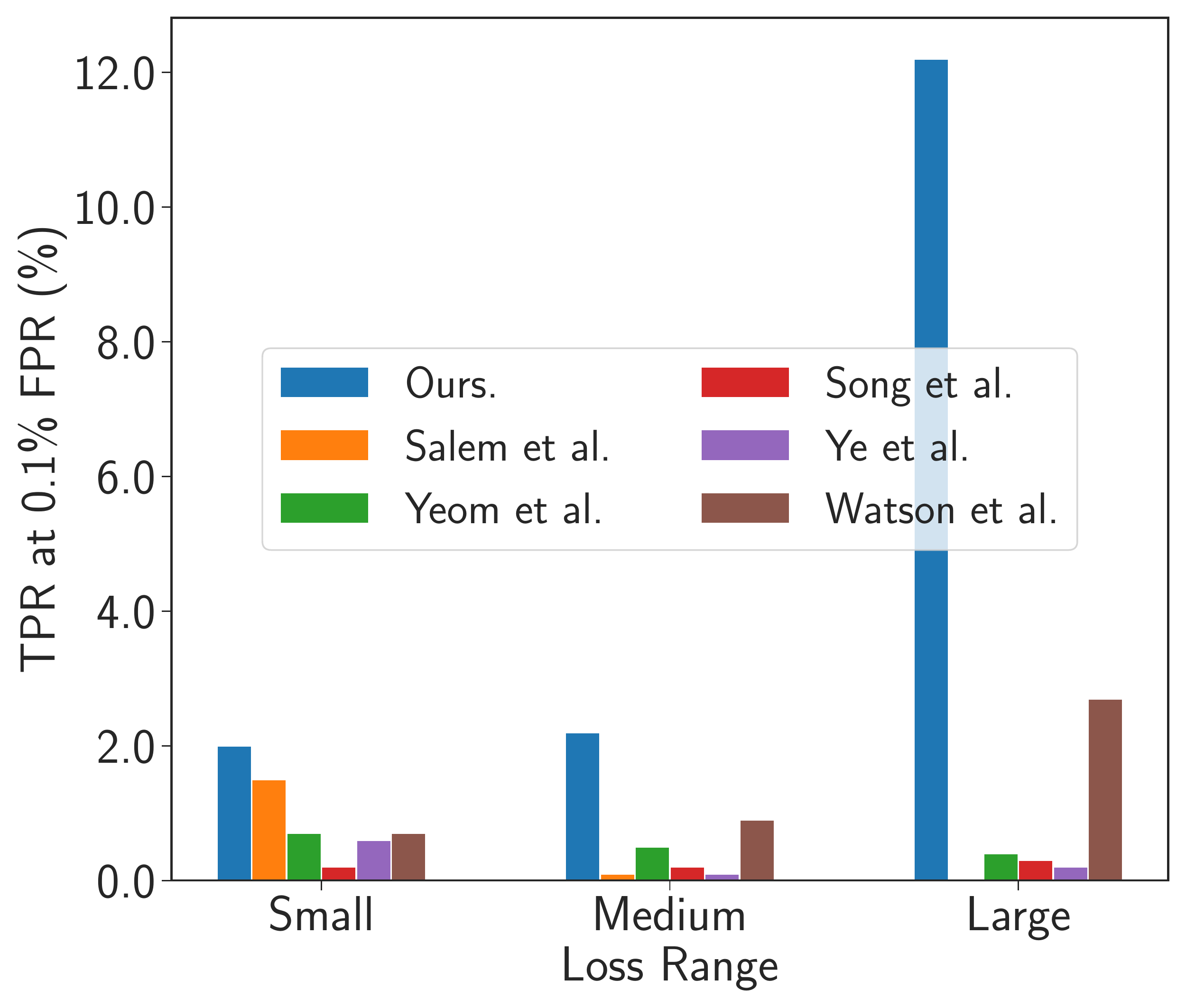}}
\caption{TPR at 0.1\% FPR of different attacks for MLP trained on three datasets (Purchase, Location, and News) for samples with different losses obtained from the target model. 
Here we consider three loss ranges: 'small' [0.0,0.02), 'medium' [0.02,0.2), and 'large' [0.2,$+\infty$].}
\label{fig:mlp_tpr_at_each_range}
\end{figure*}

\end{document}